\documentclass[a4paper,11pt]{article}
\pdfoutput=1 

\usepackage{jheppub} 

\usepackage[T1]{fontenc} 

\usepackage{graphics}
\usepackage{amssymb}
\usepackage{hyperref}
\usepackage{latexsym}
\usepackage{bm}
\usepackage{slashed}
\usepackage{datetime}
\usepackage{mciteplus}
\usepackage{multirow}
\usepackage{color}
\usepackage[utf8]{inputenc}
\usepackage[normalem]{ulem}

\def\nostrocostruttino#1\over#2{\mathrel{\mathop{\kern 0pt \rlap {\hbox{$#1$}}} \hbox{\kern-.125em $#2$}}}
\def\sumint{\nostrocostruttino \sum \over {\displaystyle\int}}
\newcommand{\be}{\begin{equation}}
\newcommand{\ee}{\end{equation}}
\newcommand{\bea}{\begin{eqnarray}}
\newcommand{\eea}{\end{eqnarray}}

\title{General helicity formalism for two-hadron production in $e^+e^-$ annihilation within a TMD approach}

\author[a,b,1]{Umberto D'Alesio,\note{Corresponding author.}}
\author[b]{Francesco Murgia,}
\author[a,b]{and Marco Zaccheddu}

\affiliation[a]{Dipartimento di Fisica, Universit\`a di Cagliari, Cittadella Universitaria, I-09042 Monserrato (CA), Italy}
\affiliation[b]{INFN, Sezione di Cagliari, Cittadella Universitaria, I-09042 Monserrato (CA), Italy}

\emailAdd{umberto.dalesio@ca.infn.it}
\emailAdd{francesco.murgia@ca.infn.it}
\emailAdd{marco.zaccheddu@ca.infn.it}

\abstract{We present the complete leading-order results for the azimuthal dependences and polarization observables in $e^+e^-\to h_1 h_2 + X$ processes, where the two hadrons are produced almost back-to-back, within a transverse momentum dependent (TMD) factorization scheme. We consider spinless (or unpolarized) and spin-1/2 hadron production and give the full set of the corresponding quark and gluon TMD fragmentation functions (TMD-FFs). By adopting the helicity formalism, which allows for a more direct probabilistic interpretation, single- and double-polarization cases are discussed in detail.  
Simplified expressions, useful for phenomenological analyses, are obtained by assuming a factorized Gaussian-like dependence on intrinsic transverse momenta for the TMD-FFs.}

\begin{document}
\maketitle
\flushbottom

\section{Introduction}
\label{intro}

The comprehension of the internal structure of nucleons in terms of partons, quarks and gluons, as well as of the partonic fragmentation mechanism into hadrons represents a key issue in hadron and, more generally, particle physics. It is now well consolidated that when we access observables sensitive to the full 3-dimensional (3D) structure of hadrons we can reach a more complete understanding of their fundamental properties, unattainable within the standard 1-dimensional leading-twist (LT) collinear formulation. If also parton and/or hadron spins are involved the information we can gather is definitely much richer, allowing to reveal important spin-momentum correlations and parton orbital angular momentum effects. In this context transverse momentum dependent parton distribution (TMD-PDFs) and fragmentation (TMD-FFs) functions, collectively referred to as TMDs, have been successfully introduced, see Refs.~\cite{DAlesio:2007bjf, Barone:2010zz, Aidala:2012mv, Anselmino:2020vlp} for a full account of the TMD approach. These distributions encode the 3D nucleon spin structure in momentum space and play a fundamental role in many unpolarized as well as spin-dependent observables.

From the phenomenological point of view, the most studied TMD-PDFs are certainly the unpolarized and the Sivers distribution functions~\cite{Sivers:1989cc,Sivers:1990fh}, the latter giving the asymmetric azimuthal distribution of unpolarized partons inside a fast-moving transversely polarized nucleon.
In the fragmentation sector a similar level of accuracy has been obtained for the unpolarized and the Collins fragmentation functions~\cite{Collins:1992kk}, the last one giving the asymmetric azimuthal distribution of unpolarized hadrons produced in the fragmentation of a transversely polarized quark.

The last years have also witnessed significant progress in the formulation of factorization theorems in terms of TMDs for a well defined class of processes~\cite{Ji:2004xq, Ji:2004wu, Collins:2011zzd, GarciaEchevarria:2011rb}, all characterized by the presence of two strongly ordered energy scales: namely, semi-inclusive deep inelastic scattering (SIDIS), Drell-Yan (DY) and $e^+e^-$ annihilation processes, where the two scales are the virtuality of the exchanged boson (the larger one) and the transverse momentum of the final hadron (SIDIS), of the lepton-pair (DY) or of the hadron-pair in $e^+e^-$ collisions (the smaller one).

The complete structure of azimuthal dependences for particle production in (un)polarized collisions, within a TMD approach, specifically for SIDIS~\cite{Mulders:1995dh, Bacchetta:2006tn, Anselmino:2011ch}, DY~\cite{Arnold:2008kf} processes and hadron-pair production in $e^+e^-$ collisions~\cite{Boer:1997mf} has been formulated in full detail. In particular, in Ref.~\cite{Anselmino:2011ch} the full decomposition in terms of TMDs was obtained for the process $\ell p \to \ell' h \, X$ adopting the helicity formalism. This, indeed, allows for a direct probabilistic interpretation within a more intuitive picture, which splits the physical process in three phases, each of them described by the corresponding helicity amplitude: the inclusive emission of a parton by the initial hadron $p \to q\,X$, the perturbatively calculable hard interaction $\ell q \to \ell' q'$, and the emission of the final hadron by the outgoing quark, $q' \to h \, X$.  We recall that in this study only spinless or unpolarized hadrons were considered.
A somehow pioneering work in this direction, even if in the context of a process for which TMD factorization is assumed only as a phenomenological ansatz, was presented in Ref.~\cite{Anselmino:2005sh}. There, the focus was on transverse single-spin asymmetries for inclusive particle production in $pp$ collisions, and, within the helicity formalism, all leading-twist TMD-PDFs, for quark and gluons, were considered together with the two TMD-FFs for spinless (or unpolarized) hadrons.

A striking feature of SIDIS processes, which have played and still play a leading role in accessing TMDs, is the strong entanglement between the transverse momentum dependence arising from the TMD-PDFs and TMD-FFs in the resulting and measurable transverse momentum of the final hadron~\cite{Anselmino:2018psi}. In this respect DY and $e^+e^-$ processes are of fundamental importance, being sensitive separately to TMD-PDFs and TMD-FFs, respectively.

Here we apply the helicity formalism to $e^+e^-\to h_1 h_2 \, X$ processes and derive the complete expressions for all leading-twist azimuthal dependences and polarization observables at leading order, adopting two commonly used reference frames. The case of single-hadron production within a jet, as pointed out in a series of recent papers \cite{Gamberg:2018fwy, Kang:2020yqw, Boglione:2020auc, Boglione:2020cwn}, requires a more dedicated study and it will be  addressed in a future work.

We will present the complete set of all leading-twist TMD-FFs for spinless (or unpolarized) and spin-1/2 hadrons both for quarks and gluons, discussing their main properties and their physical meaning. Moreover, by adopting a factorized transverse momentum dependence in terms of Gaussian-like TMD-FFs we derive simplified expressions for all single- and double-polarized observables, useful in phenomenological analyses. The results are in perfect agreement with those of Ref.~\cite{Boer:1997mf}.

The motivations behind this study are several: from the theory point of view, it represents the extension to spin-1/2 hadrons of the helicity formalism and the classification of all leading-twist quark and gluon TMD-FFs, focusing, for quarks, on their role in polarized hadron production in $e^+ e^-$ collisions; on the phenomenology side, it timely matches the renewed interest, triggered by recent data from the Belle Collaboration on the transverse $\Lambda$ polarization~\cite{Guan:2018ckx}, in the associate hadron production in $e^+e^-$ annihilation processes. In this context, indeed, the first phenomenological analysis within a TMD framework has been carried out~\cite{DAlesio:2020wjq} (see also Ref.~\cite{Callos:2020qtu} for a similar study).

The paper is organized as follows: in Section~\ref{2hadrons} we present the main formulae for the computation within the helicity formalism of the azimuthal dependences for the production of (un)polarized hadron-pairs in $e^+ e^-$ collisions. In Section~\ref{softq} we give the complete set of quark TMD-FFs for spin-1/2 hadrons at leading twist, while in Sec.~\ref{azimdep} we collect the explicit expressions for unpolarized, single- and double-polarized hadron production. In Sec.~\ref{had-conv} we present all convolutions in terms of the TMD-FFs for a specific kinematical configuration; in particular, in Sec.~\ref{gauss} we give the corresponding explicit analytical formulae assuming a factorized Gaussian dependence for the TMD-FFs. Our conclusions are gathered in Sec.~\ref{concl}. Useful results are derived and collected in the Appendices; in particular, in Appendix~\ref{hel-ampl} we give the main properties of the helicity fragmentation amplitudes, in Appendix~\ref{g-FF} we present the complete set of gluon TMD-FFs for spin-1/2 hadrons and in Appendix~\ref{Amst} the relation with other common notations is worked out. In Appendix~\ref{tensor} we discuss the tensorial analysis adopted to extract the observed azimuthal dependences. Useful relations and definitions on the helicity frames are given in Appendix~\ref{hel-frames}.

\section{Production of two hadrons in $e^+e^-$ annihilation}
\label{2hadrons}

We consider the production of two almost back-to-back hadrons in the $e^+ e^- \rightarrow h_1(S_1)\, h_2(S_2)
+ X$ process, where $S_{1,2}$ are the hadron spins, within a TMD approach and adopting the helicity formalism. Since we will be also interested in heavy baryon, like $\Lambda$'s, production, we will pay special attention to hadron mass effects. Two reference frames are usually adopted for such a study: the ``thrust frame'', where one identifies the direction of the two opposite jets and measure the corresponding azimuthal distributions of the two hadrons within the jets; the ``hadron frame'', where one measures only the momenta of the two hadrons and the azimuthal distribution of one hadron with respect to the other.

\begin{figure}
\centering
\includegraphics[width=12cm]{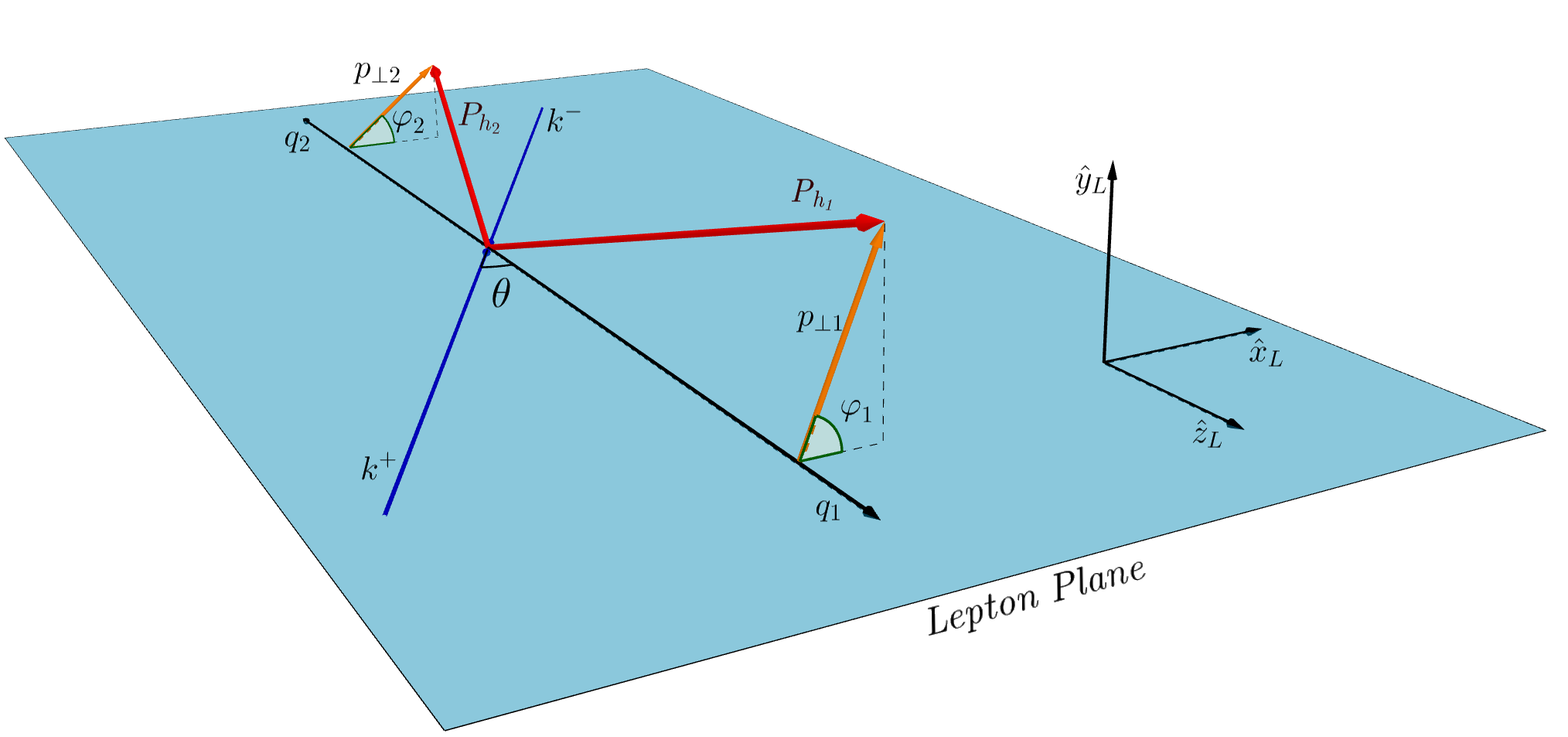}
    \caption{Kinematics for the the thrust-frame configuration.}
    \label{fig:kin1}
\end{figure}

More precisely, the thrust frame is chosen so that the $e^+e^-\to q \, \bar q$ scattering occurs in the $\widehat{xz}$ plane, with the back-to-back quark and antiquark moving along the $\hat{z}$-axis. This choice requires, experimentally, the reconstruction of the jet thrust axis, but it involves a very simple kinematics.
In the configuration of Fig.~\ref{fig:kin1}, the four-momenta of the $e^+, e^-$ leptons (respectively $k^+, k^-$) and of the $q, \bar q$ pair ($q_1, q_2$ respectively) are
\bea
q_1 = \frac{\sqrt{s}}{2}(1,0,0,1) &\quad\quad&
q_2 = \frac{\sqrt{s}}{2}(1,0,0,-1)\\
k^- = \frac{\sqrt{s}}{2}(1,-\sin\theta,0,\cos\theta) &\quad\quad&
k^+ = \frac{\sqrt{s}}{2}(1,\sin\theta,0,-\cos\theta) \label{kpkm}\>,
\eea
where $s$ is the center-of-mass energy squared.
These define the lepton plane. We will use the generic notation $\hat{\bm{x}}_L$, $\hat{\bm{y}}_L$, $\hat{\bm{z}}_L$ for the axes in this frame.

The final hadrons $h_1$ and $h_2$ carry light-cone momentum fractions $z_1$ and $z_2$ and have intrinsic transverse momenta $\bm{p}_{\perp 1}$ and $\bm{p}_{\perp 2}$ with respect to the direction of the corresponding fragmenting quarks,
\be
\bm{p}_{\perp 1} = p_{\perp 1}(\cos\varphi_1,\sin\varphi_1,0) \quad\quad\quad
\bm{p}_{\perp 2} = p_{\perp 2}(\cos\varphi_2,\sin\varphi_2,0)\,,
\ee
with $p_\perp= |\bm{p}_\perp|$, so that their four-momenta can be expressed as
\bea
&& P_{h_1}=\left( z_1 \frac{\sqrt{s}}{2}\Big(1+\frac{a_{h_1}^2}{z_1^2}\Big),p_{\perp 1} \cos\varphi_1, p_{\perp 1} \, \sin\varphi_1,
        \;z_1 \frac{\sqrt{s}}{2}\Big(1-\frac{a_{h_1}^2}{z_1^2}\Big) \right) \label{Ph1}\\
&& P_{h_2}=\left( z_2 \frac{\sqrt{s}}{2}\Big(1+\frac{a_{h_2}^2}{z_2^2}\Big), p_{\perp 2} \, \cos\varphi_2, p_{\perp 2} \, \sin\varphi_2,
        -z_2 \frac{\sqrt{s}}{2}\Big(1-\frac{a_{h_2}^2}{z_2^2}\Big) \right) \label{Ph2}\,,
\eea
where
\be
a^2_{h_{1,2}} = \frac{p_{\perp 1,2}^2 + M_{h_{1,2}}^2}{s} = \eta_{\perp 1,2}^2 + \frac {M_{h_{1,2}}^2}{s}\,.
\ee

At large center of mass energies and not too small values of $z$, one can keep only the lowest-order contribution in $\eta_\perp$, with the whole mass dependence, and work with the much simpler kinematics:
\bea
&& P_{h_1}\simeq \left( z_1 \frac{\sqrt{s}}{2}\Big(1+\frac{M_{h_1}^2}{z_1^2s}\Big), p_{\perp 1} \cos\varphi_1, p_{\perp 1} \, \sin\varphi_1,
        z_1 \frac{\sqrt{s}}{2}\Big(1-\frac{M_{h_1}^2}{z_1^2s}\Big) \right) \\
&& P_{h_2}\simeq \left( z_2 \frac{\sqrt{s}}{2}\Big(1+\frac{M_{h_2}^2}{z_2^2s}\Big), p_{\perp 2} \, \cos\varphi_2, p_{\perp 2} \, \sin\varphi_2,
        -z_2 \frac{\sqrt{s}}{2}\Big(1-\frac{M_{h_2}^2}{z_2^2s}\Big) \right) \,.
\eea
It is important to stress that we will keep terms in $p_\perp/\sqrt{s}$ only when this is essential to give a non zero result.

For massive hadrons, two further scaling variables are usually introduced: the energy fraction (often adopted in the experimental analyses)
\be
z_h = 2E_h/\sqrt s
\ee
and the momentum fraction
\be
z_{p} = 2 |\bm{P}_{h}|/\sqrt s\,.
\ee
These are related to the light-cone momentum fractions as follows:
\bea
z_{h} & = & z\,\Big(1 + \frac{a_h^2}{z^2}\Big) \simeq  z\,\Big(1 + \frac{M_h^2}{z^2s}\Big) \label{zh}\\
z_{p} & = & z\,\Big[\Big(1 - \frac{a_h^2}{z^2}\Big)^2 + 4 \frac{\eta_\perp^2}{z^2}\Big]^{1/2} \simeq  z\,\Big(1 - \frac{M_h^2}{z^2s}\Big) 
\label{zp}\\
z_p & = & z_h \Big( 1 - 4\,\frac{M_h^2}{z_h^2s}\Big)^{1/2} \label{zpzh}\,.
\eea
For later use we also define the following quantity:
\bea
\beta_{1,2}^2 &= & 1-4\,\frac{\eta_{\perp 1,2}^2}{z_{p_{1,2}}^2}\,.
\eea
Notice that, adopting the above variables, one can express the two hadron three-momenta as
\bea
\bm{P}_{h_1} & = & |\bm{P}_{h_1}| \bigg( \frac { p_{\perp 1} } { |\bm{P}_{h_1}| } \,\cos\varphi_1, \frac { p_{\perp 1} } { |\bm{P}_{h_1}| }\sin\varphi_1, \sqrt{ 1-\frac { p_{\perp 1}^2 } { |\bm{P}_{h_1}|^2}} \bigg) \nonumber\\
&=&
|\bm{P}_{h_1}|\bigg(  \frac { 2\eta_{\perp 1} } { z_{p_1} }\cos\varphi_1,\frac { 2\eta_{\perp 1} } { z_{p_1} } \sin\varphi_1, \beta_1 \bigg) \label{P1}\\
\bm{P}_{h_2} &=& |\bm{P}_{h_2}|\bigg( \frac{ p_{\perp 2} } {|\bm{P}_{h_2}|} \,\cos\varphi_2, \frac{ p_{\perp 2}}{|\bm{P}_{h_2}|}\sin\varphi_2, -\sqrt{1-\frac {p_{\perp 2}^2}{|\bm{P}_{h_2}|^2}} \bigg) \nonumber\\
&=&
|\bm{P}_{h_2}|\bigg(  \frac{2\eta_{\perp 2}} {z_{p_2}}\cos\varphi_2,\frac{2\eta_{\perp 2}}{z_{p_2}}\sin\varphi_2, -\beta_2 \bigg)\label{P2}\,.
\eea

\begin{figure}
\centering
\includegraphics[width=12cm]{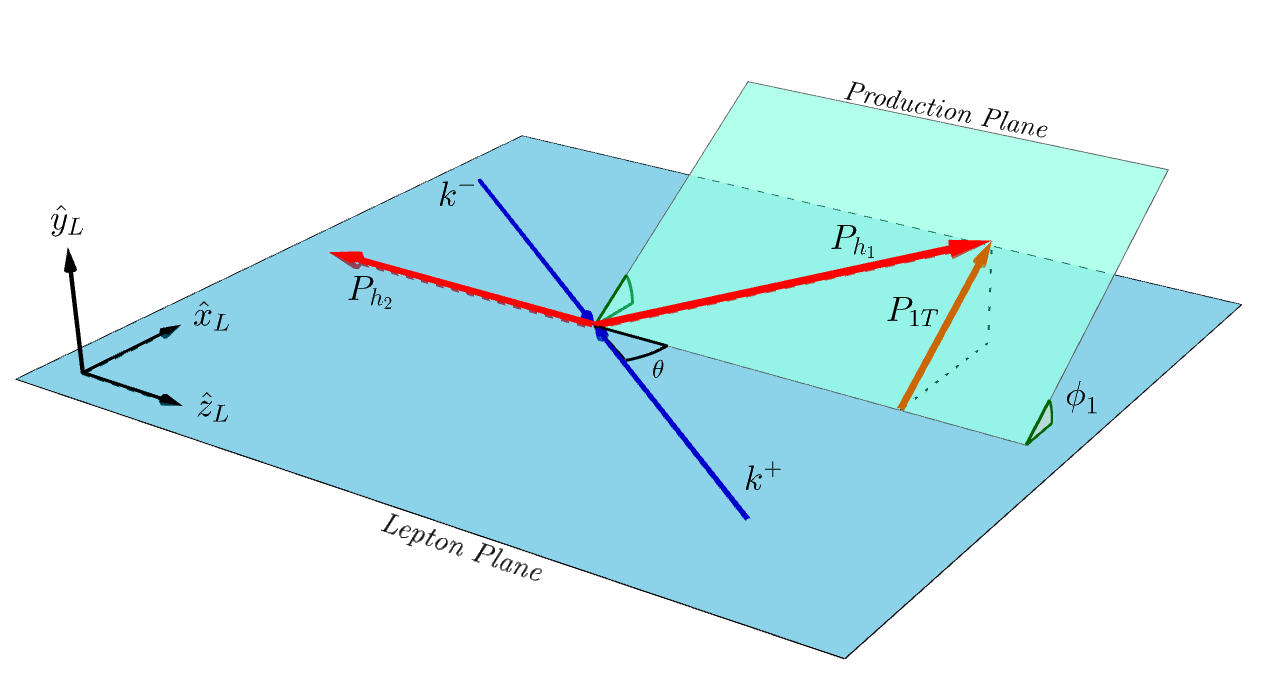}
    \caption{Kinematics for the the hadron-frame configuration.}
    \label{fig:kin2}
\end{figure}

Moving to the hadron frame, here one can fix the $\hat z$-axis in the opposite direction with respect to the observed hadron $h_2$ and the $\widehat{xz}$ lepton plane as determined by the lepton and the $h_2$ directions (with the $e^+e^-$ axis at angle $\theta$)\footnote{We use the same angle as in the thrust frame for the sake of simplicity.}. There will be another relevant plane, the production one, determined by $\hat z$ and the direction of the other observed hadron $h_1$, at an angle $\phi_1$ with respect to the $\widehat{xz}$ plane. This kinematical configuration is shown in Fig.~\ref{fig:kin2}: it has the advantage that it does not require the reconstruction of the thrust axis.
Also in this case we will use the notation $\hat{\bm{x}}_L$, $\hat{\bm{y}}_L$, $\hat{\bm{z}}_L$ for the axes in this frame, to which we refer generically to as the laboratory frame.
Notice that the incoming particles entering the partonic process $e^+e^-\to q\bar q$ are still in a c.m.~frame with $\bm{k}_1=-\bm{k}_2$ and $\bm{q}_1=-\bm{q}_2$ and stay on a plane, even if not coinciding with the lepton plane and depending on the intrinsic transverse momenta. In order to give it in a compact form we will use the relations among the three scaling variables defined above, as follows:
\bea
P_{h_2}&=&(E_2,0,0,-|\bm{P}_{h_2}|) = \frac{\sqrt s}{2} z_{h_2}\bigg(1, 0,0, -\sqrt{ 1 - 4\,\frac{M_{h_2}^2}{z_{h_2}^2s}}\bigg) =
\frac{\sqrt s}{2} (z_{h_2}, 0,0, - z_{p_2})\nonumber\\
\label{Ph2had}
\eea
\bea
q_2&=&\left( \frac{\sqrt{s}}{2},-\frac{p_{\perp 2}}{z_{p_{2}}} \, \cos\varphi_2,
       -\frac{p_{\perp 2}}{z_{p_{2}}} \, \sin\varphi_2, -\frac{\sqrt{s}}{2}\beta_2 \right) \simeq \left( \frac{\sqrt{s}}{2},-\frac{p_{\perp 2}}{z_{p_{2}}} \, \cos\varphi_2,
       -\frac{p_{\perp 2}}{z_{p_{2}}} \, \sin\varphi_2, -\frac{\sqrt{s}}{2} \right)\nonumber\\
       &&\label{q2}\\
q_1&=&\left( \frac{\sqrt{s}}{2}, \frac{p_{\perp 2}}{z_{p_{2}}} \, \cos\varphi_2,
       \frac{p_{\perp 2}}{z_{p_{2}}} \, \sin\varphi_2, \frac{\sqrt{s}}{2}\beta_2 \right) \simeq
       \left( \frac{\sqrt{s}}{2}, \frac{p_{\perp 2}}{z_{p_{2}}} \, \cos\varphi_2,
       \frac{p_{\perp 2}}{z_{p_{2}}} \, \sin\varphi_2, \frac{\sqrt{s}}{2} \right) \label{q1}
\eea
\bea
\bm{P}_{h_1}&=&\left( P_{1T}\cos\phi_1,P_{1T}\sin\phi_1,z_{p_1}\frac{\sqrt{s}}{2}\sqrt{1-4\frac{\eta_{1T}^2}{z_{p_1}^2}} \right) \simeq
\left( P_{1T}\cos\phi_1,P_{1T}\sin\phi_1,z_{p_1}\frac{\sqrt{s}}{2} \right) \nonumber\\
&&\label{Ph1had}
\eea
\bea
\bm{p}_{\perp 1} & = & \bm{P}_{h_1} - z_{p_1}\,\beta_1\, \bm{q}_1
= \Bigg( P_{1T}\cos\phi_1-\frac{z_{p_1}}{z_{p_2}}\,p_{\perp 2}\, \beta_1\,\cos\varphi_2, P_{1T}\sin\phi_1-\frac{z_{p_1}}{z_{p_2}}\,p_{\perp 2}\, \beta_1\,\sin\varphi_2,\nonumber \\
&&  z_{p_1}\frac{\sqrt s}{2} \Bigg(\sqrt{1-4\frac{\eta_{1T}^2}{z_{p_1}^2}} -\beta_1\beta_2 \Bigg) \Bigg)\nonumber \\
& \simeq &\Bigg( P_{1T}\cos\phi_1-\frac{z_{p_1}}{z_{p_2}}\,p_{\perp 2}\,
\cos\varphi_2, P_{1T}\sin\phi_1-\frac{z_{p_1}}{z_{p_2}}\,p_{\perp 2}\,\sin\varphi_2, 0 \Bigg)\,,
\eea
with $\eta_{1T} = \frac{P_{1T}}{\sqrt s}$ and where in the leading-order approximation in $\eta_\perp$ one consistently neglects second order corrections in $\eta_{1T}$ as well. Notice that the approximate expressions in Eqs.~(\ref{q2}) and (\ref{q1}) are valid for all scaling variables defined in Eqs.~(\ref{zh})-(\ref{zpzh}).

In both configurations the master formula, at leading order (LO), allowing to calculate the most general spin configuration in the helicity formalism is given by
\bea
&&\rho^{h_1,S_1}_{\lambda_{h_1},\lambda^{'}_{h_1}} \rho^{h_2, S_2}_{\lambda_{h_2},\lambda^{'}_{h_2}} 
\frac{d\sigma^{e^+e^-\rightarrow h_1 h_2 X}}{d \cos \theta dz_1 d^2\bm{p}_{\perp 1} dz_2 d^2\bm{p}_{\perp 2}}\nonumber\\
&&= \sum_{q} \sum_{\{ \lambda \}} \frac{1}{32 \pi s}\frac{1}{4} 
 \hat{M}_{\lambda_{q}\lambda_{\bar q},\lambda_{+}\lambda_{-}}\hat{M}^{\ast}_{\lambda^{'}_{q}\lambda^{'}_{\bar q},\lambda_{+}\lambda_{-}}  \hat{D}^{\lambda_{h_1},\lambda^{'}_{h_1}}_{\lambda_{q},\lambda^{'}_{q}}(z_1,\bm{p}_{\perp 1})  \hat{D}^{\lambda_{h_2},\lambda^{'}_{h_2}}_{\lambda_{\bar q},\lambda^{'}_{\bar q}}(z_2,\bm{p}_{\perp 2})\,.
\label{ee_hh}
\eea

Let us clarify the physical meaning of Eq.~(\ref{ee_hh}) -- our starting point -- by making detailed comments on its notation
and contents.

The formula is written according to the TMD factorization theorem, separating the soft, long distance, from the hard, short distance, contributions. The hard part is computable in perturbative QED, while information on the soft one has to be extracted from experiments and/or parameterized. More explicitly:
\begin{enumerate}
\item  $\rho^{h, S_h}_{\lambda_{h},\lambda^{'}_{h}}$ is the helicity density matrix of the hadron $h$ with spin $S_h$. It describes the spin orientation of the particle in \emph{its helicity frame}; e.g.~for a spin-1/2 particle, Tr$(\sigma_i \rho) = P_i$ is the $i$-th component of the polarization vector $\bm{P}$ in the helicity frame of the particle;
\item the notation $\{ \lambda \}$ implies a sum over all repeated helicity indices; 
\item the $\hat{M}_{\lambda_{q}\lambda_{\bar q},\lambda_{+}\lambda_{-}} $'s  are the helicity scattering amplitudes for the  elementary process $e^+(\lambda_+) + e^-(\lambda_-) \rightarrow q(\lambda_q) + \bar q(\lambda_{\bar q})$, $q=u,d,s,\bar u,\bar d, \bar s$ (neglecting heavy flavours). They represent the hard contribution to the cross section and can be calculated perturbatively within QED.
\item the $\hat{D}^{\lambda_{h},\lambda^{'}_{h}}_{\lambda_{q},\lambda^{'}_{q}}(z,\bm{p}_{\perp}) $ is the product of the helicity fragmentation amplitudes for the $q \rightarrow h +X$ process, and is directly related to the leading-twist TMD fragmentation functions for the hadron $h$. It represents the soft component of the cross section, as it will be discussed in detail in the next section. Here we recall only, referring now to a generic parton $c$, a quark or a gluon,  that it is defined as
\begin{equation}
\begin{split}
\hat{D}^{\lambda_{h},\lambda^{'}_{h}}_{\lambda_{c},\lambda^{'}_{c}} (z,\bm{p}_{\perp}) = \sumint_{{X},\lambda_{X}}
\hat{\mathcal{D}}_{\lambda_{h},\lambda_{X};\lambda_c }(z,\bm{p}_\perp) \hat{\mathcal{D}}^{\ast}_{\lambda^{'}_{h},\lambda_{X};\lambda^{'}_c }(z,\bm{p}_\perp)\,,
\label{frag_mod}
\end{split}
\end{equation}
where the $\sumint_{{X},\lambda_{X}}$ stands for a spin sum and phase space integration over all undetected particles, considered as a system $X$. The usual unpolarized fragmentation function $D_{h/c}(z)$, i.e.~the number density of hadrons $h$ resulting from the fragmentation of an unpolarized parton $c$ and carrying a light-cone momentum fraction $z$, is given by
\be
D_{h/c}(z) = \frac{1}{2} \sum_{\lambda_c, \lambda_h} \int d^2\bm{p}_\perp \hat{D}_{\lambda_c,\lambda_c}^{\lambda_h,\lambda_h}(z,\bm{p}_\perp)\,.
\ee
\end{enumerate}

The  expression in Eq.~(\ref{ee_hh}) has to be suitably integrated over the unobserved variables according to the chosen kinematical set-up. In the thrust frame, as discussed for instance in the context of spin-zero meson production for the study of the Collins effect \cite{Anselmino:2007fs}, one changes the angular variables from $(\varphi_1,\varphi_2)$ to $(\varphi_1,\varphi_1+\varphi_2)$, integrates over the azimuthal angle $\varphi_1$ and, eventually, over the moduli of the intrinsic transverse momenta.

In the hadron-frame configuration, where the parton momenta are not measurable, one has to integrate over them. For the most general case, i.e.~the  unpolarized, single-polarized, double-polarized hadron production, we will have
\be
\label{desahdfr}
\frac{d\sigma^{e^+e^-\to h_1(S_1) h_2(S_2)\, X}}{d\cos\theta dz_1dz_2d^2\bm{P}_{1T}} = \int d^2\bm{p}_{\perp 1} \,d^2\bm{p}_{\perp 2}\, \delta^{(2)}(\bm{p}_{\perp 1}- \bm{P}_{h_1} + z_{p_1}\,\beta_1\,\bm{q}_1 )\,
\frac{d\sigma^{e^+e^-\to h_1(S_1) h_2(S_2)\, X}}{d \cos \theta dz_1 d^2\bm{p}_{\perp 1} dz_2 d^2\bm{p}_{\perp 2}}\,,
\ee
where $\bm{q}_1$ is given in terms of $\bm{p}_{\perp 2}$, see Eq.~(\ref{q1}). At the lowest order in the transverse momentum dependence the above equation reads
\bea
\label{desahdfr2}
&&\frac{d\sigma^{e^+e^-\to h_1(S_1) h_2(S_2)\, X}}{d\cos\theta dz_1dz_2d^2\bm{P}_{1T}} \nonumber\\
&& = \int d^2\bm{p}_{\perp 1} \,d^2\bm{p}_{\perp 2}\, \delta^{(2)}(\bm{p}_{\perp 1}- \bm{P}_{1T} + \bm{p}_{\perp 2}\,z_{p_1}/z_{p_2})\,
\frac{d\sigma^{e^+e^-\to h_1(S_1) h_2(S_2)\, X}}{d \cos \theta dz_1 d^2\bm{p}_{\perp 1} dz_2 d^2\bm{p}_{\perp 2}}\,.
\eea

In the next section we will discuss in details the soft contributions to the master formula.

\section{Soft Physics}  
\label{softq}
The way the parton spin is transferred to the hadrons can be formally described, in general, by bilinear combinations of the helicity fragmentation amplitudes for the process $c \to h+X$. The hadron polarizations are, indeed, related to their parent parton polarizations. In this sense, one could equally well interpret Eq.~(\ref{ee_hh}) either in terms of hadron polarizations or in terms of the fragmentation amplitudes.

We present here our results starting from the former approach, which is somewhat more direct and allows us to give a complete classification of the TMD fragmentation functions at leading twist. However, the latter approach offers a deeper understanding of some of the basic properties of our factorized scheme (e.g.~the parity properties) and allows for a direct comparison with other formalisms used to describe the same spin effects.

For these reasons and for their relevance in the present study we will discuss the quark case in full detail here below. The corresponding gluon TMD-FFs can be found in Appendix~\ref{g-FF}. The derivation of the explicit relations between our formalism and that of the Amsterdam group is given in Appendix~\ref{Amst}.

\subsection{Quark TMD fragmentation functions for spin-1/2 hadrons at leading twist}

\label{qFF}

We start by introducing the TMD fragmentation function for a polarized quark, $q$, with spin $s_q$ and polarization vector $\bm{P}^q$, fragmenting into an unpolarized hadron $h$: $\hat{D}_{h/q,s_q}(z, \bm{p}_\perp)$. This, together with the helicity density matrices of the quark and the hadron and the generalized fragmentation amplitudes defined in Eq.~(\ref{frag_mod}), allows us to give the complete set of leading-twist quark TMD-FFs, as follows:
\begin{equation}
\rho^{h,S_h}_{\lambda_{h},\lambda^{'}_{h}} \hat{D}_{h/q,s_q}(z,\bm{p}_{\perp}) = \sum_{\lambda_{q},\lambda^{'}_{q}} \rho^{q}_{\lambda_{q},\lambda^{'}_{q}} \hat{D}^{\lambda_{h},\lambda^{'}_{h}}_{\lambda_{q},\lambda^{'}_{q}}(z,\bm{p}_{\perp})\,.
\label{hadron2quark}
\end{equation}

The hadron helicity density matrix for a spin-1/2 hadron can be expressed in terms of the components of its polarization vector $\bm{P}^h = (P^h_X,P^h_Y,P^h_Z) = (P^h_T\cos\phi_{S_h},P^h_T\sin\phi_{S_h},P^h_L)$ as \cite{Anselmino:2005sh}
\begin{equation}
\begin{split}
\rho^{h,S_h}_{\lambda_{h},\lambda^{'}_{h}} &= \frac{1}{2}
\begin{pmatrix}
\rho^{h}_{++} &\rho^{h}_{+-} \\
\rho^{h}_{-+} & \rho^{h}_{--} \\
\end{pmatrix} \\
 &= \frac{1}{2}
\begin{pmatrix}
1+ P^{h}_Z & P^{h}_X -iP^{h}_Y \\
P^{h}_X + iP^{h}_Y & 1- P^{h}_Z \\
\end{pmatrix} \\
%
&= \frac{1}{2}
\begin{pmatrix}
1+ P^{h}_L & P^{h}_T e^{-i\phi_{S_{h}}} \\
P^{h}_T e^{i\phi_{S_{h}}} & 1- P^{h}_L \\
\end{pmatrix} \, ,
\end{split}
\label{mat_h}
\end{equation}
where $X,Y,Z$ are the directions of the axes of the hadron helicity reference frame. The above matrix elements have the following properties
\bea
\rho^{h}_{++} + \rho^{h}_{--}  &=& 1 \label{rr}\\
\rho^{h}_{++} - \rho^{h}_{--}  &=& P^{h}_Z = P^{h}_L \label{rrz} \\
2{\rm Re} \rho^{h}_{-+} =  2{\rm Re}\rho^{h}_{+-} &=& P^{h}_X = P^h_T\cos\phi_{S_h} \label{rrx}\\
2{\rm Im} \rho^{h}_{-+} = -2{\rm Im}\rho^{h}_{+-} &=& P^{h}_Y = P^h_T\sin\phi_{S_h}\, . \label{rry}
\eea
By using the above relations, we can write products of the form
\begin{equation}
 P^{h}_J \hat{D}_{h/q,s_q} = \hat{D}^{h/q}_{S_J/s_q}  -  \hat{D}^{h/q}_{-S_J/s_q} \equiv \Delta \hat{D}^{h/q}_{S_J/s_q}\,
\label{defP}
\end{equation}
where $J=X,Y,Z$.

We will use the notations:
\bea
(P^h_J \, \hat D_{h/q,s_T}) &=&  \hat D_{S_J/s_T}^{h/q} - \hat D_{-S_J/s_T}^{h/q} \equiv
\Delta \hat D_{S_J/s_T}^{h/q}(z, \bm{p}_{\perp}) \label{DxY}\\
(P^h_J \, \hat D_{h/q,s_z}) &=&  \hat D_{S_J/+}^{h/q} -  \hat D_{-S_J/+}^{h/q} \equiv \Delta \hat D_{S_J/+}^{h/q}(z, \bm{p}_{\perp}) =\Delta \hat D_{S_J/s_z}^{h/q}  \label{DxZ}\\
\hat D_{h/q,s_T} &=&  D_{h/q}(z, p_{\perp}) + \frac{1}{2}\, \Delta \hat D_{h/s_T}(z, \bm{p}_{\perp})\,,\label{Dunp}
\label{main-tableq}
\eea
where $s_T$ ($s_z$) stands for the quark transverse (longitudinal) spin component in its helicity frame.

These amount to eight independent quantities, which represent the $\bm{p}_\perp$-unintegrated fragmentation functions of hadron $h$ with polarization $\bm{P}^h$ (defined in the hadron helicity frame) coming from a quark $q$ with spin $s_q$ (specified in the partonic helicity frame).

All of these functions have a simple direct physical meaning: for instance, the $X$ component of Eq.~(\ref{DxY}), $(P^h_X \, \hat D_{h/q,s_T})$ represents the amount of polarization along the $X$ axis (in the hadronic helicity frame) carried by hadrons $h$ coming from the fragmentation of a transversely polarized quark $q,s_T$; $(P^h_Y \, \hat D_{h/q,s_T})$ is related to the $\bm{p}_\perp$-dependent transversity fragmentation function, which survives upon integration over $d^2\bm{p}_\perp$. Similarly, the $Z$ component of Eq.~(\ref{DxZ}) $(P^h_Z \, \hat D_{h/q,s_z})$ is the unintegrated helicity fragmentation function, which, once integrated over the transverse momentum, gives the TMD quark helicity fragmentation function $\Delta{D}^{h/q}_{S_Z/s_z}$.

Notice that two independent fragmentation functions appear in the definition of $\hat D_{h/q,s_T}$, Eq.~(\ref{Dunp}), which is the only term in the sum over $\lambda_h$, $\lambda_h'$ which corresponds to hadron $h$ being unpolarized: $D_{h/q}(z, p_{\perp})$, the unintegrated number density of unpolarized hadrons coming from the fragmentation of an unpolarized quark $q$, and $\Delta \hat D_{h/s_T}(z, \bm{p}_{\perp})$, the Collins function \cite{Collins:1992kk}. The latter permits the distribution of unpolarized hadrons $h$ to depend upon the transverse polarization of the parent quark $q$. In general, for a quark in a pure transverse spin state $\bm{s}_T$ and corresponding unit polarization vector, $\bm{P}^q$, we could also rewrite it as
\be
\hat D_{h/q,s_T}
=  D_{h/q}(z,p_\perp) + \frac{1}{2}\Delta^{N}\! D_{h/q^{\uparrow}}(\hat{\bm{p}}_q \times \hat{\bm{p}}_{\perp})\cdot {\bm{P}}^q\,,
\label{Collins}
\ee
where we have explicitly extracted the angular dependence, according to the so-called ‘‘Trento conventions’’ \cite{Bacchetta:2004jz}, and where $\hat{\bm{p}}_q$ is the unit vector along the quark three-momentum.

To deepen our understanding of the above eight TMD-FFs we could exploit Eq.~(\ref{hadron2quark}) with the help of the helicity density matrix for quarks. In close analogy to the hadron case we have
\begin{equation}
\begin{split}
\rho^{q}_{\lambda_{q},\lambda^{'}_{q}} &= \frac{1}{2}
\begin{pmatrix}
\rho^{q}_{++} &\rho^{q}_{+-} \\
\rho^{q}_{-+} & \rho^{q}_{--} \\
\end{pmatrix} \\
& = \frac{1}{2}
\begin{pmatrix}
1+ P^q_z & P^q_x -iP^q_y \\
P^q_x + iP^q_y & 1- P^q_z \\
\end{pmatrix} \\
%
&= \frac{1}{2}
\begin{pmatrix}
1+ P^q_L & P^q_T e^{-i\phi_{s_q}} \\
P^q_T e^{i\phi_{s_q}} & 1- P^q_L \\
\end{pmatrix}\, ,
\end{split}
\label{mat_c}
\end{equation}
where $x,y,z$ are the axes of the helicity reference frame of the quark and $\phi_{s_q}$ is the azimuthal angle of the quark's spin.

By summing over the quark helicity indices on the right-hand side of Eq.~(\ref{hadron2quark}) and keeping fixed those of the hadron density matrix  on the left-hand side, we obtain:
\bea
\rho^{h,S_h}_{++} \hat{D}_{h/q,s_q} & = & \frac{1}{2}(1+ P^{h}_Z) \hat{D}_{h/q,s_q} \nonumber\\
&=&\frac{1}{2}(D^{++}_{++} +D^{++}_{--}) + \frac{1}{2}P^q_L(D^{++}_{++}- D^{++}_{--}) \nonumber\\
&&+\, P^q_T [{\rm Re}D^{++}_{+-}\cos{(\phi_{s_q} - \phi_{h})} + {\rm Im}D^{++}_{+-}\sin{(\phi_{s_q} - \phi_{h})}] \label{h2q_1}
\eea
\bea
\rho^{h,S_h}_{--} \hat{D}_{h/q,s_q} & = &\frac{1}{2}(1 -  P^{h}_Z) \hat{D}_{h/q,s_q} \nonumber\\
&=&\frac{1}{2}(D^{++}_{++} +D^{++}_{--}) - \frac{1}{2}P^q_L(D^{++}_{++}- D^{++}_{--}) \nonumber\\
&& -\, P^q_T [{\rm Re}D^{++}_{+-}\cos{(\phi_{s_q} - \phi_{h})} - {\rm Im}D^{++}_{+-}\sin{(\phi_{s_q} - \phi_{h})}]\label{h2q_2}
\eea
\bea
\rho^{h,S_h}_{+-} \hat{D}_{h/q,s_q} & = &\frac{1}{2}(P^{h}_X -iP^{h}_Y) \hat{D}_{h/q,s_q} \nonumber\\
&=& i{\rm Im}D^{+-}_{++} + P^q_L {\rm Re}D^{+-}_{++} \nonumber\\
&& +\,  \frac{1}{2}P^q_T [(D^{+-}_{+-} +D^{+-}_{-+})\cos{(\phi_{s_q} - \phi_{h})} -i (D^{+-}_{+-}-D^{+-}_{-+})\sin{(\phi_{s_q} - \phi_{h})}] \,,\nonumber\\
&&
\label{h2q_3}
\eea
where we have used the properties of $\hat{D}^{\lambda_{h},\lambda^{'}_{h}}_{\lambda_{q},\lambda^{'}_{q}}(z,\bm{p}_{\perp})$ collected in Appendix~\ref{hel-ampl}.

To better clarify the above as well as the following expressions, we recall that according to our ``hat-convention'' the quantities like $\hat{D}$ (or $\Delta \hat{D}$) depend on $\bm{p}_\perp$, including its phase, while quantities like ${D}$ (or $\Delta {D}$) do not depend on phases anymore, as such dependence has been explicitly factorized out.


By combining these expressions we are able to find the eight fragmentation functions discussed above. For instance, by summing and subtracting Eqs.~(\ref{h2q_1}) and (\ref{h2q_2}) we have respectively
\bea
\hat{D}_{h/q,s_q} & = &
 (D^{++}_{++} + D^{++}_{--}) + 2P^q_T {\rm Im} D^{++}_{+-}\sin{(\phi_{s_q} -\phi_{h})}\label{piupiu}\\
P_Z^h\hat{D}_{h/q,s_q} & = &
 P^q_L(D^{++}_{++}- D^{++}_{--}) + 2P^q_T {\rm  Re}D^{++}_{+-}\cos{(\phi_{s_q} - \phi_{h})}\label{piumeno}\,.
\eea
Meanwhile taking from Eq.~(\ref{h2q_3}) the real and imaginary part we find respectively the fragmentation for a hadron polarized along its $X$ and $Y$ helicity axes
\bea
P_X^h\hat{D}_{h/q,s_q} &
= & 2P^q_L {\rm Re}D^{+-}_{++} + P^q_T (D^{+-}_{+-} +D^{+-}_{-+})\cos{(\phi_{s_q} - \phi_{h})} \label{real}\\
P_Y^h\hat{D}_{h/q,s_q} &=&
-2 {\rm Im} D^{+-}_{++} + P^q_T (D^{+-}_{+-} -D^{+-}_{-+})\sin{(\phi_{s_q} - \phi_{h})}\label{imma} \,.
\eea
The two above relations can be expressed in a more compact form in terms of the transverse polarization of the final hadron
\be
P_X^h = P_T^h \cos\phi_{S_h} \hspace{1cm}
P_Y^h = P_T^h \sin\phi_{S_h}\,.
\ee
By multiplying Eq.~(\ref{real}) by $\cos\phi_{S_h}$ and Eq.~(\ref{imma}) by $\sin\phi_{S_h}$ and summing them up, we get
\bea
P_T^h\hat{D}_{h/q,s_q} &=& -2 \,{\rm Im} D^{+-}_{++} \sin\phi_{S_h} +
2P^q_L\, {\rm Re}D^{+-}_{++} \cos\phi_{S_h}  \nonumber \\
&&+\,  P_T^q  \left[ D^{+-}_{+-} \cos(\phi_{S_h} - \phi_{s_q} + \phi_h)
+ D^{-+}_{+-} \cos(\phi_{S_h} + \phi_{s_q} - \phi_h) \right] \,. \label{PTSh}
\eea
Moreover, it is easy to show that the azimuthal angle of $\bm{P}^h$ in its helicity frame, $\phi_{S_h}$, and the same angle measured in the quark helicity frame, $\phi'_{S_h}$, are related as
\be
\label{phiangles}
\phi_{S_h} = \phi'_{S_h} - \phi_h + {\cal O}(p_\perp^2/s)\,,
\ee
so that Eq.~(\ref{PTSh}) can be recast as
\bea
P_T^h\hat{D}_{h/q,s_q} &=& -2 \,{\rm Im} D^{+-}_{++} \sin(\phi'_{S_h}-\phi_h) +
2\,P^q_L {\rm Re}D^{+-}_{++} \cos(\phi'_{S_h}-\phi_h)  \nonumber \\
&&+\,  P_T^q \left[ D^{+-}_{+-} \cos(\phi'_{S_h} - \phi_{s_q})
+ D^{-+}_{+-} \cos(\phi'_{S_h} + \phi_{s_q} - 2\phi_h) \right] \label{PTSh-ap}\,.
\eea

From the above equations, giving the hadron polarization in terms of the fragmentation amplitudes and the quark polarization, we can define the eight TMD-FFs as follows.

Through Eq.~(\ref{piupiu}) we have then
\bea
\hat{D}_{h/q} & = & \hat{D}_{h/q,s_L} = (D^{++}_{++}+ D^{++}_{--})  \equiv D_{h/q} \label{hq_nop_1}\\
\hat{D}_{h/q,s_T} &= & 
(D^{++}_{++}+ D^{++}_{--}) + 2\, {\rm Im}D^{++}_{+-}\sin{(\phi_{s_q} - \phi_{h})} \,.\label{hq_nop}
\eea
The first expression gives the fragmentation function for an unpolarized hadron generated by an unpolarized quark. This is equal by parity to the fragmentation function of a longitudinally polarized quark into an unpolarized hadron. In the second expression we have the TMD-FF of an unpolarized hadron generated by a transversely polarized quark.
By using the following identity
\begin{equation}
\sin{(\phi_{s_q} - \phi_{h})}= (\hat{\bm{p}}_q \times \hat{\bm{p}}_{\perp})\cdot {\bm{P}}^q  ,
\label{triple}
\end{equation}
we recover the expression in Eq.~(\ref{Collins}) and, at the same time, we find the relation between $ {\rm Im}D^{++}_{+-}$ and $ \Delta^{N}\! {D}_{h/q^{\uparrow}}$.

From Eq.~(\ref{piumeno}) we have
\bea
P_Z^h\hat{D}_{h/q,s_L}
&=&  (D^{++}_{++}- D^{++}_{--})= \Delta D^{h/q}_{S_Z/s_L} \label{hq_pz_1} \\
P_Z^h\hat{D}_{h/q,s_T}&=& 2\,{\rm Re} D^{++}_{+-}\cos{(\phi_{s_q} - \phi_{h})} =  \Delta D^{h/q}_{S_Z/s_T}\cos{(\phi_{s_q} - \phi_{h})}\,,\label{hq_pz}
\eea
giving the FF for a longitudinally polarized hadron produced, respectively, by a longitudinally and a transversely polarized quark. Analogously, from Eq.~(\ref{real}) we get
\bea
P_X^h\hat{D}_{h/q,s_L}
&=& 2\, {\rm Re}D^{+-}_{++} = \Delta D^{h/q}_{S_X/s_L} \label{hq_px_1}\\
P_X^h\hat{D}_{h/q,s_T}
&=&
(D^{+-}_{+-} + D^{+-}_{-+})\cos{(\phi_{s_q} - \phi_{h})} =  \Delta D^{h/q}_{S_X/s_T} \cos{(\phi_{s_q} - \phi_{h})} \,, \label{hq_px}
\eea
where we have the FF for a transversely polarized hadron along the $X$ axis produced by, respectively, a longitudinally and a transversely polarized quark. Finally, from Eq.~(\ref{imma}) we have
\bea
P_Y^h\hat{D}_{h/q,s_L} &=& P_Y^h\hat{D}_{h/q} =
-2 {\rm Im}D^{+-}_{++}  = \Delta D^{h}_{S_Y/q}\label{hq_py_1}\\
P_Y^h\hat{D}_{h/q,s_T}
&  = & -2 {\rm Im}D^{+-}_{++}  + (D^{+-}_{+-} -D^{+-}_{-+})\sin{(\phi_{s_q} - \phi_{h})} \nonumber\\
& = & \Delta {D}^{h}_{S_Y/q} + \Delta^{-}\hat{D}^{h/q}_{S_Y/s_T} =  \Delta {D}^{h}_{S_Y/q} + \Delta^{-}D^{h/q}_{S_Y/s_T}\sin{(\phi_{s_q} - \phi_{h})} \,, \label{hq_py}
\eea
giving the TMD-FF for an unpolarized quark and for a transversely polarized quark fragmenting into a hadron transversely polarized along the $Y$ axis.
The first expression gives the so-called polarizing fragmentation function, introduced in Refs.~\cite{Boer:1997mf, Boer:1997nt} and considered in a phenomenological study of transverse $\Lambda$ polarization in inclusive hadron collisions~\cite{Anselmino:2000vs} and in semi-inclusive deep inelastic processes~\cite{Anselmino:2001js}. It is also worth mentioning that the function $\Delta \hat{D}^{h/q}_{S_Y/s_T}\equiv P_Y^h\hat{D}_{h/q,s_T}$ entering the second equation can be decomposed into two terms, the polarizing FF term which is independent of the quark transverse polarization, and a term which changes sign when the quark polarization direction is reversed:
\begin{equation}
\Delta^{-}\!\hat{D}^{h/q}_{S_Y/s_T}  = \frac{1}{2} [\Delta\hat{D}^{h/q}_{S_Y/s_T}  - \Delta\hat{D}^{h/q}_{S_Y/-s_T}] = - \Delta^{-}\!\hat{D}^{h/q}_{S_Y/-s_T}.
\label{change_sign}
\end{equation}

Before concluding this section, we notice that the azimuthal dependence of the term involving the polarizing FF in Eq.~(\ref{PTSh-ap}) can be expressed as
\begin{equation}
\sin{(\phi'_{S_h} - \phi_{h})}= (\hat{\bm{p}}_q \times \hat{\bm{p}}_{\perp})\cdot \bm{P}^h  \,.
\label{triple2}
\end{equation}
Therefore, we can define, in close analogy to the Collins FF,
\be
P_T^h\hat{D}_{h/q} = \Delta \hat{D}^{h}_{S_T/q} = \Delta^N\!D_{h^\uparrow/q}\, (\hat{\bm{p}}_q \times \hat{\bm{p}}_{\perp})\cdot \bm{P}^h \,,
\ee
for the fragmentation of an unpolarized quark into a transversely polarized hadron.

All the above results can be collected as follows:
\begin{equation}
\begin{split}
\hat{D}_{h/q}(z,\bm{p}_{\perp}) & =D_{h/q}= (D^{++}_{++} +D^{++}_{--})\\
\Delta \hat{D}_{h/q,s_T}(z,\bm{p}_{\perp}) &  = \Delta^{N}\! D_{h/q^{\uparrow}}\sin{(\phi_{s_q} - \phi_{h})} =  4 {\rm Im} D^{++}_{+-}\sin{(\phi_{s_q} - \phi_{h})}   \quad [{\rm Collins \ FF}]\\
\Delta \hat{D}^{h/q}_{S_Z/s_L}(z,\bm{p}_{\perp}) & = \Delta D^{h/q}_{S_Z/s_L} =   (D^{++}_{++}- D^{++}_{--}) \\
\Delta \hat{D}^{h/q}_{S_Z/s_T} (z,\bm{p}_{\perp})&= \Delta D^{h/q}_{S_Z/s_T}\cos{(\phi_{s_q} - \phi_{h})}= 2{\rm Re}D^{++}_{+-}\cos{(\phi_{s_q} - \phi_{h})} \\
\Delta \hat{D}^{h/q}_{S_X/s_L} (z,\bm{p}_{\perp})& =\Delta D^{h/q}_{S_X/s_L}  = 2 {\rm Re}D^{+-}_{++}  \\
\Delta \hat{D}^{h/q}_{S_X/s_T}(z,\bm{p}_{\perp}) & = \Delta D^{h/q}_{S_X/s_T}\cos{(\phi_{s_q} - \phi_{h})}= (D^{+-}_{+-} +D^{+-}_{-+})\cos{(\phi_{s_q} - \phi_{h})}\\
\Delta \hat{D}^{h}_{S_Y/q}(z,\bm{p}_{\perp}) & =\Delta D^{h}_{S_Y/q} = \Delta^{N}\! D_{h^{\uparrow}/q} =
-2 {\rm Im}D^{+-}_{++}  \quad [{\rm Polarizing \ FF}] \\
\Delta^{-}\hat{D}^{h/q}_{S_Y/s_T}(z,\bm{p}_{\perp}) &=   \Delta^{-}D^{h/q}_{S_Y/s_T} \sin{(\phi_{s_q} - \phi_{h})}=  (D^{+-}_{+-}-D^{+-}_{-+})\sin{(\phi_{s_q} - \phi_{h})}\,.
\end{split}
\label{tab_ff_quark}
\end{equation}

\section{Azimuthal dependences and polarization observables in $e^+e^-\to h_1 h_2 + X$}
\label{azimdep}

We present here all possible azimuthal dependences and polarization observables for the case of unpolarized, single-polarized and double-polarized hadron production in $e^+e^-\to h_1 h_2 + X$.
To start with, we give the partonic helicity scattering amplitudes appearing in Eq.~(\ref{ee_hh}).
From helicity conservation (with massless leptons and quarks), the only nonzero amplitudes for the process $ab\to cd$, are:
\begin{equation}
\hat{M}_{\lambda_{c}\lambda_{d},\lambda_{a}\lambda_{b}} : \Big\{\hat{M}_{+-,+-}; \hat{M}_{-+,-+}; \hat{M}_{+-,-+}; \hat{M}_{-+,+-} \Big\}\,,
\label{}
\end{equation}
where only two are independent, since
\be
\hat{M}_{+-,+-}=\hat{M}_{-+,-+}^\ast\equiv \hat{M}_2
\quad\quad\quad
\hat{M}_{-+,+-}=\hat{M}_{+-,-+}^\ast \equiv  \hat{M}_3\,.
\ee
Once again the two kinematical configurations have to be treated differently. In the thrust frame, the partonic scattering process occurs on the $xz$ plane, in the center of mass frame and the helicity scattering amplitudes are the canonical ones. These are given as
\begin{equation}
\begin{split}
\hat{M}_{+-,+-} &= \hat{M}^0_{+-,+-} = e^2 e_q \sqrt{3} \,(1+\cos\theta) \\
\hat{M}_{-+,+-} &= \hat{M}^0_{-+,+-} = e^2 e_q \sqrt{3} \,(1-\cos\theta)\,,
\end{split}
\label{ampl_0}
\end{equation}
where we have  included the color factor.

The corresponding helicity amplitudes for the hadron frame configuration are more complicate, since, even if still in the c.m.~frame, the partonic scattering process occurs out of the $\widehat{xz}$ plane (containing the lepton and the hadron $h_2$ momenta). One could relate these to the canonical ones following the procedure described in Ref.~\cite{Anselmino:2005sh} (somehow simplified here since only rotations are involved). Instead, we prefer to give them as obtained by an explicit calculation in terms of the helicity spinors, using the kinematics as given in Eqs.~(\ref{q2}) and (\ref{q1}) (with the help of Eq.~(\ref{kpkm}), where now $\theta$ refers to the relative angle between the lepton-axis and the direction of the hadron $h_2$):
\bea
\hat{M}_{+-,+-} & = & e^2 e_q \sqrt{3} \bigg[ \cos\varphi_2 (1 + \beta_2\cos\theta) - 2\frac{\eta_{\perp 2}}{z_{p_2}}\sin\theta + i \sin\varphi_2 (\beta_2 + \cos\theta)  \bigg]\nonumber\\
\hat{M}_{-+,+-} & = & e^2 e_q \sqrt{3} \bigg[ \cos\varphi_2 (1 - \beta_2\cos\theta) + 2\frac{\eta_{\perp 2}}{z_{p_2}}\sin\theta  -i \sin\varphi_2 (\beta_2 - \cos\theta)    \bigg]\,.
\eea

At the lowest order in $p_\perp/\sqrt s$, in which $\eta_{\perp 2}=0$ and $\beta_2=1$, they simplify as
\bea
\hat{M}_{+-,+-} & \simeq & e^2 e_q \sqrt{3}\, ( 1+ \cos\theta)\, e^{i\varphi_2} \;\; = \hat{M}^0_{+-,+-}\, e^{i\varphi_2} \nonumber\\
\hat{M}_{-+,+-} & \simeq & e^2 e_q \sqrt{3}\, ( 1-\cos\theta )\, e^{-i\varphi_2} = \hat{M}^0_{-+,+-}\, e^{-i\varphi_2}\,.
\label{Mhf}
\eea

\subsection{Unpolarized hadron production}
\label{unp2had}

By summing over the diagonal indices of the helicity density matrices appearing on the left hand side of Eq.~(\ref{ee_hh}) we get the unpolarized cross section:
\bea
&&\frac{d\sigma^{e^+e^-\rightarrow h_1 h_2 X}}{d\cos\theta d{\rm PS}_{12}}=\frac{1}{128 \pi s}\nonumber\\
& & \times \sum_q \Big\{ 2\Big(|\hat{M}_2|^2+ |\hat{M}_3|^2 \Big)   D_{h_1/q}(z_1, p_{\perp 1}) D_{h_2/\bar{q}}(z_2, p_{\perp 2}) \nonumber\\
&& \quad + \Big[ {\rm Re}[\hat{M}_2\hat{M}_3^{\ast}]\cos\Delta\phi - {\rm Im}[\hat{M}_2\hat{M}_3^{\ast}]\sin\Delta\phi \Big]
\Delta^{N}\! D_{h_1/q^{\uparrow}}(z_1,p_{\perp 1})   \Delta^{N}\! D_{h_2/\bar{q}^{\uparrow}} (z_2,p_{\perp 2}) \Big\}\,,\nonumber\\
&&
\label{eq:unph1h2}
\eea
where we have introduced the shorthand notation
\be
d{\rm PS}_{12} = dz_1 d^2\bm{p}_{\perp 1} dz_2 d^2\bm{p}_{\perp 2}
\ee
and where $\Delta\phi=\phi_{h_1} -\phi_{h_2}$, with $\phi_{h_{1,2}}$ the azimuthal angles of the two hadrons in the corresponding helicity frames of their parent quarks.

In the thrust frame we directly get
\bea
\frac{d\sigma^{e^+e^-\rightarrow h_1 h_2 X}}{d \cos \theta d{\rm PS}_{12}}
& = & \frac{3\pi\alpha^2}{2 s}\sum_q e_q^2\Big\{(1+\cos^2\theta) D_{h_1/q}(z_1, p_{\perp 1}) D_{h_2/\bar{q}}(z_2, p_{\perp 2}) \nonumber \\
&&+\, \frac{1}{4}\sin^2\theta \Delta^{N}\! D_{h_1/q^{\uparrow}}(z_1,p_{\perp 1}) \Delta^{N}\! D_{h_2/\bar{q}^{\uparrow}} (z_2,p_{\perp 2}) \cos(\varphi_1+\varphi_2) \Big\}
\label{eq:unph1h2thr}\,,
\eea
where we have used the relations in Eq.~(\ref{ampl_0}) and the fact that $\phi_{h_1}=\varphi_1$ and $\phi_{h_2}=2\pi-\varphi_2$ (see Appendix~\ref{hel-frames}).

For the hadron-frame configuration, by using the expressions of the helicity amplitudes in Eq.~(\ref{Mhf}) and the fact that $\phi_{h_2}=0$ (see Appendix~\ref{hel-frames}), we obtain
\bea
\frac{d\sigma^{e^+e^-\rightarrow h_1 h_2 X}}{d \cos \theta d{\rm PS}_{12} }
& = & \frac{3\pi\alpha^2}{2 s}\sum_q e_q^2\Big\{(1+\cos^2\theta) D_{h_1/q}(z_1, p_{\perp 1}) D_{h_2/\bar{q}}(z_2, p_{\perp 2}) \nonumber \\
&&+\, \frac{1}{4}\sin^2\theta \Delta^{N}\! D_{h_1/q^{\uparrow}}(z_1,p_{\perp 1}) \Delta^{N}\! D_{h_2/\bar{q}^{\uparrow}} (z_2,p_{\perp 2}) \cos(2\varphi_2+\phi_{h_1}) \Big\}
\label{eq:unph1h2had}\,.
\eea
This has then to be integrated according to Eq.~(\ref{desahdfr}). In such a case the angle $\phi_{h_1}$ can be expressed in terms of the integration variables as (see Eqs.~(\ref{cosfih1}) and (\ref{sinfih1}) and their derivation)
\bea
\cos\phi_{h_1} & = & \frac{P_{1T}}{p_{\perp 1}}\beta_2 \cos(\phi_1 - \varphi_2) - \frac{z_{p_1}}{z_{p_2}} \frac{p_{\perp 2}}{p_{\perp 1}}  \sqrt{  1- 4\,\frac{\eta^2_{1T}}{z_{p_1}^2}}
\simeq
\frac{P_{1T}}{p_{\perp 1}}\cos(\phi_1 - \varphi_2) - \frac{z_{p_1}}{z_{p_2}} \frac{p_{\perp 2}}{p_{\perp 1}}\nonumber\\
&&
\label{cosfih11}\\
\sin\phi_{h_1} & =& \frac{P_{1T}}{p_{\perp 1}}\sin(\phi_1 - \varphi_2)  \,,
\label{sinfih11}
\eea
where we have also given the lowest-order expressions in $\eta_{\perp, 1T}/z_p$. Notice that $\phi_1$, not to be confused with $\phi_{h_1}$, is the observed azimuthal angle of the hadron momentum, $\bm{P}_1$, in the hadron frame.

In the expression of the unpolarized cross section, in agreement with Ref.~\cite{Anselmino:2007fs}, we recognize, beside the ordinary contribution from the unpolarized TMD FFs, the azimuthal dependence coming from the Collins effect.

\subsection{Single-polarized hadron production}
\label{onepolh1h2}

In order to compute the polarization state for a single hadron, let us refer to $h_1$, one has to consider the diagonal part of the helicity density matrix for the second hadron, $h_2$, and take the proper combination of the matrix elements of hadron $h_1$, see Eq.~(\ref{ee_hh}). By an analogous calculation one obtains:
\bea
&& P^{h_1}_X \frac{d\sigma^{e^+e^-\rightarrow h_1 h_2 X}}{d \cos \theta d{\rm PS}_{12} }= \frac{1}{64 \pi s}\nonumber\\
& & \times
 \sum_q
\Big[ {\rm Im}[\hat{M}_2\hat{M}_3^{\ast}]\cos\Delta\phi + {\rm Re}[\hat{M}_2\hat{M}_3^{\ast}]\sin\Delta\phi \Big]\,
\Delta D^{h_1/q}_{S_X/s_T} (z_1,p_{\perp 1}) \Delta^N\!D_{h_2/\bar{q}^{\uparrow}}(z_2,p_{\perp 2})\nonumber\\
&&\label{eq:PX}
\eea
\bea
&&P^{h_1}_Y\frac{d\sigma^{e^+e^-\rightarrow h_1 h_2 X}}{d \cos \theta d{\rm PS}_{12}} = \frac{1}{64 \pi s}\nonumber\\
 & & \times
\sum_q \Big\{ \Big(|\hat{M}_2|^2+|\hat{M}_3|^2 \Big)
\Delta^N\! D_{h_1^\uparrow/q}(z_1,p_{\perp 1})  D_{h_2/\bar{q} }(z_2,p_{\perp 2}) \nonumber \\
&& \quad + \Big[ {\rm Re}[\hat{M}_2\hat{M}_3^{\ast}]\cos\Delta\phi - {\rm Im}[\hat{M}_2\hat{M}_3^{\ast}]\sin\Delta\phi \Big] \,\Delta^{-}\!D^{h_1/q}_{S_Y/s_T}(z_1,p_{\perp 1})\Delta^N\! D_{h_2/\bar{q}^{\uparrow}}(z_2,p_{\perp 2})
\Big\}\nonumber\\
&&
\label{eq:PY}
\eea
\bea
&&P^{h_1}_Z \frac{d\sigma^{e^+e^-\rightarrow h_1 h_2 X}}{d \cos \theta d{\rm PS}_{12} } =
\frac{1}{64 \pi s} \nonumber\\
&&\times \sum_q
\Big[ {\rm Im}[\hat{M}_2\hat{M}_3^{\ast}]\cos\Delta\phi + {\rm Re}[\hat{M}_2\hat{M}_3^{\ast}]\sin\Delta\phi \Big]\,\Delta D^{h_1/q}_{S_Z/s_T} (z_1,p_{\perp 1}) \Delta^N\!D_{h_2/\bar{q}^{\uparrow}}(z_2,p_{\perp 2})\,,\nonumber\\
&&\label{eq:PZ}
\eea
where we can recognize the contributions coming from a transversely polarized quark-antiquark pair, i.e.~the coupling of the Collins function with $\Delta D^{h_1/q}_{S_X/s_T}$, Eq.~(\ref{eq:PX}), $\Delta^{-}\!D^{h_1/q}_{S_Y/s_T}$, Eq.~(\ref{eq:PY}) (second line), and $\Delta D^{h_1/q}_{S_Z/s_T}$, Eq.~(\ref{eq:PZ}), as well as from an unpolarized quark-antiquark pair, first line of Eq.~(\ref{eq:PY}), involving the polarizing FF.

To get the expressions in the two adopted frames it is enough to replace (as already done in the previous Section)
\bea
 {\rm Re}[\hat{M}_2\hat{M}_3^{\ast}]\cos\Delta\phi - {\rm Im}[\hat{M}_2\hat{M}_3^{\ast}]\sin\Delta\phi & = &
3e_q^2 (4\pi\alpha)^2\, \sin^2\theta\cos(\varphi_1+\varphi_2) \label{comb1}\\
{\rm Im}[\hat{M}_2\hat{M}_3^{\ast}]\cos\Delta\phi + {\rm Re}[\hat{M}_2\hat{M}_3^{\ast}] \sin\Delta\phi & = &
3e_q^2 (4\pi\alpha)^2\, \sin^2\theta\sin(\varphi_1+\varphi_2)\label{comb2}
\eea
for the thrust frame, and
\bea
 {\rm Re}[\hat{M}_2\hat{M}_3^{\ast}]\cos\Delta\phi - {\rm Im}[\hat{M}_2\hat{M}_3^{\ast}]\sin\Delta\phi & = &
3e_q^2 (4\pi\alpha)^2\, \sin^2\theta\cos(2\varphi_2+\phi_{h_1}) \label{comb3}\\
{\rm Im}[\hat{M}_2\hat{M}_3^{\ast}]\cos\Delta\phi + {\rm Re}[\hat{M}_2\hat{M}_3^{\ast}] \sin\Delta\phi & = &
3e_q^2 (4\pi\alpha)^2\, \sin^2\theta\sin(2\varphi_2+\phi_{h_1}) \label{comb4}
\eea
for the hadron frame. In both cases
\be
|\hat{M}_2|^2+|\hat{M}_3|^2 = 6e^2_q\,  (4\pi\alpha)^2 \, (1+\cos^2\theta)\,.
\label{comb-unp}
\ee
Once again one has to properly integrate over the unobserved kinematical quantities, with $\phi_{h_1}$ given in Eqs.~(\ref{cosfih11}) and (\ref{sinfih11}). For completeness we give also the explicit expressions, for the thrust-frame configuration,
\bea
P^{h_1}_X \frac{d\sigma^{e^+e^-\rightarrow h_1 h_2 X}}{d \cos \theta d{\rm PS}_{12}}
& = & \frac{3\pi\alpha^2}{4 s}\sum_q e_q^2\sin^2\theta \Delta D^{h_1/q}_{S_X/s_T}(z_1,p_{\perp 1}) \Delta^{N}\! D_{h_2/\bar{q}^{\uparrow}} (z_2,p_{\perp 2}) \sin(\varphi_1+\varphi_2)\nonumber\\
&&
\label{eq:PXthr}\\
P^{h_1}_Y\frac{d\sigma^{e^+e^-\rightarrow h_1 h_2 X}}{d \cos \theta d{\rm PS}_{12}}
& = & \frac{3\pi\alpha^2}{2 s}\sum_q e_q^2\Big\{(1+\cos^2\theta) \Delta^N\! D_{h_1^\uparrow/q}(z_1,p_{\perp 1}) D_{h_2/\bar{q}}(z_2, p_{\perp 2}) \nonumber \\
&& +\,\frac{1}{2}\sin^2\theta \Delta^{-}D^{h_1/q}_{S_Y/s_T}(z_1,p_{\perp 1}) \Delta^{N}\! D_{h_2/\bar{q}^{\uparrow}} (z_2,p_{\perp 2}) \cos(\varphi_1+\varphi_2) \Big\}
\label{eq:PYthr}\\
P^{h_1}_Z \frac{d\sigma^{e^+e^-\rightarrow h_1 h_2 X}}{d \cos \theta d{\rm PS}_{12}}
& = & \frac{3\pi\alpha^2}{4 s}\sum_q e_q^2\sin^2\theta \Delta D^{h_1/q}_{S_Z/s_T}(z_1,p_{\perp 1}) \Delta^{N}\! D_{h_2/\bar{q}^{\uparrow}} (z_2,p_{\perp 2}) \sin(\varphi_1+\varphi_2)\nonumber\\
&&
\label{eq:PZthr}\,,
\eea
and for the hadron-frame case
\bea
P^{h_1}_X \frac{d\sigma^{e^+e^-\rightarrow h_1 h_2 X}}{d \cos \theta d{\rm PS}_{12}}
& = & \frac{3\pi\alpha^2}{4 s}\sum_q e_q^2\sin^2\theta \Delta D^{h_1/q}_{S_X/s_T}(z_1,p_{\perp 1}) \Delta^{N}\! D_{h_2/\bar{q}^{\uparrow}} (z_2,p_{\perp 2}) \sin(2\varphi_2+\phi_{h_1})\nonumber\\
&&\label{eq:PXhadf}\\
P^{h_1}_Y\frac{d\sigma^{e^+e^-\rightarrow h_1 h_2 X}}{d \cos \theta d{\rm PS}_{12}}
& = & \frac{3\pi\alpha^2}{2 s}\sum_q e_q^2\Big\{(1+\cos^2\theta) \Delta^N\! D_{h_1^\uparrow/q}(z_1,p_{\perp 1}) D_{h_2/\bar{q}}(z_2, p_{\perp 2}) \nonumber \\
&&+ \,\frac{1}{2}\sin^2\theta \Delta^{-}D^{h_1}_{S_Y/s_T}(z_1,p_{\perp 1}) \Delta^{N}\! D_{h_2/\bar{q}^{\uparrow}} (z_2,p_{\perp 2}) \cos(2\varphi_2+\phi_{h_1}) \Big\}\nonumber\\
&&
\label{eq:PYhadf}\\
P^{h_1}_Z \frac{d\sigma^{e^+e^-\rightarrow h_1 h_2 X}}{d \cos \theta d{\rm PS}_{12}}
& = & \frac{3\pi\alpha^2}{4 s}\sum_q e_q^2\sin^2\theta \Delta D^{h_1/q}_{S_Z/s_T}(z_1,p_{\perp 1}) \Delta^{N}\! D_{h_2/\bar{q}^{\uparrow}} (z_2,p_{\perp 2}) \sin(2\varphi_2+\phi_{h_1})\nonumber\\
&&\label{eq:PZhadf}\,.
\eea

\subsection{Double-polarized hadron production}

In a very similar way, we can compute all possible double spin configurations for the production of two spin-1/2 hadrons. The following expressions are simply obtained by considering proper combinations of the helicity density matrices of the two hadrons and then exploiting the sum over the helicity indices of quarks and antiquarks in Eq.~(\ref{ee_hh}). The general formulas so obtained are:
\bea
&& P^{h_1}_X P^{h_2}_X\frac{d\sigma^{e^+e^-\rightarrow h_1 h_2 X}}{d \cos \theta d{\rm PS}_{12}} =
\frac{1}{64 \pi s} \nonumber\\
&&\times \sum_q \Big\{  - \Big(|\hat{M}_2|^2+|\hat{M}_3|^2 \Big)  \Delta D^{h_1/q}_{S_X/s_L}(z_1,p_{\perp 1})\Delta D^{h_2/\bar{q}}_{S_X/s_L}(z_2, p_{\perp 2}) \nonumber\\
&&\quad + 2 \, \Big[ {\rm Re}[\hat{M}_2\hat{M}_3^{\ast}]\cos\Delta\phi - {\rm Im}[\hat{M}_2\hat{M}_3^{\ast}]\sin\Delta\phi \Big]  \Delta D^{h_1/q}_{S_X/s_T}(z_1,p_{\perp 1}) \Delta D^{h_2/\bar{q}}_{S_X/s_T} (z_2,p_{\perp 2})\Big\} \nonumber\\
\eea
\bea
&&P^{h_1}_X P^{h_2}_Y\frac{d\sigma^{e^+e^-\rightarrow h_1 h_2 X}}{d \cos \theta d{\rm PS}_{12}} =
\frac{1}{32 \pi s} \nonumber\\
&&\times \sum_q \Big\{ \Big[ {\rm Im}[\hat{M}_2\hat{M}_3^{\ast}]\cos\Delta\phi + {\rm Re}[\hat{M}_2\hat{M}_3^{\ast}]\sin\Delta\phi \Big]\,
\Delta D^{h_1/q}_{S_X/s_T}(z_1,p_{\perp 1}) \Delta^-D^{h_2/\bar{q}}_{S_Y/s_T} (z_2,p_{\perp 2}) \,  \Big\}\nonumber\\
&&
\eea
\bea
&&P^{h_1}_X P^{h_2}_Z\frac{d\sigma^{e^+e^-\rightarrow h_1 h_2 X}}{d \cos \theta d{\rm PS}_{12}} =
\frac{1}{64 \pi s} \nonumber\\
&&\times \sum_q \Big\{  - \Big(|\hat{M}_2|^2+|\hat{M}_3|^2 \Big)  \Delta D^{h_1/q}_{S_X/s_L}(z_1,p_{\perp 1})\Delta D^{h_2/\bar{q}}_{S_Z/s_L}(z_2, p_{\perp 2}) \nonumber\\
&& \quad + 2 \, \Big[ {\rm Re}[\hat{M}_2\hat{M}_3^{\ast}]\cos\Delta\phi - {\rm Im}[\hat{M}_2\hat{M}_3^{\ast}]\sin\Delta\phi \Big]  \Delta D^{h_1/q}_{S_X/s_T}(z_1,p_{\perp 1})
\Delta D^{h_2/\bar{q}}_{S_Z/s_T} (z_2,p_{\perp 2})\Big\}\nonumber\\
\eea
\bea
&&P^{h_1}_Y P^{h_2}_X\frac{d\sigma^{e^+e^-\rightarrow h_1 h_2 X}}{d \cos \theta d{\rm PS}_{12}} =
\frac{1}{32 \pi s} \nonumber\\
&&\times \sum_q \Big\{- \Big[ {\rm Im}[\hat{M}_2\hat{M}_3^{\ast}]\cos\Delta\phi + {\rm Re}[\hat{M}_2\hat{M}_3^{\ast}]\sin\Delta\phi \Big]\,
\Delta^-D^{h_1/q}_{S_Y/s_T}(z_1,p_{\perp 1}) \Delta D^{h_2/\bar{q}}_{S_X/s_T} (z_2,p_{\perp 2}) \,  \Big\}\nonumber\\
&&
\eea
\bea
&& P^{h_1}_Y P^{h_2}_Y \frac{d\sigma^{e^+e^-\rightarrow h_1 h_2 X}}{d \cos \theta d{\rm PS}_{12}}  =
\frac{1}{64 \pi s} \nonumber\\
&&\times
\sum_q \Big\{\Big(|\hat{M}_2|^2+|\hat{M}_3|^2 \Big)  \Delta^N\! D_{h_1^\uparrow/q}(z_1,p_{\perp 1})\Delta^N\! D_{h_2^\uparrow/\bar q}(z_2, p_{\perp 2}) \nonumber\\
&&+ 2 \, \Big[ {\rm Re}[\hat{M}_2\hat{M}_3^{\ast}]\cos\Delta\phi - {\rm Im}[\hat{M}_2\hat{M}_3^{\ast}]\sin\Delta\phi \Big]  \Delta^- D^{h_1/q}_{S_Y/s_T}(z_1,p_{\perp 1})
\Delta^- D^{h_2/\bar{q}}_{S_Y/s_T} (z_2,p_{\perp 2})\Big\} \nonumber\\
\eea
\bea
&&P^{h_1}_Y P^{h_2}_Z\frac{d\sigma^{e^+e^-\rightarrow h_1 h_2 X}}{d \cos \theta d{\rm PS}_{12}}  =
\frac{1}{32 \pi s} \nonumber\\
&&\times  \sum_q
\Big\{-\Big[ {\rm Im}[\hat{M}_2\hat{M}_3^{\ast}]\cos\Delta\phi + {\rm Re}[\hat{M}_2\hat{M}_3^{\ast}] \sin\Delta\phi \Big]  \Delta^- D^{h_1/q}_{S_Y/s_T}(z_1,p_{\perp 1})
\Delta D^{h_2/\bar{q}}_{S_Z/s_T} (z_2,p_{\perp 2})\Big\}\nonumber\\
\eea
\bea
&&P^{h_1}_Z P^{h_2}_X\frac{d\sigma^{e^+e^-\rightarrow h_1 h_2 X}}{d \cos \theta d{\rm PS}_{12}}  =
\frac{1}{64 \pi s} \nonumber\\
&&\times \sum_q \Big\{  - \Big(|\hat{M}_2|^2+|\hat{M}_3|^2 \Big)  \Delta D^{h_1/q}_{S_Z/s_L}(z_1,p_{\perp 1})\Delta D^{h_2/\bar{q}}_{S_X/s_L}(z_2, p_{\perp 2}) \nonumber\\
&&+ 2 \, \Big[ {\rm Re}[\hat{M}_2\hat{M}_3^{\ast}]\cos\Delta\phi - {\rm Im}[\hat{M}_2\hat{M}_3^{\ast}]\sin\Delta\phi \Big]  \Delta D^{h_1/q}_{S_Z/s_T}(z_1,p_{\perp 1})
\Delta D^{h_2/\bar{q}}_{S_X/s_T} (z_2,p_{\perp 2})\Big\}\nonumber\\
\eea
\bea
&&P^{h_1}_Z P^{h_2}_Y\frac{d\sigma^{e^+e^-\rightarrow h_1 h_2 X}}{d \cos \theta d{\rm PS}_{12}}  =
\frac{1}{32 \pi s} \nonumber\\
&&\times \sum_q
\Big\{\Big[ {\rm Im}[\hat{M}_2\hat{M}_3^{\ast}]\cos\Delta\phi + {\rm Re}[\hat{M}_2\hat{M}_3^{\ast}] \sin\Delta\phi \Big]  \Delta D^{h_1/q}_{S_Z/s_T}(z_1,p_{\perp 1})
\Delta^- D^{h_2/\bar{q}}_{S_Y/s_T} (z_2,p_{\perp 2})\Big\}\nonumber\\
&&
\eea
\bea
&& P^{h_1}_Z P^{h_2}_Z \frac{d\sigma^{e^+e^-\rightarrow h_1 h_2 X}}{d \cos \theta d{\rm PS}_{12}}  =
\frac{1}{64 \pi s} \nonumber\\
&& \times \sum_q \Big\{  - \Big(|\hat{M}_2|^2+|\hat{M}_3|^2 \Big)  \Delta D^{h_1/q}_{S_Z/s_L}(z_1,p_{\perp 1})\Delta D^{h_2/\bar{q}}_{S_Z/s_L}(z_2, p_{\perp 2}) \nonumber\\
&&+ 2 \, \Big[ {\rm Re}[\hat{M}_2\hat{M}_3^{\ast}]\cos\Delta\phi - {\rm Im}[\hat{M}_2\hat{M}_3^{\ast}]\sin\Delta\phi \Big]  \Delta D^{h_1/q}_{S_Z/s_T}(z_1,p_{\perp 1})
\Delta D^{h_2/\bar{q}}_{S_Z/s_T} (z_2,p_{\perp 2})\Big\}\,.\nonumber\\
\eea
By using the expressions in Eqs.~(\ref{comb1})-(\ref{comb-unp}) one can directly obtain the results for the two kinematical configurations considered here.

\section{Hadron frame: complete results}
\label{had-conv}

As already pointed out, while in the thrust frame at least in principle one can directly measure the azimuthal dependences on $\varphi_{1,2}$ entering the formulas of the previous section, in the hadron frame the partonic variables are not accessible and one has to properly extract the azimuthal dependences on $\phi_1$.
This can be done by employing proper projection techniques thanks to a simple tensorial analysis, as described in Appendix~\ref{tensor}, without formulating any particular assumption on the $p_\perp$ dependence of the fragmentation functions. This will also allow for a more direct comparison with the results of Ref.~\cite{Boer:1997mf}.

As discussed in Section~\ref{2hadrons} we will have to compute the following expression
\bea
&&\frac{d\sigma^{e^+e^-\to h_1(S_1) h_2(S_2)\, X}}{d\cos\theta dz_1dz_2d^2\bm{P}_{1T}} \nonumber\\
&&= \int d^2\bm{p}_{\perp 1} \,d^2\bm{p}_{\perp 2}\, \delta^{(2)}(\bm{p}_{\perp 1}- \bm{P}_{1T} + \bm{p}_{\perp 2}\,z_{p_1}/z_{p_2})\,
\frac{d\sigma^{e^+e^-\to h_1(S_1) h_2(S_2)\, X}}{d \cos \theta dz_1 d^2\bm{p}_{\perp 1} dz_2 d^2\bm{p}_{\perp 2}}\,.
\eea
To this aim, we define, for a generic couple of (un)polarized TMD-FFs, one referring to the hadron $h_1$ ($\Delta D^{h_1}$) and one to the hadron $h_2$ ($\Delta D^{h_2}$), the following convolution over the intrinsic transverse momenta
\bea
\mathcal{C}[w \Delta D^{h_1}\Delta D^{h_2}] &= &  \sum_q e^2_q  \int d^2\bm{p}_{\bot 1} d^2\bm{p}_{\bot 2}\,\delta^{(2)} \big(\bm{p}_{\bot 1}  - \bm{P}_{ 1 T}  + \bm{p}_{\bot 2}\, z_{p_{1}} / z_{p_{2}}  \big)\nonumber\\
&& \times   w(\bm{p}_{\bot 2},\bm{P}_{1T})\Delta  D_{h_1/q}(z_1,p_{\bot 1}) \Delta D_{h_2/\bar q }(z_2,p_{\bot 2})  \,,\label{conv}
\eea
where the TMD-FFs depend only on the moduli of the intrinsic transverse momenta and the weight $w(\bm{p}_{\bot 2},\bm{P}_{1T}) $ depends on the specific azimuthal structure considered, as illustrated below. Notice that this expression differs from the corresponding one in Ref.~\cite{Boer:1997mf} by a factor $1/z_2^2$, and by the definition of the parton momenta, that we recall here below:
\bea
(-z \, \bm{k}_T)_{\rm Amsterdam}  &=& \bm{p} _\perp \label{a-noi2} \\
(\hat{\bm{h}})_{\rm Amsterdam} &=& \frac{\bm{P}_{1T}}{P_{1T}} = \hat{\bm{P}}_{1T} \>.
\label{a-noi3}
\eea

 By means of this procedure one obtains the following hadron-frame azimuthal dependences for the (un)polarized cross sections.

\subsection{Unpolarized case}
\begin{equation}
\frac{d\sigma^{e^+e^-\rightarrow h_1 h_2 X}}{d \cos \theta dz_1  dz_2 d^2\bm{P}_{1T}} =  \frac{3 \pi \alpha^2}{2 s} \bigg\{ \big( 1 + \cos^2\theta \big)  F_{UU}   +  \sin^2\theta \cos(2 \phi_1) F^{\cos(2\phi_1)}_{UU} \bigg \}\,,
\end{equation}
with
\begin{equation}
F_{UU} =  \sum_q e^2_q  \int  d^2\bm{p}_{\bot 2}\, D_{h_1/q}(z_1, p_{\bot 1}) D_{h_2/\bar{q}}(z_2, p_{\bot 2})
= \mathcal{C}[D_1 \bar{D}_1]
\label{FUU1}
\end{equation}
\bea
&&\cos(2\phi_1)\,F^{\cos(2\phi_1)}_{UU}\nonumber\\
&&= \sum_q e^2_q \int d^2\bm{p}_{\bot 2} \frac{1}{4} \Big[\frac{P_{1T}}{p_{\bot1}}  \cos(\phi_1 + \varphi_2) - \frac{ z_{p_1}}{z_{p_2}} \frac{p_{\bot2}}{ p_{\bot1}}
\cos(2 \varphi_2) \Big] \Delta^N\! D_{h_1/q^\uparrow}   \Delta^N\! D_{h_2/\bar{q}^\uparrow} \nonumber\\
& &= \cos(2\phi_1)\,\mathcal{C}\bigg[\frac{1}{4} \bigg\{\frac{P_{1T}}{p_{\perp 1}}\,\hat{\bm{p}}_{\bot 2} \cdot \hat{\bm{P}}_{1T}
- \frac{z_{p_1}}{z_{p_2}} \frac{p_{\perp 2}}{p_{\perp 1}}\,\bigg[ 2 \big(\hat{\bm{p}}_{\bot 2} \cdot \hat{\bm{P}}_{1T} \big)^2  - 1 \bigg]\bigg\} \Delta^N\! D_{h_1/q^\uparrow} \Delta^N\! D_{h_2/\bar{q}^\uparrow}\bigg] \nonumber\\
& & =\cos(2\phi_1)\,\mathcal{C}\bigg[ \bigg\{\frac{\bm{p}_{\bot 2} \cdot \bm{P}_{1T}} {z_1 z_2} -
\bigg[ 2 \bigg(\frac{\bm{p}_{\bot 2} \cdot \hat{\bm{P}}_{1T} }{z_2} \bigg)^2  - \frac{p^2_{\bot 2}}{z^2_2}\bigg]\bigg\} \frac{H^{\bot}_1 \bar{H}^{\bot}_1}{ M_{h_1} M_{h_2}}\bigg]   \,,
\label{FUUcos}
\eea
where the second line is obtained by using Eqs.~(\ref{cosfih1}) and (\ref{sinfih1}), and in the third line, valid neglecting terms in $M_h/\sqrt s$, we have switched to the Amsterdam notation (see Appendix~\ref{Amst}) for a more direct comparison with Ref.~\cite{Boer:1997mf}.

\subsection{Single-polarized case}

Let us start with the longitudinal polarization
\be
P^{h_1}_Z \frac{d\sigma^{e^+e^-\rightarrow h_1 h_2 X}}{d \cos \theta dz_1  dz_2 d^2\bm{P}_{1T}} =  \frac{3 \pi \alpha^2}{2 s}  \sin^2\theta \sin(2\phi_1)  F^{\sin(2\phi_1)}_{LU}
	\label{}
\ee
\bea
&&\sin(2\phi_1)\,F^{\sin(2\phi_1)}_{LU} \nonumber\\
&&= \sum_q e^2_q \int d^2\bm{p}_{\bot 2} \frac{1}{2} \Big[\frac{P_{1T}}{p_{\bot1}}  \sin(\phi_1 + \varphi_2) -
\frac{ z_{p_1}}{z_{p_2}}\frac{p_{\bot2}}{ p_{\bot1}} \sin(2 \varphi_2)\Big] \Delta D^{h_1/q}_{S_Z/s_T}   \Delta^{N}\! D_{h_2/\bar{q}^\uparrow} \nonumber\\
&&= \sin(2\phi_1)\,\mathcal{C}\bigg[\frac{1}{2} \bigg\{\frac{P_{1T}}{p_{\perp 1}}\, \hat{\bm{p}}_{\bot 2} \cdot \hat{\bm{P}}_{1T}
- \frac{z_{p_1}}{z_{p_2}} \frac{p_{\perp 2}}{p_{\perp 1}}\,\bigg[ 2 \big(\hat{\bm{p}}_{\bot 2} \cdot \hat{\bm{P}}_{1T} \big)^2  - 1 \bigg]\bigg\} \Delta D^{h_1/q}_{S_Z/s_T}   \Delta^{N}\! D_{h_2/\bar{q}^\uparrow} \bigg]\nonumber\\
&& = - \sin(2\phi_1)\,\mathcal{C}\bigg[\bigg\{ \frac{\bm{p}_{\bot 2} \cdot \bm{P}_{1T}}{z_1 z_2} - \bigg[2 \bigg(\frac{\bm{p}_{\bot 2} \cdot \hat{\bm{P}}_{1T} }{z_2} \bigg)^2  - \frac{p^2_{\bot 2}}{z^2_2}\bigg]\bigg\} \frac{   H^{\bot}_{1L} \bar{H}^{\bot}_1}{ M_{h_1} M_{h_2}}  \bigg]\,,
\eea
where once again the last equality is valid in the Amsterdam notation and  neglecting hadron mass effects.

For the transverse polarization, firstly we have to express it in the hadron frame, the laboratory ($L$) frame:
\be
\bm{P}^{h_1}_T= P^{h_1}_X \,\hat{\bm{X}}_{h_1} + P^{h_1}_Y \, \hat{\bm{Y}}_{h_1} = P^{h_1}_{x_L} \,\hat{\bm{x}}_{L} + P^{h_1}_{y_L} \, \hat{\bm{y}}_{L} = P^{h_1}_T (\cos\phi_{S_1}^L\,\hat{\bm{x}}_{L} + \sin\phi_{S_1}^L\,\hat{\bm{y}}_{L} )
\ee
adopting Eqs.~(\ref{eq:xh1}) and  (\ref{eq:yh1}). By combining the transverse polarization components in the lab frame as
\be
P_T^{h_1} = P^{h_1}_{x_L}\cos\phi_{S_1}^L + P^{h_1}_{y_L}\sin\phi_{S_1}^L\,,
\ee
we then get
\bea
&&P^{h_1}_T\, \frac{d\sigma^{e^+e^-\rightarrow h_1 h_2 X}}{d \cos \theta dz_1  dz_2 d^2\bm{P}_{1T}}= \frac{3\pi \alpha^2 }{2s}  \nonumber\\
&&\times \bigg\{ \Big( 1 + \cos^2\theta \Big)  \sin(\phi_1 - \phi_{S_1}^L) F^{\sin(\phi_1 - \phi_{S_1}^L)}_{TU}\nonumber\\
&&\quad +\, \sin^2\theta \bigg(  \sin(\phi_1 + \phi_{S_1}^L) F^{ \sin(\phi_1 + \phi_{S_1}^L)}_{TU} +   \sin(3\phi_1 - \phi_{S_1}^L) F^{\sin(3\phi_1 - \phi_{S_1}^L) }_{TU} \bigg) \bigg\}\,,\label{PTlab}
\eea
where
\bea
&&\sin(\phi_1 - \phi_{S_1}^L)\,F^{\sin(\phi_1 - \phi_{S_1}^L)}_{TU} \nonumber\\
&& =  \sum_q e^2_q \int d^2\bm{p}_{\bot 2} \Big[\frac{ z_{p_1}}{z_{p_2}}\,\frac{p_{\bot2}}{ p_{\bot1}} \sin{( \varphi_2-\phi_{S_1}^L)} - \frac{P_{1T}}{p_{\bot1}}  \sin(\phi_1 - \phi_{S_1}^L) \Big] \Delta^N\! D_{h_1^\uparrow/q}   D_{h_2/\bar{q}} \nonumber\\
& &= \sin(\phi_1 - \phi_{S_1}^L)\,
\mathcal{C}\bigg[ \bigg( \frac{z_{p_1}}{z_{p_2}} \frac{p_{\bot2}}{ p_{\bot1}} \,\hat{\bm{p}}_{\bot 2} \cdot \hat{\bm{P}}_{1T} - \frac{P_{1T}}{p_{\perp 1}} \bigg) \Delta^N\! D_{h_1^\uparrow/q} D_{h_2/\bar{q}} \bigg]\nonumber\\
&&= \sin(\phi_1 - \phi_{S_1}^L)\,
\mathcal{C}\bigg[  \bigg( \frac{\bm{p}_{\bot 2} \cdot \hat{\bm{P}}_{1T}}{z_2} - \frac{P_{1T}}{z_1} \bigg) \frac{D^{\bot}_{1T} \bar{D}_1}{M_{h_1}}  \bigg]
\label{FTU1}
\eea
\bea
&&2 \sin(\phi_1 + \phi_{S_1}^L)\,F^{ \sin(\phi_1 + \phi_{S_1}^L)}_{TU} \nonumber\\
&&=\sum_q e^2_q \int d^2\bm{p}_{\bot 2} \sin( \varphi_2+\phi_{S_1}^L) \frac{1}{2} \big(\Delta D^{h_1/q}_{S_X/s_T} + \Delta^-\! D^{h_1/q}_{S_Y/s_T}\big)\, \Delta^{N}\! D_{h_2/\bar{q}^{\uparrow}}
\nonumber \\
&&=
\sin(\phi_1 + \phi_{S_1}^L)\,\mathcal{C}\bigg [ \big( \hat{\bm{p}}_{\bot 2} \cdot \hat{\bm{P}}_{1T} \big) \frac{1}{2}\big(\Delta D^{h_1/q}_{S_X/s_T} + \Delta^-\! D^{h_1/q}_{S_Y/s_T}\big) \Delta^{N}\! D_{h_2/\bar{q}^{\uparrow}} \bigg ]\nonumber\\
&&=
2\sin(\phi_1 + \phi_{S_1}^L)\,\mathcal{C}\bigg [ \bigg( \frac{\bm{p}_{\bot 2} \cdot \hat{\bm{P}}_{1T}}{z_2} \bigg)\frac{H_1 \bar{H}^{\bot}_{1}}{ M_{h_2}}  \bigg ]
\label{FTU2}
\eea
\bea
&&2\sin(3\phi_1 - \phi_{S_1}^L)\,F^{\sin(3\phi_1 - \phi_{S_1}^L) }_{TU} \nonumber\\
&&=\sum_q e^2_q \int d^2\bm{p}_{\bot 2}\bigg\{ \frac{P_{1T}^2}{p_{\perp 1}^2} \sin(\varphi_2+2\phi_1-\phi_{S_1}^L) - 2 \frac{z_{p_1}}{z_{p_2}}\frac{P_{1T}p_{\perp 2}}{p_{\perp 1}^2}\sin(2\varphi_2+\phi_1-\phi_{S_1}^L) \nonumber\\
&& \quad + \frac{z_{p_1}^2}{z_{p_2}^2}\frac{p_{\perp 2}^2}{p_{\perp 1}^2} \sin(3\varphi_2-\phi_{S_1}^L) \bigg\} \frac{1}{2} \bigg(\Delta D^{h_1/q}_{S_X/s_T} - \Delta^-\! D^{h_1/q}_{S_Y/s_T}\bigg)\, \Delta^{N}\! D_{h_2/\bar{q}^{\uparrow}}\nonumber\\
& &=
\sin(3\phi_1 - \phi_{S_1}^L) \,\mathcal{C} \Bigg[\bigg\{ \frac{z_{p_1}^2}{z_{p_2}^2} \frac{p_{\perp 2}^2}{p_{\perp 1}^2}\bigg[
4\big( \hat{\bm{p}}_{\bot 2} \cdot \hat{\bm{P}}_{1T} \big)^3  - 3\, \big( \hat{\bm{p}}_{\bot 2} \cdot \hat{\bm{P}}_{1T} \big)\bigg]  + \frac{P_{1T}^2}{p_{\perp 1}^2}\,
 \big( \hat{\bm{p}}_{\bot 2} \cdot \hat{\bm{P}}_{1T} \big)  \nonumber \\
&&\quad - 2\, \frac{z_{p_1}}{z_{p_2}}\frac{p_{\perp 2} P_{1T}}{p_{\perp 1}^2}\bigg[ 2\big( \hat{\bm{p}}_{\bot 2} \cdot \hat{\bm{P}}_{1T} \big)^2 - 1 \bigg]  \bigg\}
\frac{1}{2} \bigg(\Delta D^{h_1/q}_{S_X/s_T} - \Delta^-\! D^{h_1/q}_{S_Y/s_T}\bigg)\, \Delta^{N}\! D_{h_2/\bar{q}^{\uparrow}}  \Bigg]\nonumber\\
& &=
2\sin(3\phi_1 - \phi_{S_1}^L) \,\mathcal{C} \Bigg[\bigg\{ \bigg[
4\bigg( \frac{\bm{p}_{\bot 2} \cdot \hat{\bm{P}}_{1T}}{z_2} \bigg)^3  - 3\,\frac{p^2_{\bot 2}}{z^2_2} \bigg( \frac{\bm{p}_{\bot 2} \cdot \hat{\bm{P}}_{1T}}{z_2} \bigg)\bigg]   \nonumber \\
&&\quad +   \bigg( \frac{\bm{p}_{\bot 2} \cdot \bm{P}_{1T}}{z_1 z_2} \bigg) \frac{P_{1T}}{z_1} - 2\bigg[ 2\bigg( \frac{\bm{p}_{\bot 2} \cdot \hat{\bm{P}}_{1T}}{z_2} \bigg)^2 - \frac{p^2_{\bot 2}}{z^2_2} \bigg] \frac{P_{1T}}{z_1} \bigg\} \frac{   H^{\bot}_{1T} \bar{H}^{\bot}_{1} }{2 M^2_{h_1} M_{h_2}} \Bigg]\,.\label{FTU3}
\eea

A relevant projection, often adopted in experimental analyses, is the polarization orthogonal to the hadron plane, that is along
\be
\hat{\bm{n}} \equiv (\cos\phi_n, \sin\phi_n,0)= \frac{- \bm{P}_2  \times  \bm{P}_1}{|\bm{P}_2  \times  \bm{P}_1|} = -\sin\phi_1 \hat{\bm{x}}_L + \cos\phi_1 \hat{\bm{y}}_L\,,
\ee
implying
\be
 \phi_n = \phi_1 + \frac{\pi}{2}\,.
\ee
The polarization along $\hat{\bm{n}}$ can be directly obtained by identifying $\phi_{S_1}^L=\phi_n$ in Eq.~(\ref{PTlab}) as follows
\bea
&& P^{h_1}_n \frac{d\sigma^{e^+e^-\rightarrow h_1 h_2 X}}{d \cos \theta dz_1  dz_2 d^2\bm{P}_{1T}}\nonumber\\
&& \equiv  P^{h_1}_T(\phi_{S_1}^L=\phi_n)\, \frac{d\sigma^{e^+e^-\rightarrow h_1 h_2 X}}{d \cos \theta dz_1  dz_2 d^2\bm{P}_{1T}}= \frac{3\pi \alpha^2 }{2s}  \nonumber\\
&&\times
 \bigg\{ -\big( 1 + \cos^2\theta \big)  F^{\sin(\phi_1 - \phi_{S_1}^L)}_{TU}+ \sin^2\theta\cos(2\phi_1) \bigg(   F^{ \sin(\phi_1 + \phi_{S_1}^L)}_{TU} -   F^{\sin(3\phi_1 - \phi_{S_1}^L) }_{TU} \bigg) \bigg\}\,.\nonumber \\
\eea
In such a case, by integrating over $\bm{P}_{1T}$, and therefore over the azimuthal angle $\phi_1$, one is sensitive only to the structure function $F^{\sin(\phi_1 - \phi_{S_1}^L)}_{TU}$, Eq.~(\ref{FTU1}), that is given as a convolution of the polarizing TMD-FF for hadron $h_1$ with the unpolarized TMD-FF for hadron $h_2$. Notice that also in the denominator only the term $F_{UU}$, Eq.~(\ref{FUU1}), survives.
The final expression, simplifying a common factor $(1+\cos^2\theta)$, reads
\be
P^{h_1}_n(z_1,z_2) = -\frac{F^{\sin(\phi_1 - \phi_{S_1}^L)}_{TU}}{F_{UU}} \,.
\label{Pn1}
\ee
This is indeed the strategy adopted in the recent phenomenological analysis performed in Ref.~\cite{DAlesio:2020wjq}.

\subsection{Double-polarized case}

In the following cases we give directly the expressions in terms of convolutions.
\begin{equation}
P^{h_1}_Z P^{h_2}_Z\frac{d\sigma^{e^+e^-\rightarrow h_1 h_2 X}}{d \cos \theta dz_1  dz_2 d^2\bm{P}_{ 1 T}} =\frac{3\pi\alpha^2 }{2s}\Bigg\{ -  \Big( 1 + \cos^2\theta \Big) F_{LL} + \sin^2 {\theta} \cos(2\phi_1) F^{\cos(2\phi_1)}_{LL}  \Bigg\} \,,
\label{}
\end{equation}
where
\bea
F_{LL} & = &\mathcal{C}\big[G_{1L}\bar{G}_{1L}\big]\\
F^{\cos(2\phi_1)}_{LL} &=& \mathcal{C}\Bigg[\bigg\{\frac{P_{1T}}{p_{\perp 1}} \,\hat{\bm{p}}_{\bot 2} \cdot \hat{\bm{P}}_{1T}
- \frac{z_{p_1}}{z_{p_2}} \frac{p_{\bot 2}}{p_{\perp 1}} \bigg[2 \big(\hat{\bm{p}}_{\bot 2} \cdot \hat{\bm{P}}_{1T} \big)^2  - 1 \bigg]\bigg\}
\Delta D^{h_1/q}_{S_Z/s_T}\Delta D^{h_2/\bar q}_{S_Z/s_T}   \Bigg]\nonumber \\
&=& \mathcal{C}\Bigg[\bigg\{ \frac{\bm{p}_{\bot 2} \cdot \bm{P}_{1T}}{z_1 z_2}
- \bigg[2 \bigg(\frac{\bm{p}_{\bot 2} \cdot \hat{\bm{P}}_{1T} }{z_2} \bigg)^2  - \frac{p^2_{\bot 2}}{z^2_2}\bigg]\bigg\}\frac{H^{\bot}_{1L}\bar{H}^{\bot}_{1L}}{M_{h_1} M_{h_2}}   \Bigg]
\,.
\eea

For the case of a transversely-longitudinally polarized hadron pair we have
\bea
&& P^{h_1}_T  P^{h_2}_Z \frac{d\sigma^{e^+e^-\rightarrow h_1 h_2 X}}{d \cos \theta dz_1  dz_2 d^2\bm{P}_{ 1 T}}= \frac{3\pi \alpha^2 }{2 {s}} \nonumber\\
&&\times  \Bigg\{  \Big( 1 + \cos^2\theta \Big) \cos(\phi_1 - \phi_{S_1}^L) F^{\cos(\phi_1 - \phi_{S_1}^L )}_{TL} \nonumber\\
&&\quad + \sin^2 {\theta} \bigg[ \cos(\phi_1 + \phi_{S_1}^L ) F^{\cos(\phi_1 + \phi_{S_1}^L )}_{TL} +  \cos(3\phi_1 - \phi_{S_1}^L ) F^{\cos(3\phi_1 - \phi_{S_1}^L )}_{TL}   \bigg] \Bigg\}\,,
\eea
where
\bea
F^{\cos(\phi_1 - \phi_{S_1}^L )}_{TL} &=&  \mathcal{C}\Bigg[  \bigg(\frac{z_{p_1}}{z_{p_2}}\frac{p_{\perp 2}}{p_{\perp 1}}\,\hat{\bm{p}}_{\bot 2} \cdot \hat{\bm{P}}_{1T}  -\frac{P_{1T}}{p_{\perp 1}}  \bigg) \Delta D^{h_1/q}_{S_X/s_L} \Delta D^{h_2/\bar q}_{S_Z/s_L}  \Bigg]\nonumber\\
& = &  - \,\mathcal{C}\Bigg[  \bigg(\frac{\bm{p}_{\bot 2} \cdot \hat{\bm{P}}_{1T} }{z_2}   -\frac{P_{1T}}{z_1}  \bigg) \frac{G_{1T}\bar{G}_{1L}}{M_{h_1} } \Bigg]
 \eea
 \bea
F^{\cos(\phi_1 + \phi_{S_1}^L )}_{TL} & = &
\mathcal{C} \Bigg[ \big(\hat{\bm{p}}_{\bot 2} \cdot \hat{\bm{P}}_{1T} \big)  \frac{1}{2}\bigg(\Delta D^{h_1/q}_{S_X/s_T} + \Delta^-\! D^{h_1/q}_{S_Y/s_T}\bigg) \Delta D^{h_2/\bar q}_{S_Z/s_T}  \Bigg]\nonumber\\
&=& -\,
\mathcal{C}\Bigg[ \bigg(\frac{\bm{p}_{\bot 2} \cdot \hat{\bm{P}}_{1T} }{z_2} \bigg)  \frac{H_{1}\bar{H}^{\bot}_{1L}}{M_{h_2} }  \Bigg]
\eea
\bea
2\, F^{\cos(3\phi_1 - \phi_{S_1}^L )}_{TL} &=&
\mathcal{C}\Bigg[\bigg\{\frac{z_{p_1}^2}{z_{p_2}^2}
\frac{p_{\perp 2}^2}{p_{\perp 1}^2}
\bigg[  4 \big( \hat{\bm{p}}_{\bot 2} \cdot \hat{\bm{P}}_{1T} \big)^3  -3  \big( \hat{\bm{p}}_{\bot 2} \cdot \hat{\bm{P}}_{1T} \big)\bigg] +   \frac{P_{1T}^2}{p_{\perp 1}^2}\,
\big( \hat{\bm{p}}_{\bot 2} \cdot \hat{\bm{P}}_{1T}\big)  \nonumber\\
&-& 2\,\frac{z_{p_1}}{z_{p_2}} \frac{p_{\perp 2}P_{1T}}{p^2_{\perp 1}} \bigg[ 2\big( \hat{\bm{p}}_{\bot 2} \cdot \hat{\bm{P}}_{1T} \big)^2 - 1 \bigg]  \bigg\}  \bigg(\Delta D^{h_1/q}_{S_X/s_T} - \Delta^-\! D^{h_1/q}_{S_Y/s_T}\bigg) \Delta D^{h_2/\bar q}_{S_Z/s_T}    \Bigg]\nonumber\\
&=&
-\,\mathcal{C}\Bigg[\bigg\{\bigg[  4 \bigg( \frac{\bm{p}_{\bot 2} \cdot \hat{\bm{P}}_{1T}}{z_2} \bigg)^3  -3 \frac{p^2_{\bot 2}}{z^2_2} \bigg( \frac{\bm{p}_{\bot 2} \cdot \hat{\bm{P}}_{1T}}{z_2} \bigg)\bigg] +  \bigg( \frac{\bm{p}_{\bot 2} \cdot \bm{P}_{1T}}{z_1 z_2} \bigg) \frac{P_{1T}}{z_1} \nonumber\\
&-& 2\, \bigg[ 2\bigg( \frac{\bm{p}_{\bot 2} \cdot \hat{\bm{P}}_{1T}}{z_2} \bigg)^2 - \frac{p^2_{\bot 2}}{z^2_2} \bigg] \frac{P_{1T}}{z_1} \bigg\} \frac{H^{\bot}_{1T}\bar{H}^{\bot}_{1L}}{M_{h_1}^2M_{h_2}}    \Bigg]
\,,
\eea
where the minus signs appearing when switching to the Amsterdam notation come from the relative signs between the definition of the corresponding TMD-FFs.

Finally, when both hadrons are transversely polarized we have
\bea
& & P^{h_1}_T P^{h_2}_T\frac{d\sigma^{e^+e^-\rightarrow h_1 h_2 X}}{d \cos \theta dz_1  dz_2 d^2\bm{P}_{ 1 T}} =
\frac{3 \pi \alpha^2}{2s} \nonumber\\
&&\times \bigg\{ \Big( 1 + \cos^2\theta \Big)
\frac{1}{2}  \Big[\cos(2\phi_1 - \phi_{S_1}^L- \phi_{S_2}^L)F^{\cos(2\phi_1 - \phi_{S_1}^L- \phi_{S_2}^L)}_{TT} - \cos(\phi_{S_1}^L- \phi_{S_2}^L)F^{\cos(\phi_{S_1}^L- \phi_{S_2}^L)}_{TT} \Big] \nonumber\\
&&\quad +\,\sin^2\theta \Big[  \cos(\phi_{S_1}^L + \phi_{S_2}^L)F^{\cos(\phi_{S_1}^L + \phi_{S_2}^L)}_{TT} + \cos(2\phi_1 + \phi_{S_1}^L - \phi_{S_2}^L)F^{\cos(2\phi_1 +\phi_{S_1}^L - \phi_{S_2}^L)}_{TT} \nonumber\\
&&\quad +\,\cos(2\phi_1 - \phi_{S_1}^L + \phi_{S_2}^L)F^{\cos(2\phi_1 -\phi_{S_1}^L + \phi_{S_2}^L)}_{TT} + \cos(4\phi_1 - \phi_{S_1}^L - \phi_{S_2}^L)F^{\cos(4\phi_1 -\phi_{S_1}^L - \phi_{S_2}^L)}_{TT} \Big] \bigg\}  \Bigg\} \,,\nonumber\\
\label{PTT}
\eea
where
\bea
F^{\cos( \phi_{S_1}^L + \phi_{S_2}^L ) }_{TT} &=& \mathcal{C}\bigg[
\frac{1}{2}\big(\Delta D^{h_1/q}_{S_X/s_T} + \Delta^-\! D^{h_1/q}_{S_Y/s_T}\big) \frac{1}{2}\big(\Delta D^{h_2/\bar q}_{S_X/s_T} + \Delta^-\! D^{h_2/\bar q}_{S_Y/s_T}\big) \bigg]=
\mathcal{C}\big[ H_1 \bar{H}_1 \big]\nonumber\\
\eea
\bea
F^{\cos(2\phi_1 - \phi_{S_1}^L - \phi_{S_2}^L ) }_{TT} &=&
\mathcal{C}\Bigg[ \bigg\{ \frac{P_{1T}}{p_{\perp 1}}\,\hat{\bm{p}}_{\bot 2} \cdot \hat{\bm{P}}_{1T} - \frac{z_{p_ 1}}{z_{p_ 2}} \frac{p_{\perp 2}}{p_{\perp 1}}\big[2 \big(\hat{\bm{p}}_{\bot 2} \cdot \hat{\bm{P}}_{1T} \big)^2  - 1 \big]\bigg\} \nonumber\\
&&\quad\times \bigg(\Delta^N\!D_{h_1^\uparrow/q} \Delta^N\!D_{h_2^\uparrow/\bar q} - \Delta D^{h_1/q}_{S_X/s_L} \Delta D^{h_2/\bar q}_{S_X/s_L} \bigg)  \Bigg]\nonumber\\
&=&
\mathcal{C}\Bigg[ \bigg\{ \frac{\bm{p}_{\bot 2} \cdot \bm{P}_{1T}}{z_1 z_2} - \bigg[2 \bigg(\frac{\bm{p}_{\bot 2} \cdot \hat{\bm{P}}_{1T} }{z_2} \bigg)^2  - \frac{p^2_{\bot 2}}{z^2_2}\bigg]\bigg\} \frac{ D^{\bot}_{1T} \bar{D}^{\bot}_{1T} - G_{1T}\bar{G}_{1T} }{M_{h_1} M_{h_2}} \Bigg]\nonumber\\
\eea
\bea
F^{\cos(\phi_{S_1}^L - \phi_{S_2}^L ) }_{TT} &= & \mathcal{C}\Bigg[ \bigg\{\frac{P_{1T}}{p_{\perp 1}}\,\hat{\bm{p}}_{\bot 2} \cdot \hat{\bm{P}}_{1T}  - \frac{z_{p_ 1}}{z_{p_ 2}} \frac{p_{\perp 2}}{p_{\perp 1}}  \bigg\}\nonumber\\
&& \quad\times \bigg(\Delta^N\!D_{h_1^\uparrow/q} \Delta^N\!D_{h_2^\uparrow/\bar q} + \Delta D^{h_1/q}_{S_X/s_L} \Delta D^{h_2/\bar q}_{S_X/s_L} \bigg) \Bigg]\nonumber\\
&= & \mathcal{C}\Bigg[ \bigg\{\frac{\bm{p}_{\bot 2} \cdot \bm{P}_{1T}}{z_1 z_2}  - \frac{p^2_{\bot 2}}{z^2_2}  \bigg\} \frac{ D^{\bot}_{1T} \bar{D}^{\bot}_{1T} + G_{1T}\bar{G}_{1T} }{M_{h_1} M_{h_2}} \Bigg]
\eea
\bea
F^{\cos( 2 \phi_1 + \phi_{S_1}^L - \phi_{S_2}^L ) }_{TT} &=& \mathcal{C}\Bigg[\bigg\{2 \big(\hat{\bm{p}}_{\bot 2} \cdot \hat{\bm{P}}_{1T} \big)^2 - 1 \bigg\} \nonumber\\
&& \quad\times \frac{1}{2}\bigg(\Delta D^{h_1/q}_{S_X/s_T} + \Delta^-\! D^{h_1/q}_{S_Y/s_T}\bigg)  \frac{1}{2}\bigg(\Delta D^{h_2/\bar q}_{S_X/s_T} - \Delta^-\! D^{h_2/\bar q}_{S_Y/s_T}\bigg)  \Bigg]\nonumber\\
&=&\mathcal{C}\Bigg[\bigg\{2 \bigg(\frac{\bm{p}_{\bot 2} \cdot \hat{\bm{P}}_{1T} }{z_2} \bigg)^2 - \frac{p^2_{\bot 2}}{z^2_2} \bigg\} \frac{ H_1 \bar{H}^{\bot}_{1T}}{2  M^2_{h_2}} \Bigg]
\eea
\bea
F^{\cos( 2 \phi_1 - \phi_{S_1}^L + \phi_{S_2}^L ) }_{TT} &=& \mathcal{C}\Bigg[  \bigg\{   \frac{P^2_{1T}}{p_{\perp 1}^2} + \frac{z^2_{p_1}}{z^2_{p_ 2}} \frac{p^2_{\perp 2}}{p^2_{\perp 1}}
\bigg[2 \big(\hat{\bm{p}}_{\bot 2} \cdot \hat{\bm{P}}_{1T} \big)^2 - 1\bigg] - 2\frac{z_{p_1}}{z_{p_ 2}} \frac{p_{\perp 2}P_{1T}}{p^2_{\perp 1}}  \big(\hat{\bm{p}}_{\bot 2} \cdot \hat{\bm{P}}_{1T} \big) \bigg\} \nonumber\\
&&\quad \times \frac{1}{2}\bigg(\Delta D^{h_1/q}_{S_X/s_T} - \Delta^-\! D^{h_1/q}_{S_Y/s_T}\bigg)  \frac{1}{2}\bigg(\Delta D^{h_2/\bar q}_{S_X/s_T} + \Delta^-\! D^{h_2/\bar q}_{S_Y/s_T}\bigg)
\Bigg]\nonumber\\
&=& \mathcal{C}\Bigg[  \bigg\{   \frac{P^2_{1T}}{z^2_1} +
\bigg[2 \bigg(\frac{\bm{p}_{\bot 2} \cdot \hat{\bm{P}}_{1T} }{z_2} \bigg)^2 - \frac{p^2_{\bot 2}}{z^2_2}\bigg] - 2 \bigg(\frac{\bm{p}_{\bot 2} \cdot \bm{P}_{1T} }{z_1 z_2} \bigg) \bigg\} \frac{ H^{\bot}_{1T} \bar{H}_1}{2  M^2_{h_1}} \Bigg]\nonumber\\
\eea
\bea
&&F^{\cos( 4 \phi_1 - \phi_{S_1}^L - \phi_{S_2}^L ) }_{TT} \nonumber\\
&& = \mathcal{C}\Bigg[ \bigg\{    %
   8 \frac{z^2_{p_ 1}}{z^2_{p_ 2}} \frac{p^2_{\perp 2}}{p^2_{\perp 1}}\bigg[ \big(\hat{\bm{p}}_{\bot 2} \cdot \hat{\bm{P}}_{1T} \big)^4%
   - \big(\hat{\bm{p}}_{\bot 2} \cdot \hat{\bm{P}}_{1T} \big)^2+\frac{1}{8}\bigg]%
   -8\frac{z_{p_ 1}}{z_{p_ 2}} \frac{p_{\perp 2}P_{1T}}{p^2_{\perp 1}} \big(\hat{\bm{p}}_{\bot 2} \cdot \hat{\bm{P}}_{1T} \big)^3%
   \nonumber\\ %
   &&\quad + \frac{P^2_{1T}}{p^2_{\perp 1}}\big[2 \big(\hat{\bm{p}}_{\bot 2} \cdot \hat{\bm{P}}_{1T} \big)^2%
   - 1\big] %
   + 6\frac{z_{p_1}}{z_{p_2}}\frac{P_{1T} p_{\bot 2}}{p^2_{\perp 1}} \big(\hat{\bm{p}}_{\bot 2} \cdot \hat{\bm{P}}  _{1T} \big) \bigg\} \nonumber\\
   && \quad\quad\times \frac{1}{2}\bigg(\Delta D^{h_1/q}_{S_X/s_T} - \Delta^-\! D^{h_1/q}_{S_Y/s_T}\bigg)  \frac{1}{2}\bigg(\Delta D^{h_2/\bar q}_{S_X/s_T} - \Delta^-\! D^{h_2/\bar q}_{S_Y/s_T}\bigg)
    \Bigg]\nonumber\\
  &&= \mathcal{C}\Bigg[ \bigg\{    %
   8 \bigg[ \bigg(\frac{\bm{p}_{\bot 2} \cdot \hat{\bm{P}}_{1T} }{z_2} \bigg)^4%
   -\frac{p^2_{\bot 2}}{z^2_2} \bigg(\frac{\bm{p}_{\bot 2} \cdot \hat{\bm{P}}_{1T} }{z_2} \bigg)^2+\frac{p^4_{\bot 2}}{8z^4_2}\bigg]%
   -8\frac{P_{1T}}{z_1} \bigg(\frac{\bm{p}_{\bot 2} \cdot \hat{\bm{P}}_{1T} }{z_2} \bigg)^3%
   \nonumber\\ %
   &&\quad + \frac{P^2_{1T}}{z^2_1z^2_2}\big[ 2\big(\bm{p}_{\bot 2} \cdot \hat{\bm{P}}_{1T} \big)^2%
   - p^2_{\bot 2}\big] %
   + 6\frac{P_{1T} p^2_{\bot 2}}{z_1 z^2_2} \bigg(\frac{\bm{p}_{\bot 2} \cdot \hat{\bm{P}}  _{1T} }{z_2} \bigg) \bigg\} \frac{ H^{\bot}_{1T} \bar{H}^{\bot}_{1T}}{4 M^2_{h_1}  M^2_{h_2}} \Bigg] \,.\nonumber\\
\eea

\subsection{A Gaussian model}
\label{gauss}
The above results can be cast in a more explicit and simplified form by adopting a suitable functional $p_\perp$ dependence for the TMD-FFs involved. We focus for its relevance and complexity on the hadron-frame configuration, since for the thrust-frame case the intrinsic transverse momentum dependence of the two hadrons is completely factorized. In this respect the use of a Gaussian-like dependence together with the leading-order approximation in $p_\perp/\sqrt s$ (for not too small $z$ values), allow us to carry out, analytically, all the integrations over the intrinsic transverse momenta, leading to simplified expressions.

We consider explicitly three cases that represent the TMD-FFs appearing in the present study. For a generic fragmentation function of a quark into a hadron $h$, that is the unpolarized and any of the single- or double-polarized FF will therefore assume respectively the following form
\bea
{D}^{\rm unp}_h(z,p_\perp)& =& D^{\rm unp}_h(z)\,g(p_\perp)\\
\Delta{D}_h(z,p_\perp)& = & \Delta D_h(z)\,g^{(i)}(p_\perp)\,,
\eea
with $i=0,1,2$. Our parameterizations are required to respect angular momentum conservation in the forward direction, therefore we define
\begin{equation}
g^{(i)}(p_\perp) = \Bigl(\frac{p_\perp}{M}\Bigr)^i\,h^{(i)}(p_\perp)\,.
\label{gip}
\end{equation}
In particular, $g(p_\perp)$ is a simple Gaussian normalized to unity:
\begin{equation}
g(p_\perp) = \frac{1}{\pi\langle\,p_\perp^2\,\rangle}\,
e^{-p_\perp^2/\langle p_\perp^2 \rangle}\,,
\label{g0p}
\end{equation}
while
\begin{equation}
g^{(i)}(p_\perp) = \left(\frac{p_\perp}{M}\right)^i\,h^{(i)}(p_\perp) =
K_i\,\left(\frac{p_\perp}{M}\right)^i\,e^{-p_\perp^2/\langle p_\perp^2\rangle_{_\Delta}}\,,
\qquad i=0,1,2\,,
\label{hip}
\end{equation}
where $\langle p_\perp^2\rangle_{_\Delta}$ is the $p_\perp$ width of the corresponding $\Delta{ D}$. Notice that $g^{(0)}(p_\perp)$ is still a simple Gaussian, parametrizing the transverse momentum dependence of double polarized TMD-FFs surviving in the collinear limit.

The normalization constants, $K_i$, can be chosen to properly fulfill the positivity bounds. A simple way is indeed to impose the bound separately on the $p_\perp$- and the $z$-dependent part. The first is trivial since we can use the explicit functional form. For instance, for the unpolarized and the single-polarized TMD-FF, like the Collins ($C$) and the polarizing ($p$) FF, we have
\bea
D_{h/q }(z,p_{\perp }) &  = & D_{h/q }(z)  \frac{e^{-p^2_{\perp}/ \langle p^2_{\perp} \rangle}}{\pi \langle p^2_{\perp} \rangle}\\
\Delta D_{h/q}^{C,p}(z,p_{\perp}) &=& \Delta D_{h/q}^{C,p}(z) \sqrt{2e}\,\frac{p_{\perp}}{M_{_{C,p}}} \frac{e^{-p^2_{\perp}/ \langle p^2_{\perp} \rangle_{_{C,p}}}}{\pi \langle p^2_{\perp} \rangle}\,,
\eea
where
\be
\langle p^2_{\perp} \rangle_{_{C,p}} = \frac{\langle p^2_{\perp} \rangle M_{_{C,p}}^2}{\langle p^2_{\perp} \rangle + M_{_{C,p}}^2}\,.
\ee
These expressions fulfill the positivity bounds of the TMD-FFs by simply imposing the corresponding bound on the $z$-dependent parts.

Here below we present the results for the azimuthal dependences and polarization observables in the hadron frame entering Eq.~(\ref{desahdfr}) and the convolutions defined in Sec.~\ref{had-conv}, limiting ourselves to the unpolarized and single-polarized cases. These indeed are at present the most interesting from the phenomenological point of view.

We list here all expressions:
\be
\label{FUU}
F_{UU} = D_{h_1/q}(z_1) D_{h_2/\bar{q}}(z_2)
\frac{e^{-P^2_{1T}/ \langle p^2_\perp\rangle_{_{12}}}}{\pi\langle p^2_\perp\rangle_{_{12}}}
\ee
\bea
F^{\cos(2\phi_1)}_{UU} &=& \Delta^{N}\! D_{h_1/q^{\uparrow}}(z_1)  \Delta^{N}\! D_{h_2/\bar{q}^{\uparrow}}(z_2) \nonumber\\
&& \times \frac{e}{2\,M_{_{C_1}} M_{_{C_2}}} \frac{e^{-P^2_{1T}/\langle p^2_\perp\rangle_{_{C_1,C_2}}}}{\pi \langle p^2_\perp\rangle_{_{C_1,C_2}}} \frac{z_{p_1}}{z_{p_2}} \frac{P^2_{1T}}{ \langle p^2_\perp\rangle_{_{C_1,C_2}}^2}  \frac{\langle p^2_{\perp} \rangle^2_{_{C_1}} \langle p^2_{\perp} \rangle^2_{_{C_2}}}{\langle p^2_{\perp} \rangle_{_1} \langle p^2_{\perp} \rangle_{_2}}\,,
\eea
where
\bea
\label{bt2unp}
\langle p^2_\perp\rangle_{_{12}} & = &\frac{z^2_{p_2}\langle p^2_{\perp} \rangle_{_1}   +  z^2_{p_1} \langle p^2_{\perp} \rangle_{_2} }{z^2_{p_2}} \\
\langle p^2_\perp\rangle_{_{C_1, C_2}} & = &\frac{z^2_{p_2}\langle p^2_{\perp} \rangle_{_{C_1}} + z^2_{p_1}  \langle p^2_{\perp} \rangle_{_{C_2}}}{z^2_{p_2}}\,,
\eea
and
\be
\Delta^N\! D_{h_1/q^\uparrow}(z) \equiv \Delta D^C_{h_1/q}(z)\,.
\ee

Notice that for Gaussian widths independent of the scaling variables and for hadrons of the same kind, the above formulae simplify and coincide with the results of Ref.~\cite{Anselmino:2007fs}.

For its later use, we define
\be
\langle p^2_\perp\rangle_{_{\Delta_1, \Delta_2}} = \frac{z^2_{p_2}\langle p^2_{\perp} \rangle_{_{\Delta_1}}   + z^2_{p_1}  \langle p^2_{\perp} \rangle_{_{\Delta_2}}}{z^2_{p_2}}\,,
\ee
and, in order to have more compact expressions,
\be
{\cal G}(\Delta_1,\Delta_2) \equiv  \frac{1}{\pi \langle p^2_\perp\rangle_{_{\Delta_1,\Delta_2}}}\exp{\Big[-\frac{P^2_{1T}}{\langle p^2_\perp\rangle_{_{\Delta_1,\Delta_2}}}\Big]}\,.
\ee
Moving to the single-polarized hadron production we have
\be
F^{\sin(2\phi_1)}_{LU} = \Delta D^{h_1/q}_{S_Z/s_T}(z_1) \Delta^{N}\! D_{h_2/\bar{q}^{\uparrow}}(z_2)
\frac{e\,{\cal G}(LT_1,C_2)}{M_{_{LT_1}} M_{_{C_2}}}\,
\frac{P^2_{1T}} {\langle p^2_\perp\rangle_{_{LT_1,C_2}}^2} \, \frac{z_{p_1}}{z_{p_2}}  \frac{\langle p^2_{\perp} \rangle^2_{_{LT_1}} \langle p^2_{\perp} \rangle^2_{_{C_2}}}{\langle p^2_{\perp} \rangle_{_1} \langle p^2_{\perp} \rangle_{_2}}\,,
\ee
where we have adopted the following parametrization
\be
\Delta D^{h/q}_{S_Z/s_T}(z,p_{\perp}) = \Delta D^{h/q}_{S_Z/s_T}(z) \sqrt{2e} \frac{p_{\perp}}{M_{_{LT}}}\frac{e^{-p^2_{\perp}/ \langle p^2_{\perp} \rangle_{_{LT}}}}{\pi \langle p^2_{\perp} \rangle}\,.
\ee

Notice that at the lowest order in $\eta_{1T}$ this is also the longitudinal polarization in the laboratory frame.

For the computation of the transverse polarization it is more convenient to parametrize the transversely polarized TMD-FFs as follows
\bea
D^{+-}_{+-}(z,p_\perp) & \equiv & H_1(z,p_\perp) =  H_1(z)  \frac{ e^{- p^2_{\perp}/ \langle p^2_{\perp} \rangle_{_T}}}{\pi \langle p^2_{\perp} \rangle_{_T}}\\
D^{+-}_{-+}(z,p_\perp) & \equiv & \frac{p^2_{\perp}}{2 z^2 M^2_h}H^{\perp}_{1T}(z,p_\perp) = H_{1T}^\perp (z)\, e\,\frac{p^2_{\perp}}{M^2_{_{TT}}} \frac{ e^{-p^2_{\perp}/ \langle p^2_{\perp} \rangle_{_{TT}}}}{\pi \langle p^2_{\perp} \rangle}\,,	
\eea
where in the second equation we have kept distinct the mass parameter entering the relation between the two standard notations and the one related to the $p_\perp$ dependence to be extracted from data.

This leads to the following results
\be
F^{\sin(\phi_1 - \phi^L_{S_1})}_{TU} = -\Delta^N\! D_{h_1^\uparrow/q}(z_1) D_{h_2/\bar{q}}(z_2)
\,\frac{\sqrt{2e}\,{\cal G}(p_1,2)}{M_{_{p_1}}}
\frac{P_{1T}}{\langle p^2_{\perp} \rangle_{_{p_1,2}}} \frac{\langle p^2_{\perp} \rangle^2_{_{p_1}}}{\langle p^2_{\perp} \rangle_{_1}}
\label{FTUpol}
\ee
\be
F^{\sin(\phi_1 + \phi^L_{S_1})}_{TU}=\frac{1}{2}
H_1(z_1)\Delta^{N}\! D_{h_2/\bar{q}^{\uparrow}}(z_2)
\, \frac{\sqrt{2e}{\cal G}(T_1,C_2)}{M_{_{C_2}}}
\frac{P_{1T}} {\langle p^2_{\perp} \rangle_{_{T_1,C_2}}} \, \frac{\langle p^2_{\perp} \rangle^2_{_{C_2}}}{\langle p^2_{\perp} \rangle_{_2}}  \frac{z_{p_1}}{z_{p_2}}
\ee
\be
F^{\sin(3\phi_1 - \phi^L_{S_1})}_{TU} =\frac{1}{2}
\,H_{1T}^\perp(z_1)\Delta^{N}\! D_{h_2/\bar{q}^{\uparrow}}(z_2)
\, \frac{\sqrt{2e}\,e\,{\cal G}(TT_1,C_2)}{ M^2_{_{TT_1}} M_{_{C_2}}}
\,\frac{P^3_{1T}}{\langle p^2_{\perp} \rangle^3_{_{TT_1, C_2}}} \frac{z_{p_1}}{z_{p_2}} \frac{\langle p^2_{\perp} \rangle^2_{_{C_2}} \langle p^2_{\perp} \rangle^3_{_{TT_1}}}{\langle p^2_{\perp} \rangle_2 \langle p^2_{\perp} \rangle_{_1}}\,,
\ee
where we have restored the standard notation also for the polarizing FF
\be
\Delta^N\! D_{h_1^\uparrow/q}(z_1) \equiv \Delta D^p_{h_1/q}(z_1)\,.
\ee

For its relevance we give here the result for the transverse polarization along $\hat{\bm{n}}$, integrated over $\bm{P}_{1T}$, Eq.~(\ref{Pn1}), adopting the Gaussian parametrizations (see Eqs.~(\ref{FTUpol}) and (\ref{FUU})):
\bea
{\cal P}^{h_1}_n(z_1,z_2) & = & \sqrt{\frac{e\pi}{2}}\frac{1}{M_p} \frac{\langle p_\perp^2\rangle^2_{p_1}}{\langle p_{\perp}^2\rangle_1}\,\frac{z_2}{\big\{[z_1(1-M_{h_1}^2/(z_1^2s))]^2\langle p_{\perp}^2\rangle_2 +z_2^2 \langle p_\perp^2\rangle_{p_1}\big\}^{1/2}}\nonumber\\
&\times & \frac{\sum_{q} e^2_q\,\Delta^N\! D_{h_1^\uparrow/q}(z_1)D_{h_2/\bar q}(z_2)}{ \sum_{q}  e^2_q\,
 D_{h_1/q}(z_1)D_{h_2/\bar q}(z_2)}\,.
\label{Polnh}
\eea

\section{Conclusions}
\label{concl}

The study of hadron production in $e^+e^-$ collisions is definitely the most powerful tool to access the parton-to-hadron fragmentation mechanism in a direct and clear way. When azimuthal modulations are considered, together with the polarization states for the case of spin-1/2 hadrons, the information we can extract is extremely rich and, within a TMD approach, could allow to disentangle important spin-momentum correlations and shed light on interesting effects.

We have presented, within the helicity formalism, the complete structure of all leading-twist azimuthal and polarization observables for almost back-to-back hadron-pair production in $e^+e^-$ annihilation processes. This approach, extremely intuitive, gives indeed a more direct probabilistic picture and allows one to follow the underlying processes at the partonic level. It is important to stress how our helicity amplitudes for the different factorized steps lead to final results perfectly equivalent to those obtained within the TMD factorization at leading order~\cite{Boer:1997mf}.

We have considered both spinless and spin-1/2 hadrons, discussing in detail the classification and  properties of the full set of leading-twist TMD fragmentation functions for quarks and gluons. For completeness we have also shown the connection with the Amsterdam notation,  widely adopted in the literature.

We have presented our results adopting two reference frames: the thrust frame, which requires the reconstruction of the jet thrust axis and allows for the analysis of the azimuthal dependences of the hadron pair around this direction; the hadron frame, defined only in terms of the two hadron momenta.
For the latter case we have shown as, by means of a simple tensorial analysis, one can extract the measurable azimuthal dependences. All expressions for single- and double-polarized hadron production have been presented, adopting a conventional form in terms of structure functions.

Special attention has been devoted to the proper expressions of all quantities when moving from the particle helicity frames to the configurations accessible in the measurements.

Moreover, by employing a Gaussian ansatz for the transverse momentum dependence of the TMD-FFs, we have obtained simple expressions for the structure functions, useful in phenomenological analyses. In particular, we have re-derived the explicit formula for the single transverse polarization of spin-1/2 hadron production in $e^+e^-\to h_1^\uparrow h_2 +X$ processes, which has allowed for the first ever extraction of the polarizing FF for $\Lambda$ hyperons from Belle data~\cite{DAlesio:2020wjq}.

This work follows analogous studies already developed by some of us for the SIDIS processes~\cite{Anselmino:2011ch}, where once again the presence of two ordered energy scales guarantees the validity of a TMD factorization scheme. There the complete expressions for all the SIDIS spin asymmetries and the cross section azimuthal dependences were presented. We also recall that this approach was actually first introduced for processes with a single large scale~\cite{Anselmino:2005sh}, like $pp\to \pi X$, where the factorization is only assumed as a starting point. There, the complete classification of TMD-PDFs, within the helicity formalism, both for quarks and gluons was originally introduced.
In this respect the present work represents a sort of complementary study in the fragmentation sector, leading to an overall and complete picture within the helicity formalism of the realm of leading-twist TMDs and the ways to access them.

\section*{Acknowledgments}
We thank M.~Anselmino, E.~Leader and S.~Melis for useful discussions at various stages of this work.
This project has received funding from the European Union’s Horizon 2020 research and innovation programme under grant agreement N.~824093 (STRONG-2020).

\appendix

\section{Fragmentation amplitudes for spin-1/2 hadrons and their properties}
\label{hel-ampl}

As mentioned in Section~\ref{2hadrons}, we can define for a fragmentation process a hadron fragmentation density matrix or a generalized fragmentation function. Following the approach adopted in Ref.~\cite{Anselmino:2005sh}, we define
\begin{equation}
\begin{split}
\hat{D}^{\lambda_{h},\lambda'_{h}}_{\lambda_{c},\lambda'_{c}} (z,\bm{p}_{\perp}) = \sumint_{{X},\lambda_{X}}\!
\hat{\mathcal{D}}_{\lambda_{h},\lambda_{X};\lambda_c }(z,\bm{p}_\perp) \hat{\mathcal{D}}^{\ast}_{\lambda'_{h},\lambda_{X};\lambda'_c }(z,\bm{p}_\perp)\,,
\label{frag_mod2}
\end{split}
\end{equation}
where $\hat{\mathcal{D}}_{\lambda_{h},\lambda_{X};\lambda_c }$ is the fragmentation amplitude for the process $c \rightarrow h +X$, with  $z =\frac{P^+_h}{p^+_c}$ the  hadron light-cone momentum fraction, and  $\bm{p}_\perp$ the transverse momentum of the hadron with respect to the parton.

Defining $\phi_h$ as the azimuthal angle of the hadron $h$ in the helicity reference frame of the parton $c$ we can rewrite the fragmentation function as  \cite{Leader:2001gr, Anselmino:2005sh}
\begin{equation}
\begin{split}
\hat{\mathcal{D}}_{\lambda_{h},\lambda_{X};\lambda_c }(z,\bm{p}_\perp) = \mathcal{D}_{\lambda_{h},\lambda_{X};\lambda_c }(z,p_\perp)e^{i\lambda_c \phi_h}\,,
\label{frag_amp}
\end{split}
\end{equation}
that allows us to write the generalized fragmentation function in the following way
\begin{equation}
\hat{D}^{\lambda_{h},\lambda'_{h}}_{\lambda_{c},\lambda'_{c}}(z,\bm{p}_\perp) = D^{\lambda_{h},\lambda'_{h}}_{\lambda_{c},\lambda'_{c}}(z,p_\perp) \exp[{i(\lambda_c - \lambda'_c)\phi_{h}}] \,.
\label{frag_fase}
\end{equation}

The function $D^{\lambda_{h},\lambda'_{h}}_{\lambda_{c},\lambda'_{c}}(z,p_{\perp})$, modulus of  $\hat {D}^{\lambda_{h},\lambda'_{h}}_{\lambda_{c},\lambda'_{c}}(z,\bm{p}_{\perp})$,
has the same definition of Eq.~(\ref{frag_mod2}), with $\hat{\mathcal{D}}$ replaced by $ \mathcal{D}$, but without any phase dependence.

From Eq.~(\ref{frag_mod2}) we can get the following relation, valid both for quarks and gluons,
\bea
\big(\hat{D}^{\lambda'_{h},\lambda_{h}}_{\lambda'_{c},\lambda_{c}}\big)^{\ast} & = & \hat{D}^{\lambda_{h},\lambda'_{h}}_{\lambda_{c},\lambda'_{c}}\,,
\label{frag_cc0}
\eea
which, in particular, gives
\bea
(\hat{D}^{+-}_{+-})^{\ast} & = & \hat{D}^{-+}_{-+}\quad\quad\quad
(\hat{D}^{+-}_{-+})^{\ast}  =  \hat{D}^{-+}_{+-}\\
(D^{++}_{+-})^{\ast} &= & D^{++}_{-+} \quad\quad\quad
(D^{+-}_{++})^{\ast} =  D^{-+}_{++}\,.
\label{frag_cc1}
\eea
Concerning the parity properties of the $\mathcal{D}$ amplitudes they are the usual ones valid for the helicity amplitudes in the $\phi_h =0$ plane, that is \cite{Leader:2001gr}
\begin{equation}
\mathcal{D}_{-\lambda_{h},-\lambda_{X};-\lambda_c } = \eta\, 
e^{i\pi (s_c - S_h - S_X)} \,
e^{i \pi(\lambda_c -\lambda_h + \lambda_X)}\,\mathcal{D}_{\lambda_{h},\lambda_{X};\lambda_c }\,,
\label{parity_amp}
\end{equation}
where $\eta$ is an intrinsic parity factor such that $\eta^2=1$. This implies that
\begin{equation}
D^{-\lambda_{h},-\lambda^{'}_{h}}_{-\lambda_{c},-\lambda^{'}_{c}}(z,p_{\perp}) = e^{i \pi [(\lambda_c -\lambda^{'}_c ) - (\lambda_{h} - \lambda^{'}_{h})]}\,
D^{\lambda_{h},\lambda^{'}_{h}}_{\lambda_{c},\lambda^{'}_{c}} (z,p_{\perp})\,.
\label{frag_parity}
\end{equation}
Notice that the extra factor makes a difference between quarks and gluons.

Indeed, by exploiting the above relation, we get for quark (upper signs) e gluon (lower signs) fragmentation functions:
\bea
D^{++}_{++} &= & \mbox{} D^{--}_{--}      \quad\quad\quad  D^{++}_{--} = D^{--}_{++}\\
D^{++}_{+-} &= & \mp D^{--}_{-+} \quad\quad \, D^{++}_{-+} = \mp D^{--}_{+-}\\
D^{+-}_{++} &= &- D^{-+}_{--} \quad\quad \,    D^{-+}_{++} = - D^{+-}_{--}\\
D^{+-}_{+-} &= & \pm D^{-+}_{-+} \quad\quad \, D^{+-}_{-+} = \pm D^{-+}_{+-}\label{pmpm}\,.
\label{qg_parity}
\eea

By using the above relations we see that there are six independent quantities
\begin{equation}
D^{++}_{++} , D^{++}_{--} , D^{++}_{+-}, D^{+-}_{++}, D^{+-}_{+-}, D^{+-}_{-+}.
\label{six_ff}
\end{equation}
These are in principle complex quantities, however:  $D^{++}_{++}$ and $D^{++}_{--}$ are real since, as we can see from Eq.~(\ref{frag_mod2}), they are moduli squared; $D^{+-}_{+-}$ and $D^{+-}_{-+}$ are purely real for quarks and purely imaginary for gluons, see Eqs.~(\ref{frag_cc1}) and (\ref{pmpm}); and, eventually, $D^{++}_{+-}$  and $D^{+-}_{++}$ are both complex, giving us four additional real quantities. This leaves us with eight independent real quantities.

\section{Gluon TMD Fragmentation Functions for spin-1/2 hadrons at leading twist}
\label{g-FF}

Even if not relevant for the present study that at LO involves only quark contributions, for the sake of completeness it is worth and interesting to present the gluon TMD-FFs for spin-1/2 hadrons, following the same procedure as for the quark ones. We start again with
\begin{equation}
\rho^{h,S_h}_{\lambda_{h},\lambda^{'}_{h}}\hat{D}_{h/g,P_g}(z,\bm{p}_{\perp})  = \sum_{\lambda_{g},\lambda^{'}_{g}} \rho^{g}_{\lambda_{g},\lambda^{'}_{g}}\hat{D}^{\lambda_{h},\lambda^{'}_{h}}_{\lambda_{g},\lambda^{'}_{g}}(z,\bm{p}_{\perp}) \, ,
\label{hadron2gluon}
\end{equation}
where now we have defined the fragmentation function for a massless gluon with polarization state $P_g$, $\hat{D}_{h/g,P_g}(z,\bm{p}_{\perp})$, fragmenting into an unpolarized hadron. Notice that for a spin-1 parton, like a gluon, it is more convenient to refer to the polarization states (i.e.~to circular and/or linear polarizations).

Indeed for the gluon helicity density matrix we have
\begin{equation}
\begin{split}
\rho^{g}_{\lambda_{g},\lambda^{'}_{g}}&= \frac{1}{2}
\begin{pmatrix}
\rho^{g}_{++} &\rho^{g}_{+-} \\
\rho^{g}_{-+} & \rho^{g}_{--} \\
\end{pmatrix} \\
& = \frac{1}{2}
\begin{pmatrix}
1+ P^g_z    & \mathcal{T}^g_1 -i \mathcal{T}^g_2 \\
\mathcal{T}^g_1 + i \mathcal{T}^g_2 & 1 - P^g_z \\
\end{pmatrix} \\
&= \frac{1}{2}
\begin{pmatrix}
1+ P^g_{\rm circ} & -P^g_{\rm lin} e^{-i2\phi} \\
-P^g_{\rm lin} e^{i2\phi} & 1- P^g_{\rm circ} \\
\end{pmatrix}\,.
\end{split}
\label{mat_g}
\end{equation}
In this case we can still define longitudinally, $P^g_{z}$, or circularly, $P^g_{\rm circ}$, polarization states, keeping in mind that the off-diagonal elements are now related to the linear polarization in the $\widehat{xy}$ plane with an angle $\phi$ with respect to the $x$ axis~\cite{Anselmino:2005sh}.

Once again the $x,y,z$ axes are those of the helicity frame of the gluon, where its four-momentum is $p^{\mu}_g = (p,0,0,p)$. $P^g_{\rm lin}$ is expressed in terms of the parameters $\mathcal{T}^g_1$ and $\mathcal{T}^g_2$, which are closely related to the Stokes' parameters used in classical optics; formally their role is analogous to that played by the $x$ and $y$ components of the quark polarization vectors. The use of the $\mathcal{T}^g_1$ and $\mathcal{T}^g_2$ parameters makes the gluon fragmentation functions formally similar to those of the quarks and simplifies all formulas for the spin asymmetries.

Before exploiting the sum over the gluon helicity indices on the right hand side of Eq.~(\ref{hadron2gluon}), let us show how one can define the eight gluon TMD-FFs in close analogy with the quark case:
\begin{equation}
 P^{h}_J \hat{D}_{h/q,P_g} = \hat{D}^{h/g}_{S_J/P_g}  -  \hat{D}^{h/g}_{-S_J/P_g} \equiv \Delta \hat{D}^{h/g}_{S_J/P_g}\,,
\label{defPg}
\end{equation}
where $J=X,Y,Z$. We will use again the notations:
\bea
(P^h_J \, \hat D_{h/g,P_{\rm lin}}) &=& \Delta \hat D_{S_J/P_{\rm lin}}^{h/g} = \hat D_{S_J/P_{\rm lin}}^{h/g} - \hat D_{-S_J/P_{\rm lin}}^{h/g} \equiv \Delta \hat D_{S_J/P_{\rm lin}}^{h/g}(z, \bm{p}_{\perp}) \label{DxYg}\\
(P^h_J \, \hat D_{h/g,s_z}) &=& \Delta \hat D_{S_J/s_z}^{h/g} = \hat D_{S_J/+}^{h/g} -  \hat D_{-S_J/+}^{h/g} \equiv
\Delta \hat D_{S_J/+}^{h/g}(z, \bm{p}_{\perp}) \label{DxZg}\\
\hat D_{h/g,P_{\rm lin}} &=& \hat D_{h/g}(z, p_{\perp}) + \frac{1}{2}\, \Delta \hat D_{h/P_{\rm lin}}(z, \bm{p}_{\perp})\,,\label{Dunpg}
\label{main-tableg}
\eea
where, in the second line, for the circular or longitudinal polarization, we have used $s_z$, fixed as $+$.

These amount to eight gluon TMD-FFs. Analogously to what has been done for quarks, we now exploit the sum in the right hand side of Eq.~(\ref{hadron2gluon}). Thus we can obtain the following three expressions:
\bea
\rho^{h,S_h}_{++} \hat{D}_{h/g,P_g} & = & \frac{1}{2}(1+ P^{h}_Z) \hat{D}_{h/g,P_g} \nonumber \\
&=& \frac{1}{2}(D^{++}_{++} +D^{++}_{--}) + \frac{1}{2}P^g_{\rm circ}(D^{++}_{++}- D^{++}_{--}) \nonumber \\
&& - P^g_{\rm lin} [{\rm Re}D^{++}_{+-}\cos{[2(\phi - \phi_{h})]} + {\rm Im}D^{++}_{+-}\sin{[2(\phi - \phi_{h})]}]
\label{h2g_1}
\eea
\bea
\rho^{h,S_h}_{--}  \hat{D}_{h/g,P_g}  & = & \frac{1}{2}(1- P^{h}_Z) \hat{D}_{h/g,P_g} \nonumber \\
&= & \frac{1}{2}(D^{++}_{++} +D^{++}_{--}) - \frac{1}{2}P^g_{\rm circ}(D^{++}_{++}- D^{++}_{--}) \nonumber \\
&& - P^g_{\rm lin} [{\rm Re} D^{++}_{+-}\cos{[2(\phi - \phi_{h})]} - {\rm Im}D^{++}_{+-}\sin{[2(\phi - \phi_{h})]}]
\label{h2g_2}
\eea
\bea
\rho^{h,S_h}_{+-}  \hat D_{h/g,P_g} & = & \frac{1}{2}(P^{h}_X -iP^{h}_Y) \hat D_{h/g,P_g} \nonumber \\
&= &i{\rm Im}D^{+-}_{++} + P^g_{\rm circ} {\rm Re}D^{+-}_{++} \nonumber \\
&& - \frac{1}{2}P^g_{\rm lin} [ i({\rm Im}D^{+-}_{+-} + {\rm Im}D^{+-}_{-+})\cos{[2(\phi - \phi_{h})]} \nonumber \\
&& + ({\rm Im}D^{+-}_{+-} - {\rm Im} D^{+-}_{-+})\sin{[2(\phi - \phi_{h})]}] \,,
\label{h2g_3}
\eea
where once again we have used the properties of the generalized fragmentation functions,  $\hat{D}^{\lambda_{h},\lambda^{'}_{h}}_{\lambda_{g},\lambda^{'}_{g}}(z,\bm{p}_{\perp})$, discussed in Appendix~\ref{hel-ampl}.

By suitably combining the above expressions, we can find the eight gluon TMD-FFs for a spin-1/2 hadron. For instance, by summing or subtracting Eqs.~(\ref{h2g_1}) and (\ref{h2g_2}) we obtain respectively the TMD-FF for an unpolarized and a longitudinally polarized hadron
\begin{equation}
\hat{D}_{h/g,P_g} 
= (D^{++}_{++} + D^{++}_{--})  - 2P^g_{\rm lin}{\rm Re}D^{++}_{+-}\cos{[2(\phi - \phi_{h})]}
\label{piupiu_g}
\end{equation}
\bea
P^{h}_Z  \hat{D}_{h/g,P_g}
& = & P^g_{\rm circ}(D^{++}_{++}- D^{++}_{--}) - 2P^g_{\rm lin} {\rm Im}D^{++}_{+-}\sin{[2(\phi - \phi_{h})]}.
\label{piumeno_g}
\eea
As we can see from Eq.~(\ref{piupiu_g}), we can have an unpolarized hadron coming from the fragmentation of an unpolarized or a linearly polarized gluon: this last case, in analogy with the quark one, is referred to as the \emph{Collins-like} gluon TMD-FF. By taking the real or the imaginary part of Eq.~(\ref{h2g_3}) we get the FF for a hadron transversely polarized along, respectively, its $X$ or $Y$ helicity axis as coming from a polarized gluon
\bea
P^{h}_X\hat{D}_{h/g,P_g}
& = & 2P^g_{\rm circ} {\rm Re}D^{+-}_{++}    - P^g_{\rm lin}({\rm Im}D^{+-}_{+-}- {\rm Im}D^{+-}_{-+})\sin[2(\phi - \phi_h)]
\label{real_g}
\eea
\bea
P^{h}_Y \hat{D}_{h/g,P_g} &=& 
-2 {\rm Im} D^{+-}_{++}  + P^g_{\rm lin}({\rm Im}D^{+-}_{+-} +{\rm Im}D^{+-}_{-+})\cos{[2(\phi - \phi_{h})]}\,.
\label{imma_g}
\eea
Once again we can combine the two above expressions as follows
\bea
P_T^h \hat{D}_{h/g,P_g} & = & -2 {\rm Im} D^{+-}_{++} \sin\phi_{S_h} + 2P^g_{\rm circ} {\rm Re}D^{+-}_{++} \cos\phi_{S_h} \nonumber \\
&& + P^g_{\rm lin} [{\rm Im}D^{+-}_{+-} \sin(\phi_{S_h}-2(\phi - \phi_{h}) ) + {\rm Im}D^{+-}_{-+} \sin(\phi_{S_h}+2(\phi - \phi_{h}) )
]\, ,\nonumber\\
\eea
and by using Eq.~(\ref{phiangles}) we have
\bea
P_T^h \hat{D}_{h/g,P_g} & = & -2 {\rm Im} D^{+-}_{++} \sin(\phi'_{S_h}-\phi_h) + 2P^g_{\rm circ} {\rm Re}D^{+-}_{++} \cos(\phi'_{S_h}-\phi_h) \nonumber \\
&&+ P^g_{\rm lin} [{\rm Im}D^{+-}_{+-} \sin(\phi'_{S_h}-2\phi + \phi_{h}) ) + {\rm Im}D^{+-}_{-+} \sin(\phi'_{S_h}+2\phi - 3\phi_{h}) )
]\,.\nonumber\\
\eea

By fixing now the gluon polarization we can recover the eight TMD-FFs discussed above. Using Eq.~(\ref{piupiu_g}) we have
\bea
\hat{D}_{h/g} & = & \hat{D}_{h/g,P_{\rm circ}} = D^{++}_{++} +D^{++}_{--} \equiv D_{h/g}\,,
\label{hg_nop_1}
\eea
giving the TMD-FF for an unpolarized (or circularly polarized) gluon fragmenting into an unpolarized hadron, and
\bea
\hat{D}_{h/g,P_{\rm lin}} & = &
\hat{D}_{h/g} - 2{\rm Re}D^{++}_{+-}\cos{[2(\phi - \phi_{h})]} \\
& = & \hat{D}_{h/g} + \frac{1}{2}\Delta^N D_{h/g,P_{\rm lin}}\cos{[2(\phi - \phi_{h})]} \,,
\label{hg_nop}
\eea
giving the TMD-FF for a linearly polarized gluon fragmenting into an unpolarized hadron.

Through Eq.~(\ref{piumeno_g}) we get
\bea
P_Z^h \hat{D}_{h/g,P_{\rm circ}} &=& D^{++}_{++}- D^{++}_{--}= \Delta {D}^{h/g}_{S_Z/P_{\rm circ}}
\label{hg_pz_1}\\
P_Z^h \hat{D}_{h/g,P_{\rm lin}} &=&
-2 {\rm Im} D^{++}_{+-}\, \sin{[2(\phi - \phi_{h})]}
= \Delta D^{h/g}_{S_Z/P_{\rm lin}}\sin{[2(\phi - \phi_{h})]}\,,
\label{hg_pz}
\eea
giving, respectively, the TMD-FFs for a circularly and for a linearly polarized gluon fragmenting into a longitudinally polarized hadron.
Analogously from Eq.~(\ref{real_g}) we have the FF for a hadron transversely polarized, along its $X$ helicity axis, produced by a circularly and a linearly polarized gluon
\bea
P_X^h \hat{D}_{h/g,P_{\rm circ}} &=&
 2\, {\rm Re}D^{+-}_{++} = \Delta {D}^{h/g}_{S_X/P_{\rm circ}}
\label{hg_px_1}\\
P_X^h \hat{D}_{h/g,P_{\rm lin}} &=&
 - ({\rm Im}D^{+-}_{+-}- {\rm Im}D^{+-}_{-+})\sin{[2(\phi - \phi_{h})]}
=  \Delta D^{h/g}_{S_X/P_{\rm lin}} \sin{[2(\phi - \phi_{h})]}\,.\nonumber\\
\label{hg_px}
\eea

Finally, from Eq.~(\ref{imma_g}) we have the TMD-FF for a hadron transversely polarized along its $Y$ direction coming from, respectively, a circularly and linearly polarized gluon
\bea
P_Y^h \hat{D}_{h/g,P_{\rm circ}} &=&
  -2 {\rm Im}D^{+-}_{++} = \Delta {D}^{h}_{S_Y/g}
\label{hg_py_1}\\
P_Y^h \hat{D}_{h/g,P_{\rm lin}} &=&
-2 {\rm Im}D^{+-}_{++} + ({\rm Im}D^{+-}_{+-} +{\rm Im}D^{+-}_{-+})\cos{[2(\phi - \phi_{h})]}  \\
&= & \Delta {D}^{h}_{S_Y/g} + \Delta^{-} \hat{D}^{h/g}_{S_Y/P_{\rm lin}} = \Delta {D}^{h}_{S_Y/g} + \Delta^{-}D^{h/g}_{S_Y/P_{\rm lin}}\cos{[2(\phi- \phi_{h})]} \, .\nonumber\\
\label{hg_py}
\eea
Also in this case we have introduced the function $\Delta^{-} \hat{D}^{h/g}_{S_Y/P_{\rm lin}}$, analogously to Eq.~(\ref{change_sign}), that changes sign if the gluon linear polarization has an off-set of $\pi/2$. Thus, as for the quark case, we are able to collect all TMD-FFs as follows
\begin{equation}
\begin{split}
  \hat{D}_{h/g} (z,\bm{p}_{\perp}) & = D_{h/g} = (D^{++}_{++} +D^{++}_{--})\\
%
%
\Delta \hat{D}_{h/g,P_{\rm lin}} (z,\bm{p}_{\perp}) & =  \Delta^{N}\! D_{h/g,P_{\rm lin}} \cos{[2(\phi - \phi_{h})]} = - 4\,{\rm Re} D^{++}_{+-}\,\cos{[2(\phi - \phi_{h})]}  \\
%
%
\Delta \hat{D}^{h/g}_{S_Z/P_{\rm circ}} (z,\bm{p}_{\perp}) & =  \Delta D^{h/g}_{S_Z/P_{\rm circ}} =  (D^{++}_{++}- D^{++}_{--})\\
 \Delta \hat{D}^{h/g}_{S_Z/P_{\rm lin}} (z,\bm{p}_{\perp})  &=  \Delta D^{h/g}_{S_Z/P_{\rm lin}}  \sin{[2(\phi - \phi_{h})]} = -  2{\rm Im} D^{++}_{+-} \sin{[2(\phi - \phi_{h})]} \\
 \Delta \hat{D}^{h/g}_{S_X/P_{\rm circ}} (z,\bm{p}_{\perp}) &=\Delta D^{h/g}_{S_X/P_{\rm circ}} =2 {\rm Re}D^{+-}_{++}\\
 \Delta \hat{D}^{h/g}_{S_X/P_{\rm lin}} (z,\bm{p}_{\perp})&=  \Delta D^{h/g}_{S_X/P_{\rm lin}}\sin{[2(\phi - \phi_{h})]}=- ({\rm Im} D^{+-}_{+-}- {\rm Im}D^{+-}_{-+})\sin{[2(\phi - \phi_{h})]}\\
%
  \Delta \hat{D}^{h}_{S_Y/g} (z,\bm{p}_{\perp})  &= \Delta D^{h}_{S_Y/g}= -2 {\rm Im}D^{+-}_{++}  \\
\Delta^{-}\hat{D}^{h/g}_{S_Y/P_{\rm lin}} (z,\bm{p}_{\perp}) &= \Delta^{-}D^{h/g}_{S_Y/P_{\rm lin}}  \cos{[2(\phi - \phi_{h})]} =  ({\rm Im}D^{+-}_{+-} +{\rm Im} D^{+-}_{-+}) \cos{[2(\phi - \phi_{h})]}\,.
\end{split}
\end{equation}

\section{Comparison with other notations}
\label{Amst}

Here we present a comparison of the eight leading-twist quark TMD-FFs as given in the helicity formalism with those defined in the notation of the Amsterdam group~\cite{Boer:1997nt,Boer:1997mf}.
There the main quantity, corresponding to our  $\hat{D}^{\lambda_{h},\lambda'_{h}}_{\lambda_{q},\lambda'_{q}}(z,\bm{p}_{\perp})$, is the quark-hadron correlator $\Delta(z,\bm{k}_T)$ and all mass effects are neglected.
It is important to notice that 
$\bm{k}_T$ is the transverse three-momentum of the fragmenting quark w.r.t.~the produced hadron, which implies $\bm{k}_T \simeq - \bm{p}_\perp/z$ (valid in the massless-hadron limit). We will also use the spin-polarization vector $\bm{P}^h= (P_L^h, \bm{P}_T^h)$
to allow for a more direct comparison.

For the case of an unpolarized quark fragmenting into a spin-1/2 hadron and a transversely polarized quark fragmenting into an unpolarized or spinless hadron we recover the results of the ``Trento Conventions'', see Ref.~\cite{Bacchetta:2004jz}.

Following Ref.~\cite{Boer:1997nt} we have
\bea
\Delta(z,\bm{k}_T) &=& \frac{1}{2} \bigg\{ D_1 \slashed{n}^{-} + D^{\bot}_{1T} \frac{\epsilon^{\mu \nu \rho \sigma} \gamma_{\mu} n^{-}_{\nu} k_{T \rho} P^h_{T \sigma}}{M_h} + \Big(P^h_L G_{1L} + G_{1T}\, \frac{{\bm{k}}_T \cdot \bm{P}^h_{T} }{M_h}\Big) \gamma_5 \slashed{n}^{-}\nonumber\\
 &+&  H_{1T} i \sigma^{\mu \nu} \gamma_5  n^{-}_{\mu} P^h_{T \nu}
 + \Big(P^h_L H^\perp_{1L} + H^\perp_{1T}\, \frac{\bm{k}_T \cdot \bm{P}^h_{T} }{M_h} \Big) \frac{i \sigma^{\mu \nu} \gamma_5  n^{-}_{\mu} k_{ T \nu}}{M_h} +
 H^{\bot}_{1} \frac{\sigma^{\mu \nu}k_{ T \mu} n^{-}_{\nu} }{M_h} \bigg \}\,,\nonumber\\
\eea
where $D_1 = D_1(z,{p}_\perp)$, with $p_\perp=|\bm{p}_\perp|$ as adopted through the whole paper.

By appropriate Dirac projection, $\Delta^{[\Gamma]} = \rm{Tr}[\Gamma \Delta]$, one can single out the various sectors of the fragmentation functions. In particular, $\Gamma =\slashed{n}^{+}/2 $ projects out the $D$ sector, relative to an unpolarized quark, that is the unpolarized $D_1$ and the polarizing  $D^{\bot}_{1T}$ fragmentation functions:
\bea
{\rm Tr}\bigg[\frac{\slashed{n}^{+}}{2} \, \Delta(z,\bm{k}_T)\bigg]   & = & D_1 - D^{\bot}_{1T}\, \frac{\epsilon^{\mu \nu \rho \sigma} n^{-}_{\mu} k_{T \nu} P^h_{T \rho} n^{+}_{\sigma} }{M_h} = D_1 +  D^{\bot}_{1T} \, \frac{( \hat{\bm{p}}_q \times \bm{p}_{\bot}) \cdot \bm{P}^h }{ z M_h}\,,\nonumber\\
\label{Dsec}
\eea
where $n_\pm$ are two auxiliary lightlike vectors, and the second equality holds in frames where $\bm{n}^-$ and the direction $\hat{\bm{p}}_q$ of the quark momentum point in opposite directions.

By choosing $\Gamma = \slashed{n}^{+} \gamma_5/2$ we select the contribution from a longitudinal polarized quark, the $G$ sector, that is
\bea
{\rm Tr}\bigg[\frac{\slashed{n}^{+}}{2} \gamma_5 \, \Delta(z,\bm{k}_T)\bigg]  & = & P^h_L\, G_{1L} + G_{1T}\, \frac{{\bm{k}}_T \cdot \bm{P}^h_{T} }{M_h}
= P^h_L\,  G_{1L} - P^h_{T} \,\frac{p_{\bot} }{z M_h} \, G_{1T}\, \cos{(\phi'_{S_h} - \phi_{h}) }\,,\nonumber\\
\label{Gsec}
\eea
where the angle $\phi'_{S_h}$ is the azimuthal angle of the hadron spin in the quark helicity frame.

Finally, to obtain the $H$ sector, relative to the fragmentation of a transversely polarized quark, we use the projector $\Gamma =i\sigma^{\rho \sigma} \gamma_5\, n^{+}_{\rho} P^q_{ \sigma}/2$, with $P^{q \mu}=(0, \cos\phi_{s_q}, \sin\phi_{s_q}, 0)$:
\bea
&&{\rm Tr}\bigg[\frac{1}{2} i \sigma^{\rho \sigma} \gamma_5\, n^{+}_{\rho} P^q_{\sigma} \, \Delta(z,\bm{k}_T)\bigg]\nonumber\\
&& = H_{1T}\, \bm{P}^h_{T} \cdot \bm{P}^{q} +  \Big(P^h_L H^\perp_{1L} + H^\perp_{1T}\, \frac{\bm{k}_T \cdot \bm{P}^h_{T} }{M_h} \Big) \, \frac{\bm{k}_T \cdot \bm{P}^{q} }{M_h}
- H^{\bot}_1 \frac{\epsilon^{\mu \nu \rho \sigma} n^{-}_{\mu} k_{T \nu} P^{q}_\rho n^{+}_{\sigma}} {M_h}\nonumber\\
&& = {P}^h_{T}\,\Big[ H_{1}\, \cos(\phi'_{S_h}-\phi_{s_q} )+ \frac{{p}_\perp^2}{2z^2 M_h^2} H_{1T}^\perp \cos(\phi_{s_q}+ \phi'_{S_{h}} - 2\phi_h)\Big] \nonumber \\
&& \quad - P^h_L \,  \frac{{p}_{\bot} }{zM_h} H^\perp_{1L} \, \cos(\phi_{s_q}-\phi_h) + H^{\bot}_{1} \, \frac{( \hat{\bm{p}}_q \times \bm{p}_{\bot}) \cdot \bm{P}^q }{ z M_h}\,,
\label{Hsec}
\eea
with
\be
\label{H1H1Tp}
H_1 = H_{1T} + \frac{{p}_\perp^2 }{2z^2 M_h^2}\,H^{\bot}_{1T}\,.
\ee
To obtain the relation between the $D_{\lambda_q\lambda'_q}^{\lambda_h, \lambda'_h}$ fragmentation functions, including the TMD-FFs in the notation adopted here, and those of the Amsterdam group, one has to take into account that, while the first ones are given fixing the polarization vector of the hadron, the others are given at fixed quark polarization. By properly exploiting Eqs.~(\ref{piupiu}), (\ref{piumeno}) and (\ref{PTSh-ap})
and Eqs.~(\ref{Dsec})-(\ref{Hsec}), we get
\bea
D_{h/q}   =  D^{++}_{++} + D^{++}_{--} &=& D_{1}(z,{p}_\perp) \\
\Delta D^h_{S_Y/q}  =  \Delta^{N}\! D_{h^\uparrow/q} = - 2\, {\rm Im} D^{+-}_{++} &=& \frac{{p}_\perp}{zM_h} D_{1T}^\perp (z,{p}_\perp)
\label{polAm}\\
\Delta D^{h/q}_{S_Z/s_L}  =  D^{++}_{++} - D^{++}_{--} & = & G_{1L}(z,{p}_\perp)  \\
\Delta D^{h/q}_{S_X/s_L}   =  2\,{\rm Re}D^{++}_{+-} &=& {-}\frac{{p}_\perp}{zM_h}\,G_{1T}(z,{p}_\perp)
\\
\Delta D^{h/q}_{S_Z/s_T}  = 2\, {\rm Re}D^{+-}_{++} &=&  {-}\frac{{p}_\perp}{zM_h}\,H^{\perp}_{1L}(z,{p}_{\perp})
\\
\Delta^{N}\! D_{h/q^{\uparrow}}  =  4\, {\rm Im} D^{++}_{+-} &=& \frac{2{p}_\perp}{zM_h} H_{1}^\perp(z,{p}_\perp)\\
\frac{1}{2} [\Delta D_{S_X/s_T} + \Delta^- D_{S_Y/s_T}]  =  D^{+-}_{+-} &=&  H_1(z,{p}_{\perp}) \label{H1}\\
\frac{1}{2} [\Delta D_{S_X/s_T} - \Delta^- D_{S_Y/s_T}]  = D^{+-}_{-+}  &=& \frac{{p}_\perp^2 }{2z^2 M_h^2} \, H_{1T}^\perp(z,{p}_{\perp})\label{H1Tp}
\,.
\eea

To complete this comparison we observe that by inserting Eqs.~(\ref{H1H1Tp}),  (\ref{polAm}), (\ref{H1}) and (\ref{H1Tp}) into Eqs.~(\ref{hq_px}) and (\ref{hq_py}) we get
\bea
P_X^h\hat{D}_{h/q,s_T} & = & \Delta \hat{D}^{h/q}_{S_X/s_T}(z,\bm{p}_\perp) =
\left[ H_{1T}(z,{p}_{\perp}) + \frac{{p}_\perp^2 }{z^2 M_h^2} \,
H_{1T}^\perp(z,{p}_{\perp}) \right] \, \cos(\phi_{s_q}-\phi_h) \nonumber\\
\eea
\bea
P_Y^h\hat{D}_{h/q,s_T} & = & \Delta \hat{D}^{h/q}_{S_Y/s_T}(z,\bm{p}_\perp) =
\frac{{p}_{\perp}}{zM_h} \, D_{1T}^\perp (z,{p}_{\perp}) +
H_{1T}(z,{p}_{\perp}) \,\sin(\phi_{s_q}-\phi_h) \,,\nonumber\\
\eea
showing that $H_{1T}$ and $H_{1T}^\perp$ are combinations of hadron polarized fragmentation functions.

\section{Tensorial analysis}
\label{tensor}

Here we present the tensorial decomposition of all the integrals over $p_{\perp 2}$ involved in the convolutions ${\cal C}[w\Delta D\Delta \bar D]$ appearing  in Section \ref{had-conv}, following the procedure developed in Ref.~\cite{Anselmino:2011ch} for the general helicity formalism in SIDIS.

All the integrals can be reduced to a linear combination of the following convolutions:
\bea
T^{i} &=& \frac{1}{P_{1T}}\int d^2\bm{p}_{\bot 2} \, p^i_{\bot 2} \, \Delta D^{h_1}(z_1,p_{ \bot 1 }) \, \Delta D^{h_2}(z_2,p_{ \bot 2 })
\label{Tti}    \\
T^{ij} &=& \frac{1}{P_{1T}^2}\int d^2\bm{p}_{\bot 2} \, p^i_{\bot 2}p^j_{\bot 2} \, \Delta D^{h_1}(z_1,p_{ \bot 1 }) \, \Delta D^{h_2}(z_2,p_{ \bot 2 })
\label{Ttij}  \\
T^{ijk} &=& \frac{1}{P_{1T}^3}\int d^2\bm{p}_{\bot 2} \, p^i_{\bot 2}p^j_{\bot 2}p^k_{\bot 2} \, \Delta D^{h_1}(z_1,p_{ \bot 1 }) \, \Delta D^{h_2}(z_2,p_{ \bot 2 })
\label{Ttijk} \\
T^{ijkl} &=& \frac{1}{P_{1T}^4}\int d^2\bm{p}_{\bot 2} \, p^i_{\bot 2}p^j_{\bot 2}p^k_{\bot 2}p^l_{\bot 2} \, \Delta D^{h_1}(z_1,p_{ \bot 1 }) \, \Delta D^{h_2}(z_2,p_{ \bot 2 })\,,
\label{Ttijkl}
\eea
where we have denoted by $\Delta D^{h_{1,2}}$ any fragmentation function (depending only on the moduli of the intrinsic transverse momenta) appearing in the definition of the particular structure function $F$  one is considering and where the upper index of $p_{\bot 2}$ can be $i=x,y$ (hadron frame):
\be
p^x_{\bot 2} =  p_{\bot 2} \cos \varphi_2 \, \quad\quad \; p^y_{\bot 2} =  p_{\bot 2} \sin \varphi_2\,.
\ee

Notice that we have  normalized each tensor by a suitable power of $P_{1T}$ and that
$T^i$, $T^{ij}$, $T^{ijk}$ and $T^{ijkl}$ are symmetric, rank 1, 2, 3, 4 Euclidean tensors, respectively.
Bearing in mind that $\bm{p}_{\bot 1}$ is not independent and can be expressed in terms of $\bm{p}_{\bot 2}$ and $\bm{P}_{1T}$ (see Eq.~(\ref{desahdfr2})) , we have, in a completely general way, that the convolutions depend only on  $P_{1T}$ and $\phi_1$, i.e. the
measured modulus and azimuthal phase of the final observed hadron transverse momentum:
\bea
T^{i} &=& \frac{P^i_{1T}}{P_{1T}} \,S_1(P_{1T})
\label{Ti}    \\
T^{ij} &=&  \frac{P^i_{1T} P^j_{1T}}{P_{1T}^2}\, S_2(P_{1T}) + \delta^{ij}\,S_3(P_{1T})
\label{Tij}  \\
T^{ijk} &=&  \frac{P^i_{1T} P^j_{1T} P^k_{1T}}{P_{1T}^3}\, S_4(P_{1T}) + \frac{1}{P_{1T}}\big(P^i_{1T} \delta^{jk} + P^j_{1T} \delta^{ik} + P^k_{1T} \delta^{ij} \big)\, S_5(P_{1T})
\label{Tijk}    \\
T^{ijkl} & = &  \frac{P^i_{1T} P^j_{1T} P^k_{1T} P^l_{1T}}{P_{1T}^4}\, S_6(P_{1T}) + \frac{1}{P_{1T}^2}\big(P^k_{1T} P^l_{1T} \delta^{ij} + P^l_{1T} P^j_{1T} \delta^{ik} + P^k_{1T} P^j_{1T} \delta^{il} + P^i_{1T} P^l_{1T} \delta^{jk}\nonumber\\
&&+ P^i_{1T} P^k_{1T} \delta^{jl} + P^i_{1T} P^j_{1T} \delta^{kl}\big )\,S_7(P_{1T})
+ \big(\delta^{ij}\delta^{kl} + \delta^{il}\delta^{jk} +\delta^{ik}\delta^{jl} \big)\, S_8(P_{1T})\,, \label{Tijkl}
\eea
where the tensorial structure is given by the components of $\bm{P}_{1T}$
\be
P^X_{1T} = P_{1T} \cos \phi_1 \, ; \; P^Y_{1T} =  P_{1T} \sin \phi_1
\ee
while $S_1$-$S_8$ are eight scalar functions which can only depend on $P_{1T}$ (modulus), and can be determined directly by contracting Eqs.~(\ref{Ti})-(\ref{Tijkl})  with suitable symmetric tensorial structures. One then finds
\bea
S_1(P_{1T}) &=& \frac{1}{P_{1T}}\int  d^2\bm{p}_{\bot 2} \, (\bm{p}_{\bot 2} \cdot \hat{\bm{P}}_{1T})  \, \Delta D^{h_1} \, \Delta D^{h_2}\\
S_2(P_{1T}) &=& \frac{1}{P^2_{1T}}\int  d^2\bm{p}_{\bot 2} \, [2(\bm{p}_{\bot 2} \cdot \hat{\bm{P}}_{1T})^2 - p^2_{\bot 2}]  \, \Delta D^{h_1} \, \Delta D^{h_2}\\
S_3(P_{1T}) &=& \frac{1}{P^2_{1T}}\int  d^2\bm{p}_{\bot 2} \, [p^2_{\bot 2} - (\bm{p}_{\bot 2} \cdot \hat{\bm{P}}_{1T})^2 ]  \, \Delta D^{h_1} \, \Delta D^{h_2}\\
S_4(P_{1T}) &=& \frac{1}{P^3_{1T}}\int  d^2\bm{p}_{\bot 2} \, [4(\bm{p}_{\bot 2} \cdot \hat{\bm{P}}_{1T})^3 -3 p^2_{\bot 2}(\bm{p}_{\bot 2} \cdot \hat{\bm{P}}_{1T})]  \, \Delta D^{h_1} \, \Delta D^{h_2}\\
S_5(P_{1T}) &=& \frac{1}{P_{1T}^3}\int  d^2\bm{p}_{\bot 2} \, [p^2_{\bot 2}(\bm{p}_{\bot 2} \cdot \hat{\bm{P}}_{1T}) - (\bm{p}_{\bot 2} \cdot \hat{\bm{P}}_{1T})^3 ]  \, \Delta D^{h_1} \, \Delta D^{h_2}\\
S_6(P_{1T}) &=& \frac{1}{P^4_{1T}}\int d^2\textbf{p}_{\bot 2} \Big[ 8\big(\bm{p}_{\bot 2} \cdot \hat{\bm{P}}_{1T}  \big)^4 - 8\big(\bm{p}_{\bot 2} \cdot \hat{\bm{P}}_{1T}  \big)^2\,p^2_{\bot 2} + p^4_{\bot 2} \Big]\,\Delta D^{h_1} \, \Delta D^{h_2}\nonumber \\
&&\\
S_7(P_{1T}) &=& \frac{1}{3P^4_{1T}}\int d^2\bm{p}_{\bot 2}\Big[ -4\big(\bm{p}_{\bot 2} \cdot \hat{\bm{P}}_{1T}  \big)^4 + 5\big(\bm{p}_{\bot 2} \cdot \hat{\bm{P}}_{1T}  \big)^2\,p^2_{\bot 2} - p^4_{\bot 2}   \Big]\,
\Delta D^{h_1} \, \Delta D^{h_2}\nonumber\\
&&\\
S_8(P_{1T}) &=& \frac{1}{3P^4_{1T}}\int d^2\bm{p}_{\bot 2}\Big[ \big(\bm{p}_{\bot 2} \cdot \hat{\bm{P}}_{1T} \big)^4 - 2\big(\bm{p}_{\bot 2} \cdot \hat{\bm{P}}_{1T}  \big)^2\,p^2_{\bot 2} + p^4_{\bot 2}  \Big]\,
\Delta D^{h_1} \, \Delta D^{h_2}\,.\nonumber\\
\eea
From the above relations we can get
\bea
\int d^2\bm{p}_{\perp 2} \, \cos\varphi_2 \, \Delta D^{h_1} \, \Delta D^{h_2} & =&
\cos\phi_1 \,\int d^2\bm{p}_{\perp 2}\, (\hat{\bm{p}}_{\perp 2} \cdot \hat{\bm{P}}_{1T})\;
\Delta D^{h_1} \, \Delta D^{h_2} \label{D12}
\eea
\bea
\int d^2\bm{p}_{\perp 2}\, \sin\varphi_2 \, \Delta D^{h_1} \, \Delta D^{h_2}  &=&
\sin\phi_1 \,\int d^2\bm{p}_{\perp 2}\, (\hat{\bm{p}}_{\perp 2} \cdot \hat{\bm{P}}_{1T}) \;
\Delta D^{h_1} \, \Delta D^{h_2} \label{D13}
\eea
\bea
\int d^2\bm{p}_{\perp 2}\, \cos ^2 \varphi_2 \, \Delta D^{h_1} \, \Delta D^{h_2}  &=&
\frac{1}{2} \int d^2\bm{p}_{\perp 2}\, \Big\{ 1+\cos 2\phi_1 \,
[ 2(\hat{\bm{p}}_{\perp 2} \cdot \hat{\bm{P}}_{1T})^2 -1 ] \Big\}
\;\Delta D^{h_1} \, \Delta D^{h_2} \nonumber\\
\label{D14}
\eea
\bea
\int d^2\bm{p}_{\perp 2}\, \sin ^2 \varphi_2 \, \Delta D^{h_1} \, \Delta D^{h_2}  &=&
\frac{1}{2} \int d^2\bm{p}_{\perp 2}\, \Big\{ 1 - \cos 2\phi_1 \,
[ 2(\hat{\bm{p}}_{\perp 2} \cdot \hat{\bm{P}}_{1T})^2 -1 ] \Big\}
\; \Delta D^{h_1} \, \Delta D^{h_2} \nonumber\\\label{D15}
\eea
\bea
\int d^2\bm{p}_{\perp 2}\, \cos\varphi_2 \,\sin\varphi_2 \, \Delta D^{h_1} \, \Delta D^{h_2}  &=&
\cos\phi_1 \, \sin\phi_1 \int d^2\bm{p}_{\perp 2}\,
[ 2(\hat{\bm{p}}_{\perp 2} \cdot \hat{\bm{P}}_{1T})^2 - 1 ]
\; \Delta D^{h_1} \, \Delta D^{h_2} \nonumber\\\label{D16}
\eea
\bea
\int d^2\bm{p}_{\perp 2}\, \cos ^3 \varphi_2 \, \Delta D^{h_1} \, \Delta D^{h_2}  &=&
\cos ^3 \phi_1 \int d^2\bm{p}_{\perp 2}\, [4(\hat{\bm{p}}_{\perp 2} \cdot \hat{\bm{P}}_{1T})^3 -
3(\hat{\bm{p}}_{\perp 2} \cdot \hat{\bm{P}}_{1T})] \; \Delta D^{h_1} \, \Delta D^{h_2} \nonumber \\
&& +  3 \cos\phi_1 \int d^2\bm{p}_{\perp 2}\, [(\hat{\bm{p}}_{\perp 2} \cdot \hat{\bm{P}}_{1T}) - (\hat{\bm{p}}_{\perp 2} \cdot \hat{\bm{P}}_{1T})^3] \; \Delta D^{h_1} \, \Delta D^{h_2} \nonumber\\\label{D17}
\eea
\bea
\int d^2\bm{p}_{\perp 2}\, \sin^3 \varphi_2 \, \Delta D^{h_1} \, \Delta D^{h_2}  &=&
\sin ^3 \phi_1 \int d^2\bm{p}_{\perp 2}\, [4(\hat{\bm{p}}_{\perp 2} \cdot \hat{\bm{P}}_{1T})^3 -
3(\hat{\bm{p}}_{\perp 2} \cdot \hat{\bm{P}}_{1T})] \; \Delta D^{h_1} \, \Delta D^{h_2} \nonumber \\
&& + 3 \sin\phi_1 \int d^2\bm{p}_{\perp 2}\, [(\hat{\bm{p}}_{\perp 2} \cdot \hat{\bm{P}}_{1T}) - (\hat{\bm{p}}_{\perp 2} \cdot \hat{\bm{P}}_{1T})^3] \; \Delta D^{h_1} \, \Delta D^{h_2} \nonumber\\
\label{D18}
\eea
\bea
&&\int d^2\bm{p}_{\perp 2}\, \cos ^2 \varphi_2 \sin\varphi_2 \, \Delta D^{h_1} \, \Delta D^{h_2} \nonumber\\
&&=\cos ^2 \phi_1 \sin\phi_1 \int d^2\bm{p}_{\perp 2}\,
[4(\hat{\bm{p}}_{\perp 2} \cdot \hat{\bm{P}}_{1T})^3 -
3(\hat{\bm{p}}_{\perp 2} \cdot \hat{\bm{P}}_{1T})] \; \Delta D^{h_1} \, \Delta D^{h_2} \nonumber \\
&&\quad +  \sin\phi_1 \int d^2\bm{p}_{\perp 2}\, [(\hat{\bm{p}}_{\perp 2} \cdot \hat{\bm{P}}_{1T}) -
(\hat{\bm{p}}_{\perp 2} \cdot \hat{\bm{P}}_{1T})^3] \; \Delta D^{h_1} \, \Delta D^{h_2}
\label{D19}
\eea
\bea
&&\int d^2\bm{p}_{\perp 2}\, \cos\varphi_2 \sin ^2 \varphi_2 \, \Delta D^{h_1} \, \Delta D^{h_2} \nonumber\\
&&=\cos\phi_1 \sin ^2 \phi_1 \int d^2\bm{p}_{\perp 2}\,
[4(\hat{\bm{p}}_{\perp 2} \cdot \hat{\bm{P}}_{1T})^3 -
3(\hat{\bm{p}}_{\perp 2} \cdot \hat{\bm{P}}_{1T})] \; \Delta D^{h_1} \, \Delta D^{h_2} \nonumber \\
&&\quad+ \cos\phi_1 \int d^2\bm{p}_{\perp 2}\, [(\hat{\bm{p}}_{\perp 2} \cdot \hat{\bm{P}}_{1T}) -
(\hat{\bm{p}}_{\perp 2} \cdot \hat{\bm{P}}_{1T})^3] \; \Delta D^{h_1} \, \Delta D^{h_2}  \label{D20}
\eea
\bea
&&\int  d^2\bm{p}_{\bot 2}\,\sin^4 \varphi_2 \, \Delta D^{h_1} \, \Delta D^{h_2} \nonumber\\
&&=\sin^4\phi_1   \int  d^2\bm{p}_{\bot 2}  \Big[ 8\big(\hat{\bm{p}}_{\bot 2} \cdot \hat{\bm{P}}_{1T}  \big)^4 - 8\big(\hat{\bm{p}}_{\bot 2} \cdot \hat{\bm{P}}_{1T}  \big)^2 + 1 \Big] \; \Delta D^{h_1} \, \Delta D^{h_2}\nonumber \\
&&\quad+  2 \sin^2\phi_1  \int  d^2\bm{p}_{\bot 2}  \Big[  -4\big(\hat{\bm{p}}_{\bot 2} \cdot \hat{\bm{P}}_{1T}  \big)^4 + 5\big(\hat{\bm{p}}_{\bot 2}\cdot \hat{\bm{P}}_{1T}  \big)^2 - 1 \Big]\; \Delta D^{h_1} \, \Delta D^{h_2}\nonumber \\
&&\quad + \int  d^2\bm{p}_{\bot 2} \Big[  \big(\hat{\bm{p}}_{\bot 2} \cdot \hat{\bm{P}}_{1T}  \big)^4 - 2\big(\hat{\bm{p}}_{\bot 2} \cdot \hat{\bm{P}}_{1T}  \big)^2 + 1 \Big]\; \Delta D^{h_1} \, \Delta D^{h_2}
\eea
\bea
&&\int  d^2\bm{p}_{\bot 2}\,\cos^4 \varphi_2 \, \Delta D^{h_1} \, \Delta D^{h_2} \nonumber\\
&&= \cos^4\phi_1  \int  d^2\bm{p}_{\bot 2}  \Big[ 8\big(\hat{\bm{p}}_{\bot 2} \cdot \hat{\bm{P}}_{1T}  \big)^4 - 8\big(\hat{\bm{p}}_{\bot 2} \cdot \hat{\bm{P}}_{1T}  \big)^2 + 1 \Big] \; \Delta D^{h_1} \, \Delta D^{h_2}\nonumber \\
&&\quad +  2 \cos^2\phi_1  \int  d^2\bm{p}_{\bot 2}  \Big[  -4\big(\hat{\bm{p}_{\bot 2}} \cdot \hat{\bm{P}}_{1T}  \big)^4 + 5\big(\hat{\bm{p}}_{\bot 2}\cdot \hat{\bm{P}}_{1T}  \big)^2 - 1 \Big]\; \Delta D^{h_1} \, \Delta D^{h_2}\nonumber \\
& & \quad +\int  d^2\bm{p}_{\bot 2} \Big[  \big(\hat{\bm{p}}_{\bot 2} \cdot \hat{\bm{P}}_{1T}  \big)^4 - 2\big(\hat{\bm{p}}_{\bot 2} \cdot \hat{\bm{P}}_{1T}  \big)^2 + 1 \Big]\; \Delta D^{h_1} \, \Delta D^{h_2}
\eea
\bea
&&\int  d^2\bm{p}_{\bot 2}\,\cos^2\varphi_2 \sin^2 \varphi_2 \; \Delta D^{h_1} \, \Delta D^{h_2} \nonumber\\
&&= \cos^2\phi_1 \sin^2\phi_1   \int  d^2\bm{p}_{\bot 2}  \Big[ 8\big(\hat{\bm{p}}_{\bot 2} \cdot \hat{\bm{P}}_{1T}  \big)^4 - 8\big(\hat{\bm{p}}_{\bot 2} \cdot \hat{\bm{P}}_{1T}  \big)^2 + 1\Big] \; \Delta D^{h_1} \, \Delta D^{h_2}\nonumber \\
&&\quad +  \int  d^2\bm{p}_{\bot 2}  \Big[  -\big(\hat{\bm{p}}_{\bot 2} \cdot \hat{\bm{P}}_{1T}  \big)^4 + \big(\hat{\bm{p}}_{\bot 2}\cdot \hat{\bm{P}}_{1T}  \big)^2  \Big]\; \Delta D^{h_1} \, \Delta D^{h_2}
\eea
\bea
&&\int d^2\bm{p}_{\bot 2}\,\cos\varphi_2 \sin^3 \varphi_2 \; \Delta D^{h_1} \, \Delta D^{h_2} \nonumber\\
&&= \cos\phi_1 \sin^3\phi_1   \int  d^2\bm{p}_{\bot 2}  \Big[ 8\big(\hat{\bm{p}}_{\bot 2} \cdot \hat{\bm{P}}_{1T}  \big)^4 - 8\big(\hat{\bm{p}}_{\bot 2} \cdot \hat{\bm{P}}_{1T}  \big)^2 + 1 \Big] \; \Delta D^{h_1} \, \Delta D^{h_2}\nonumber \\
&&\quad + \cos\phi_1 \sin\phi_1  \int  d^2\bm{p}_{\bot 2}  \Big[  -4\big(\hat{\bm{p}}_{\bot 2} \cdot \hat{\bm{P}}_{1T}  \big)^4 + 5\big(\hat{\bm{p}}_{\bot 2} \cdot \hat{\bm{P}}_{1T}  \big)^2 - 1\Big]\; \Delta D^{h_1} \, \Delta D^{h_2}\nonumber\\
\eea
\bea
&&\int  d^2\bm{p}_{\bot 2}\,\cos^3\varphi_2 \sin \varphi_2 \; \Delta D^{h_1} \, \Delta D^{h_2} \nonumber\\
&&= \cos^3\phi_1 \sin\phi_1   \int  d^2\bm{p}_{\bot 2}  \Big[ 8\big(\hat{\bm{p}}_{\bot 2} \cdot \hat{\bm{P}}_{1T}  \big)^4 - 8\big(\hat{\bm{p}}_{\bot 2} \cdot \hat{\bm{P}}_{1T}  \big)^2 + 1 \Big] \; \Delta D^{h_1} \, \Delta D^{h_2}\nonumber \\
&&\quad + \cos\phi_1 \sin\phi_1  \int  d^2\bm{p}_{\bot 2}  \Big[  -4\big(\hat{\bm{p}}_{\bot 2} \cdot \hat{\bm{P}}_{1T}  \big)^4 + 5\big(\hat{\bm{p}}_{\bot 2} \cdot \hat{\bm{P}}_{1T}  \big)^2 - 1 \Big]\; \Delta D^{h_1} \, \Delta D^{h_2}\,.\nonumber \\
\eea
We can then reconstruct directly the following quantities appearing in Section~\ref{had-conv}:
\bea
\int d^2\bm{p}_{\perp 2} \; \cos(2\varphi_2) \, \Delta D^{h_1} \, \Delta D^{h_2}  =
\cos(2\phi_1) \,\int d^2\bm{p}_{\perp 2} \, [ 2(\hat{\bm{p}}_{\perp 2} \cdot \hat{\bm{P}}_{1T})^2 -1 ]\;
\Delta D^{h_1} \, \Delta D^{h_2}\nonumber\\ \label{D21}
\eea
\bea
\int d^2\bm{p}_{\perp 2} \; \sin(2\varphi_2) \, \Delta D^{h_1} \, \Delta D^{h_2}  =
\sin(2\phi_1) \,\int d^2\bm{p}_{\perp 2} \, [ 2(\hat{\bm{p}}_{\perp 2} \cdot \hat{\bm{P}}_{1T})^2 -1 ] \;
\Delta D^{h_1} \, \Delta D^{h_2}\nonumber\\ \label{D22}
\eea
\bea
&&\int d^2\bm{p}_{\perp 2} \; \cos  (3\varphi_2) \, \Delta D^{h_1} \, \Delta D^{h_2} \nonumber\\
&& = \cos (3 \phi_1) \int d^2\bm{p}_{\perp 2} \, [4(\hat{\bm{p}}_{\perp 2} \cdot \hat{\bm{P}}_{1T})^3 -
3(\hat{\bm{p}}_{\perp 2} \cdot \hat{\bm{P}}_{1T})] \; \Delta D^{h_1} \, \Delta D^{h_2} \label{D23}
\eea
\bea
&& \int d^2\bm{p}_{\perp 2} \; \sin  (3\varphi_2) \, \Delta D^{h_1} \, \Delta D^{h_2} \nonumber\\
&& = \sin (3 \phi_1) \int d^2\bm{p}_{\perp 2} \, [4(\hat{\bm{p}}_{\perp 2} \cdot \hat{\bm{P}}_{1T})^3 -
3(\hat{\bm{p}}_{\perp 2} \cdot \hat{\bm{P}}_{1T})] \; \Delta D^{h_1} \, \Delta D^{h_2} \label{D24}
\eea
\bea
&&\int  d^2\bm{p}_{\bot 2}\,\cos(4\varphi_2) \; \Delta D^{h_1} \, \Delta D^{h_2} \nonumber\\
&&= \cos(4\phi_1) \int  d^2\bm{p}_{\bot 2}  \Big[ 8\big(\hat{\bm{p}}_{\bot 2} \cdot \hat{\bm{P}}_{1T}  \big)^4 - 8\big(\hat{\bm{p}}_{\bot 2} \cdot \hat{\bm{P}}_{1T}  \big)^2 + 1 \Big] \; \Delta D^{h_1} \, \Delta D^{h_2}
\eea
\bea
&& \int d^2\bm{p}_{\bot 2}\,\sin(4\varphi_2) \; \Delta D^{h_1} \, \Delta D^{h_2} \nonumber\\
&&= \sin(4\phi_1)    \int  d^2\bm{p}_{\bot 2}  \Big[ 8\big(\hat{\bm{p}}_{\bot 2} \cdot \hat{\bm{P}}_{1T}  \big)^4 - 8\big(\hat{\bm{p}}_{\bot 2} \cdot \hat{\bm{P}}_{1T}  \big)^2 + 1 \Big] \; \Delta D^{h_1} \, \Delta D^{h_2}\,.
\eea

\section{Helicity frames}
\label{hel-frames}

Our physical observables are computed in two kinematical configurations, the thrust and the hadron frames,
with axes denoted by $\hat{\bm{x}}_L$, $\hat{\bm{y}}_L$, and $\hat{\bm{z}}_L$, where we use $L$ to indicate a generic laboratory (LAB) frame. The helicity frame of a particle with momentum $\bm{p}$ along the direction $\hat{\bm{p}} =
(\sin\theta\cos\varphi, \, \sin\theta\sin\varphi, \, \cos\theta)$ -- {\it as
defined in the laboratory frame} -- can be reached by performing the rotations~\cite{Leader:2001gr}
\be
R(\varphi, \theta, 0) = R_{y'}(\theta) \, R_{z_{L}}(\varphi) \>.
\ee
The first is a rotation by an angle $\varphi$ around the $\hat{\bm{z}}_L$-axis and the second is a rotation by an angle $\theta$ around the new (that is, obtained after the first rotation) $\hat{\bm{y}}'$-axis. This means
\be
a)\;\hat{\bm{z}} = \hat{\bm{p}}\quad\quad\quad
b)\;\hat{\bm{y}} = \frac{\hat{\bm{z}}_{L} \times \hat{\bm{p}}}
{|\hat{\bm{z}}_{L} \times \hat{\bm{p}}|} =
\hat{\bm{z}}_{L} \times \hat{\bm{p}}_\perp\quad\quad\quad
c)\;\hat{\bm{x}} = \hat{\bm{y}} \times \hat{\bm{z}} \,.
 \label{helgen}
\ee

In the present study we are interested in the helicity frames of the final quark/antiquark as well as those of the two hadrons coming from their fragmentation. Let us describe them, starting from the relations between the parent quark and the corresponding hadron helicity frames and then focusing separately on the two configurations for a process $e^+(k^+) e^-(k^-) \to c(q_1)\, d(q_2) \to h_1(P_{h_1}) h_2(P_{h_2}) \, X$.

The helicity frame of a \emph{non-collinear} hadron with momentum
\[\bm{P}_h = P_h(\sin\theta_h\cos\phi_h, \sin\theta_h\sin\phi_h, \cos\theta_h)\,,\]
 as reached from the helicity frame of its parent quark, following the above procedure $a)$-$c)$ in Eq.~(\ref{helgen}), is simply given as:
\bea
 \hat{\bm{Z}}_{h} &=&   \sin\theta_h \cos\phi_{h}\, \hat{\bm{x}} + \sin\theta_h \sin\phi_{h} \,\hat{\bm{y}} + \cos\theta_h  \,\hat{\bm{z}} \equiv \hat{\bm{P}}_h \label{Zh}\\
\hat{\bm{Y}}_{h} &= &  - \sin\phi_{h} \, \hat{\bm{x}} + \cos\phi_{h} \,\hat{\bm{y}} \label{Yh} \\
\hat{\bm{X}}_{h} &= &   \cos\theta_h \cos\phi_{h}\, \hat{\bm{x}} + \cos\theta_h \sin\phi_{h} \,\hat{\bm{y}} - \sin\theta_h \, \hat{\bm{z}}  \label{Xh}\,.
\eea

\subsection{The thrust frame}
In this case the particles $c$ and $d$ move along the $\hat{\bm{z}}_L$ and $-\hat{\bm{z}}_L$ direction respectively. This results in the helicity frames with axes along the following directions in the laboratory frame:
\be
\hat{\bm{x}}_c = \hat{\bm{x}}_{L} \quad\quad\quad \hat{\bm{y}}_c =
\hat{\bm{y}}_{L} \quad\quad\quad \hat{\bm{z}}_c = \hat{\bm{z}}_{L} \label{hel-c}
\ee
for a quark/antiquark $c$ moving along +$\hat{\bm{z}}_{L}$, and
\be
\hat{\bm{x}}_d = \hat{\bm{x}}_{L} \quad\quad\quad \hat{\bm{y}}_d =
- \hat{\bm{y}}_{L} \quad\quad\quad \hat{\bm{z}}_d = - \hat{\bm{z}}_{L}
\label{hel-d}
\ee
for a quark/antiquark $d$ moving along $-\hat{\bm{z}}_{L}$. From these relations we find that the helicity axes of the hadrons $h_1$ in the LAB frame as reached from the quark helicity frames are [using Eqs.~(\ref{Zh})-(\ref{Xh}) with $h=h_1$ and Eq.~(\ref{hel-c})]
\bea
 \hat{\bm{Z}}_{h_1} & = &   \sin\theta_{h_1} \cos\phi_{h_1}\, \hat{\bm{x}}_L + \sin\theta_{h_1} \sin\phi_{h_1} \,\hat{\bm{y}}_L + \cos\theta_{h_1}  \,\hat{\bm{z}}_L\label{Zh1} \\
\hat{\bm{Y}}_{h_1} & = &  - \sin\phi_{h_1} \, \hat{\bm{x}}_L + \cos\phi_{h_1} \,\hat{\bm{y}}_L  \\
\hat{\bm{X}}_{h_2} & = &   \cos\theta_{h_1} \cos\phi_{h_1}\, \hat{\bm{x}}_L + \cos\theta_{h_1} \sin\phi_{h_1} \,\hat{\bm{y}}_L - \sin\theta_{h_1} \, \hat{\bm{z}}_L
\eea
and the ones for the hadron $h_2$ [using Eqs.~(\ref{Zh})-(\ref{Xh}) with $h=h_2$ and Eq.~(\ref{hel-d})]
\bea
 \hat{\bm{Z}}_{h_2} & = &   \sin\theta_{h_2} \cos\phi_{h_2}\, \hat{\bm{x}}_L - \sin\theta_{h_2} \sin\phi_{h_2} \,\hat{\bm{y}}_L - \cos\theta_{h_2}  \,\hat{\bm{z}}_L \label{Zh2} \\
\hat{\bm{Y}}_{h_2} & = &  - \sin\phi_{h_2} \, \hat{\bm{x}}_L - \cos\phi_{h_2} \,\hat{\bm{y}}_L  \\
\hat{\bm{X}}_{h_2} & = &   \cos\theta_{h_2} \cos\phi_{h_2}\, \hat{\bm{x}}_L - \cos\theta_{h_2} \sin\phi_{h_2} \,\hat{\bm{y}}_L + \sin\theta_{h_2} \, \hat{\bm{z}}_L\,.
\eea
On the other hand, the hadron unit three-momenta as defined in the LAB frame, see Eqs.~(\ref{P1}) and (\ref{P2}), are
\bea
\hat{\bm{P}}_{h_1}  &=& \hat{\bm{Z}}_{h_1} = \frac{2 \eta_{\bot 1}}{z_{p_1}} \cos\varphi_1\, \hat{\bm{x}}_L + \frac{2 \eta_{\bot 1}}{z_{p_1}} \sin\varphi_1 \, \hat{\bm{y}}_L + \beta_1 \, \hat{\bm{z}}_L\label{Zh11}\\
\hat{\bm{P}}_{h_2}  &=& \hat{\bm{Z}}_{h_2} = \frac{2 \eta_{\bot 2}}{z_{p_2}} \cos\varphi_2\, \hat{\bm{x}}_L + \frac{2 \eta_{\bot 2}}{z_{p_2}} \sin\varphi_2 \, \hat{\bm{y}}_L - \beta_2 \, \hat{\bm{z}}_L\label{Zh22}
\eea
By a direct comparison between Eqs.~(\ref{Zh1}) and (\ref{Zh11}) and between Eqs.~(\ref{Zh2}) and (\ref{Zh22}) we then get
\be
\phi_{h_1} = \varphi_1 \hspace*{2cm} \phi_{h_2} = 2\pi - \varphi_2 \,.
\ee

\subsection{The hadron frame}

In the hadron frame the unit vectors specifying the helicity frames of the two particles $c$ and $d$, as reached from the laboratory frame are:
\bea
\hat{\bm{z}}_{c}  &= &\frac{2 \eta_{\bot 2}}{z_{p_2}} \cos\varphi_2 \,\hat{\bm{x}}_L + \frac{2 \eta_{\bot 2}}{z_{p_2}} \sin\varphi_2 \,\hat{\bm{y}}_L + \beta_2 \,\hat{\bm{z}}_L\\
\hat{\bm{y}}_{c}  &=& - \sin\varphi_2 \,\hat{\bm{x}}_L + \cos\varphi_2 \,\hat{\bm{y}}_L  \\
\hat{\bm{x}}_{c}  &=& \beta_2 \cos\varphi_2 \, \hat{\bm{x}}_L +  \beta_2 \sin\varphi_2 \,\hat{\bm{y}}_L - \frac{2 \eta_{\bot 2}}{z_{p_2}} \,\hat{\bm{z}}_L
\eea
\bea
\hat{\bm{z}}_{d}  &=&- \frac{2 \eta_{\bot 2}}{z_{p_2}} \cos\varphi_2 \,\hat{\bm{x}}_L - \frac{2 \eta_{\bot 2}}{z_{p_2}} \sin\varphi_2 \,\hat{\bm{y}}_L - \beta_2 \,\hat{\bm{z}}_L\\
\hat{\bm{y}}_{d}  &=&  \sin\varphi_2 \, \hat{\bm{x}}_L - \cos\varphi_2 \, \hat{\bm{y}}_L  \\
\hat{\bm{x}}_{d}  &=& \beta_2 \cos\varphi_2 \,\hat{\bm{x}}_L +  \beta_2 \sin\varphi_2 \,\hat{\bm{y}}_L - \frac{2 \eta_{\bot 2}}{z_{p_2}} \,\hat{\bm{z}}_L  \,
\eea
where we started from the quark-antiquark directions in the LAB frame as defined in Eqs.~(\ref{q2}) and (\ref{q1}) and, then applied the standard procedure to identify the other two axes (see Eq.~(\ref{helgen})).

From the above relations one can get the expressions of the hadron helicity axes as reached from the parton helicity ones in the LAB  frame.
In particular, for the $Z$ axes we find
\bea
\hat{\bm{Z}}_{h_1} &=& \Big(\sin\theta_{h_1} (\beta_2\cos\phi_{h_1}\cos\varphi_2 -  \sin\phi_{h_1}\sin\varphi_2 ) + \frac{2 \eta_{\bot 2}}{z_{p_2}} \cos\theta_{h_1}\cos\varphi_2 \Big) \,\hat{\bm{x}}_L \nonumber\\
&+& \Big(\sin\theta_{h_1} (\beta_2\cos\phi_{h_1}\sin\varphi_2 +  \sin\phi_{h_1}\cos\varphi_2 ) + \frac{2 \eta_{\bot 2}}{z_{p_2}} \cos\theta_{h_1}\sin\varphi_2 \Big) \,\hat{\bm{y}}_L \nonumber\\
&-& \Big(\frac{2 \eta_{\bot 2}}{z_{p_2}}\sin\theta_{h_1}\cos\phi_{h_1} - \beta_2\cos\theta_{h_1}\Big) \,\hat{\bm{z}}_L\\
\hat{\bm{Z}}_{h_2} &=& \Big(\sin\theta_{h_2} (\beta_2\cos\phi_{h_2}\cos\varphi_2 +  \sin\phi_{h_2}\sin\varphi_2 ) - \frac{2 \eta_{\bot 2}}{z_{p_2}} \cos\theta_{h_2}\cos\varphi_2 \Big) \,\hat{\bm{x}}_L \nonumber\\
&+& \Big(\sin\theta_{h_2} (\beta_2\cos\phi_{h_2}\sin\varphi_2 -  \sin\phi_{h_2}\cos\varphi_2 ) - \frac{2 \eta_{\bot 2}}{z_{p_2}} \cos\theta_{h_2}\sin\varphi_2 \Big) \,\hat{\bm{y}}_L \nonumber\\
&-& \Big(\frac{2 \eta_{\bot 2}}{z_{p_2}}\sin\theta_{h_2}\cos\phi_{h_2} + \beta_2\cos\theta_{h_2}\Big) \,\hat{\bm{z}}_L\,,
\eea
that have to be compared with the corresponding expressions in the LAB frame, coming from Eqs.~(\ref{Ph1had}) and (\ref{Ph2had}),
\bea
\hat{\bm{Z}}_{h_1}  & = & \frac{2\eta_{1T}}{z_{p_1} } \cos\phi_1 \, \hat{\bm{x}}_L + \frac{2 \eta_{1T}}{z_{p_1}} \sin\phi_1\,
\hat{\bm{y}}_L + \sqrt{1 - \frac{4 \eta^2_{1T}}{z^2_{p_1} }} \, \hat{\bm{z}}_L \\
\hat{\bm{Z}}_{h_2} &=& - \hat{\bm{z}}_L\,.
\eea
By using
\bea
\sin\theta_{h_{1,2}} &=& \frac{2\eta_{\perp 1,2}}{z_{p_{1,2}}}\,,
\eea
after some algebra we  get
\bea
\cos\phi_{h_1} &= &\frac{\eta_{1T}}{\eta_{\bot1}}\beta_2 \cos(\phi_1 - \varphi_2) - \frac{\eta_{\bot 2}}{\eta_{\bot 1}} \frac{z_{p_1}}{z_{p_2}}\sqrt{  1- \frac{4 \eta^2_{1T}}{z^2_{p_1}}}  \simeq  \frac{P_{1T}}{p_{\bot1}} \cos(\phi_1 - \varphi_2) - \frac{p_{\bot 2}}{p_{\bot 1}} \frac{z_{p_1}}{z_{p_2}}\nonumber\\
&&  \label{cosfih1} \\
\sin\phi_{h_1} &= &\frac{\eta_{1T}}{\eta_{\bot1}}\sin(\phi_1 - \varphi_2)
    = \frac{P_{1T}}{p_{\bot1}}\sin(\phi_1 - \varphi_2) \,,
\label{sinfih1}
\eea
together with $\phi_{h_2}=0$.

For their role we give here the explicit expressions of the helicity axes of the hadrons $h_1$ and $h_2$ in terms of the variables in the LAB frame
\bea
\hat{\bm{Z}}_{h_1}  &= & \frac{2\eta_{1T}}{z_{p_1} } \cos\phi_1 \, \hat{\bm{x}}_L + \frac{2 \eta_{1T}}{z_{p_1}} \sin\phi_1 \, \hat{\bm{y}}_L + \sqrt{1 - \frac{4 \eta^2_{1T}}{z^2_{p_1} }} \, \hat{\bm{z}}_L \nonumber\\
&\simeq&  \frac{2\eta_{1T}}{z_{p_1} } \cos\phi_1 \, \hat{\bm{x}}_L + \frac{2 \eta_{1T}}{z_{p_1}} \sin\phi_1 \, \hat{\bm{y}}_L + \hat{\bm{z}}_L \simeq \hat{\bm{z}}_L
\label{eq:zh1}\\
\hat{\bm{Y}}_{h_1}  & \simeq &  \Big(\frac{z_{p_1}}{z_{p_2}} \frac{p_{\perp 2}}{p_{\perp 1}} \sin{\varphi_2}  - \frac{P_{1T}}{p_{\perp 1}}\sin{\phi_1}\Big) \, \hat{\bm{x}}_L + \Big(\frac{P_{1T}}{p_{\perp 1}}\cos{\phi_1} - \frac{z_{p_1}}{z_{p_2}} \frac{p_{\perp 2}}{p_{\perp 1}} \cos{\varphi_2}\Big) \, \hat{\bm{y}}_L \label{eq:yh1} \\
\hat{\bm{X}}_{h_1}  & \simeq & \Big(\frac{P_{1T}}{p_{\perp 1}}\cos{\phi_1} - \frac{z_{p_1}}{z_{p_2}} \frac{p_{\perp 2}}{p_{\perp 1}} \cos{\varphi_2}\Big)\, \hat{\bm{x}}_L + \Big(\frac{P_{1T}}{p_{\perp 1}}\sin{\phi_1} - \frac{z_{p_1}}{z_{p_2}} \frac{p_{\perp 2}}{p_{\perp 1}} \sin{\varphi_2}\Big) \, \hat{\bm{y}}_L\label{eq:xh1}\\
\hat{\bm{Z}}_{h_2}  &= &- \hat{\bm{z}}_L \\
\hat{\bm{Y}}_{h_2}  &= & \sin \varphi_2 \,\hat{\bm{x}}_L  -\cos \varphi_2 \,\hat{\bm{y}}_L \\
\hat{\bm{X}}_{h_2}  &= &\cos \varphi_2\, \hat{\bm{x}}_L +  \sin \varphi_2 \,\hat{\bm{y}}_L \,.
\eea

Notice that, while the $\hat{\bm{Z}}_{h_1}$  axis does not involve any intrinsic transverse momentum dependence, special care, when carrying out the integration over the unobserved variables, is due for the transverse axes. Moreover, at the lowest order in $\eta_{1T}$ the $\hat{\bm{Z}}_{h_1}$ axis almost coincides with the $\hat{\bm{z}}_L$ one.

Before concluding this section we warn the reader that the hadron helicity frames obtained directly from the LAB frame, without passing through the parton helicity frame, give a different result where, while the $\hat{\bm{Z}}$ axes coincide, the $\hat{\bm{X}}$ and $\hat{\bm{Y}}$ axes are rotated:
\bea
\hat{\bm{Z}}^{\rm{LAB}}_{h_1}  &= &\frac{2 \eta_{1T}}{z_{p_1}} \cos\phi_1 \,\hat{\bm{x}}_L + \frac{2 \eta_{ 1T}}{z_{p_1}} \sin\phi_1 \,\hat{\bm{y}}_L + \sqrt{1 - \frac{4\eta_{ 1T}^2 }{z^2_{p_1}}}\,\hat{\bm{z}}_L \\
\hat{\bm{Y}}^{\rm{LAB}}_{h_1}  &= &- \sin\phi_1 \,\hat{\bm{x}}_L + \cos\phi_1 \,\hat{\bm{y}}_L = \hat{\bm{n}} \\
\hat{\bm{X}}^{\rm{LAB}}_{h_1}  &= &\sqrt{1 - \frac{4\eta_{ 1T}^2 }{z^2_{p_1}}}\cos\phi_1 \,\hat{\bm{x}}_L + \sqrt{1 - \frac{4\eta_{ 1T}^2 }{z^2_{p_1}}}\cos\phi_1 \,\hat{\bm{y}}_L - \frac{2 \eta_{ 1T}}{z_{p_1}} \, \hat{\bm{z}}_L \\
\hat{\bm{Z}}^{\rm{LAB}}_{h_2}  & = & - \hat{\bm{z}}_L\\
\hat{\bm{Y}}^{\rm{LAB}}_{h_2}  & = & - \hat{\bm{y}}_L  \\
\hat{\bm{X}}^{\rm{LAB}}_{h_2}  &= & \hat{\bm{x}}_L \,.
\eea


\begin{thebibliography}{10}

\bibitem{DAlesio:2007bjf}
U.~D'Alesio and F.~Murgia, \emph{{Azimuthal and Single Spin Asymmetries in Hard
  Scattering Processes}},
  \href{https://doi.org/10.1016/j.ppnp.2008.01.001}{\emph{Prog. Part. Nucl.
  Phys.} {\bfseries 61} (2008) 394}
  [\href{https://arxiv.org/abs/0712.4328}{{\ttfamily 0712.4328}}].

\bibitem{Barone:2010zz}
V.~Barone, F.~Bradamante and A.~Martin, \emph{{Transverse-spin and
  transverse-momentum effects in high-energy processes}},
  \href{https://doi.org/10.1016/j.ppnp.2010.07.003}{\emph{Prog. Part. Nucl.
  Phys.} {\bfseries 65} (2010) 267}
  [\href{https://arxiv.org/abs/1011.0909}{{\ttfamily 1011.0909}}].

\bibitem{Aidala:2012mv}
C.A.~Aidala, S.D.~Bass, D.~Hasch and G.K.~Mallot, \emph{{The Spin Structure of
  the Nucleon}}, \href{https://doi.org/10.1103/RevModPhys.85.655}{\emph{Rev.
  Mod. Phys.} {\bfseries 85} (2013) 655}
  [\href{https://arxiv.org/abs/1209.2803}{{\ttfamily 1209.2803}}].

\bibitem{Anselmino:2020vlp}
M.~Anselmino, A.~Mukherjee and A.~Vossen, \emph{{Transverse spin effects in
  hard semi-inclusive collisions}},
  \href{https://doi.org/10.1016/j.ppnp.2020.103806}{\emph{Prog. Part. Nucl.
  Phys.} {\bfseries 114} (2020) 103806}
  [\href{https://arxiv.org/abs/2001.05415}{{\ttfamily 2001.05415}}].

\bibitem{Sivers:1989cc}
D.W.~Sivers, \emph{{Single Spin Production Asymmetries from the Hard Scattering
  of Point-Like Constituents}},
  \href{https://doi.org/10.1103/PhysRevD.41.83}{\emph{Phys. Rev. D} {\bfseries
  41} (1990) 83}.

\bibitem{Sivers:1990fh}
D.W.~Sivers, \emph{{Hard scattering scaling laws for single spin production
  asymmetries}}, \href{https://doi.org/10.1103/PhysRevD.43.261}{\emph{Phys.
  Rev. D} {\bfseries 43} (1991) 261}.

\bibitem{Collins:1992kk}
J.C.~Collins, \emph{{Fragmentation of transversely polarized quarks probed in
  transverse momentum distributions}},
  \href{https://doi.org/10.1016/0550-3213(93)90262-N}{\emph{Nucl. Phys. B}
  {\bfseries 396} (1993) 161}
  [\href{https://arxiv.org/abs/hep-ph/9208213}{{\ttfamily hep-ph/9208213}}].

\bibitem{Ji:2004xq}
X.-d.~Ji, J.-P.~Ma and F.~Yuan, \emph{{QCD factorization for spin-dependent
  cross sections in DIS and Drell-Yan processes at low transverse momentum}},
  \href{https://doi.org/10.1016/j.physletb.2004.07.026}{\emph{Phys. Lett. B}
  {\bfseries 597} (2004) 299}
  [\href{https://arxiv.org/abs/hep-ph/0405085}{{\ttfamily hep-ph/0405085}}].

\bibitem{Ji:2004wu}
X.-d.~Ji, J.-p.~Ma and F.~Yuan, \emph{{QCD factorization for semi-inclusive
  deep-inelastic scattering at low transverse momentum}},
  \href{https://doi.org/10.1103/PhysRevD.71.034005}{\emph{Phys. Rev. D}
  {\bfseries 71} (2005) 034005}
  [\href{https://arxiv.org/abs/hep-ph/0404183}{{\ttfamily hep-ph/0404183}}].

\bibitem{Collins:2011zzd}
J.~Collins, \emph{{Foundations of perturbative QCD}}, vol.~32, Cambridge
  University Press (11, 2013).

\bibitem{GarciaEchevarria:2011rb}
M.G.~Echevarria, A.~Idilbi and I.~Scimemi, \emph{{Factorization Theorem For
  Drell-Yan At Low $q_T$ And Transverse Momentum Distributions
  On-The-Light-Cone}},
  \href{https://doi.org/10.1007/JHEP07(2012)002}{\emph{JHEP} {\bfseries 07}
  (2012) 002} [\href{https://arxiv.org/abs/1111.4996}{{\ttfamily 1111.4996}}].

\bibitem{Mulders:1995dh}
P.J.~Mulders and R.D.~Tangerman, \emph{The complete tree-level result up to
  order $1/{Q}$ for polarized deep-inelastic leptoproduction}, {\emph{Nucl.
  Phys.} {\bfseries B461} (1996) 197}.

\bibitem{Bacchetta:2006tn}
A.~Bacchetta, M.~Diehl, K.~Goeke, A.~Metz, P.J.~Mulders and M.~Schlegel,
  \emph{{Semi-inclusive deep inelastic scattering at small transverse
  momentum}}, \href{https://doi.org/10.1088/1126-6708/2007/02/093}{\emph{JHEP}
  {\bfseries 02} (2007) 093}
  [\href{https://arxiv.org/abs/hep-ph/0611265}{{\ttfamily hep-ph/0611265}}].

\bibitem{Anselmino:2011ch}
M.~Anselmino, M.~Boglione, U.~D'Alesio, S.~Melis, F.~Murgia, E.R.~Nocera and A.~Prokudin,
  \emph{{General Helicity Formalism for Polarized Semi-Inclusive Deep
  Inelastic Scattering}},
  \href{https://doi.org/10.1103/PhysRevD.83.114019}{\emph{Phys. Rev.}
  {\bfseries D83} (2011) 114019}
  [\href{https://arxiv.org/abs/1101.1011}{{\ttfamily 1101.1011}}].

\bibitem{Arnold:2008kf}
S.~Arnold, A.~Metz and M.~Schlegel, \emph{{Dilepton production from polarized
  hadron hadron collisions}},
  \href{https://doi.org/10.1103/PhysRevD.79.034005}{\emph{Phys. Rev. D}
  {\bfseries 79} (2009) 034005}
  [\href{https://arxiv.org/abs/0809.2262}{{\ttfamily 0809.2262}}].

\bibitem{Boer:1997mf}
D.~Boer, R.~Jakob and P.J.~Mulders, \emph{Asymmetries in polarized hadron
  production in $e^+ e^-$ annihilation up to order $1/{Q}$}, {\emph{Nucl.
  Phys.} {\bfseries B504} (1997) 345}.

\bibitem{Anselmino:2005sh}
M.~Anselmino, M.~Boglione, U.~D'Alesio, E.~Leader, S.~Melis and F.~Murgia,
  \emph{{The general partonic structure for hadronic spin asymmetries}},
  \href{https://doi.org/10.1103/PhysRevD.73.014020}{\emph{Phys. Rev. D}
  {\bfseries 73} (2006) 014020}
  [\href{https://arxiv.org/abs/hep-ph/0509035}{{\ttfamily hep-ph/0509035}}].

\bibitem{Anselmino:2018psi}
M.~Anselmino, M.~Boglione, U.~D'Alesio, F.~Murgia and A.~Prokudin, \emph{{Role
  of transverse momentum dependence of unpolarized parton distribution and
  fragmentation functions in the analysis of azimuthal spin asymmetries}},
  \href{https://doi.org/10.1103/PhysRevD.98.094023}{\emph{Phys. Rev. D}
  {\bfseries 98} (2018) 094023}
  [\href{https://arxiv.org/abs/1809.09500}{{\ttfamily 1809.09500}}].

\bibitem{Gamberg:2018fwy}
L.~Gamberg, Z.-B.~Kang, D.~Pitonyak, M.~Schlegel and S.~Yoshida,
  \emph{{Polarized hyperon production in single-inclusive electron-positron
  annihilation at next-to-leading order}},
  \href{https://doi.org/10.1007/JHEP01(2019)111}{\emph{JHEP} {\bfseries 01}
  (2019) 111} [\href{https://arxiv.org/abs/1810.08645}{{\ttfamily
  1810.08645}}].

\bibitem{Kang:2020yqw}
Z.-B.~Kang, D.Y.~Shao and F.~Zhao, \emph{{QCD resummation on single hadron
  transverse momentum distribution with the thrust axis}},
  \href{https://doi.org/10.1007/JHEP12(2020)127}{\emph{JHEP} {\bfseries 12}
  (2020) 127} [\href{https://arxiv.org/abs/2007.14425}{{\ttfamily
  2007.14425}}].

\bibitem{Boglione:2020auc}
M.~Boglione and A.~Simonelli, \emph{{Factorization of $e^+e^- \to H \, X$ cross
  section, differential in $z_h$, $P_T$ and thrust, in the $2$-jet limit}},
  \href{https://doi.org/10.1007/JHEP02(2021)076}{\emph{JHEP} {\bfseries 02}
  (2021) 076} [\href{https://arxiv.org/abs/2011.07366}{{\ttfamily
  2011.07366}}].

\bibitem{Boglione:2020cwn}
M.~Boglione and A.~Simonelli, \emph{{Universality-breaking effects in $e^+e^-$
  hadronic production processes}},
  \href{https://doi.org/10.1140/epjc/s10052-020-08821-y}{\emph{Eur. Phys. J. C}
  {\bfseries 81} (2021) 96} [\href{https://arxiv.org/abs/2007.13674}{{\ttfamily
  2007.13674}}].

\bibitem{Guan:2018ckx}
{\scshape Belle} collaboration, \emph{{Observation of Transverse
  $\Lambda/\bar{\Lambda}$ Hyperon Polarization in $e^+e^-$ Annihilation at
  Belle}}, \href{https://doi.org/10.1103/PhysRevLett.122.042001}{\emph{Phys.
  Rev. Lett.} {\bfseries 122} (2019) 042001}
  [\href{https://arxiv.org/abs/1808.05000}{{\ttfamily 1808.05000}}].

\bibitem{DAlesio:2020wjq}
U.~D'Alesio, F.~Murgia and M.~Zaccheddu, \emph{{First extraction of the
  $\Lambda$ polarizing fragmentation function from Belle $e^+e^-$ data}},
  \href{https://doi.org/10.1103/PhysRevD.102.054001}{\emph{Phys. Rev. D}
  {\bfseries 102} (2020) 054001}
  [\href{https://arxiv.org/abs/2003.01128}{{\ttfamily 2003.01128}}].

\bibitem{Callos:2020qtu}
D.~Callos, Z.-B.~Kang and J.~Terry, \emph{{Extracting the transverse momentum
  dependent polarizing fragmentation functions}},
  \href{https://doi.org/10.1103/PhysRevD.102.096007}{\emph{Phys. Rev. D}
  {\bfseries 102} (2020) 096007}
  [\href{https://arxiv.org/abs/2003.04828}{{\ttfamily 2003.04828}}].

\bibitem{Anselmino:2007fs}
M.~Anselmino, M.~Boglione, U.~D'Alesio, A.~Kotzinian, F.~Murgia, A.~Prokudin and C.~T\"urk, \emph{{Transversity and Collins functions from SIDIS and
  $e^+ e^-$ data}}, \href{https://doi.org/10.1103/PhysRevD.75.054032}{\emph{Phys.
  Rev.} {\bfseries D75} (2007) 054032}
  [\href{https://arxiv.org/abs/hep-ph/0701006}{{\ttfamily hep-ph/0701006}}].

\bibitem{Bacchetta:2004jz}
A.~Bacchetta, U.~D'Alesio, M.~Diehl and C.A.~Miller, \emph{Single-spin
  asymmetries: The {T}rento conventions}, {\emph{Phys. Rev.} {\bfseries D70}
  (2004) 117504} [\href{https://arxiv.org/abs/hep-ph/0410050}{{\ttfamily
  hep-ph/0410050}}].

\bibitem{Boer:1997nt}
D.~Boer and P.~Mulders, \emph{{Time reversal odd distribution functions in
  leptoproduction}},
  \href{https://doi.org/10.1103/PhysRevD.57.5780}{\emph{Phys. Rev. D}
  {\bfseries 57} (1998) 5780}
  [\href{https://arxiv.org/abs/hep-ph/9711485}{{\ttfamily hep-ph/9711485}}].

\bibitem{Anselmino:2000vs}
M.~Anselmino, D.~Boer, U.~D'Alesio and F.~Murgia, \emph{{$\Lambda$ polarization
  from unpolarized quark fragmentation}},
  \href{https://doi.org/10.1103/PhysRevD.63.054029}{\emph{Phys. Rev. D}
  {\bfseries 63} (2001) 054029}
  [\href{https://arxiv.org/abs/hep-ph/0008186}{{\ttfamily hep-ph/0008186}}].

\bibitem{Anselmino:2001js}
M.~Anselmino, D.~Boer, U.~D'Alesio and F.~Murgia, \emph{{Transverse $\Lambda$ 
  polarization in semiinclusive DIS}},
  \href{https://doi.org/10.1103/PhysRevD.65.114014}{\emph{Phys. Rev. D}
  {\bfseries 65} (2002) 114014}
  [\href{https://arxiv.org/abs/hep-ph/0109186}{{\ttfamily hep-ph/0109186}}].

\bibitem{Leader:2001gr}
E.~Leader, \emph{{Spin in particle physics}}, {\emph{Camb. Monogr. Part. Phys.
  Nucl. Phys. Cosmol.} {\bfseries 15} (2001) 1}.

\end{thebibliography}

\providecommand{\href}[2]{#2}\begingroup\raggedright\endgroup

\end{document}